\documentclass[namedreferences]{solarphysics}
\usepackage[hyperref,optionalrh,solaromanenum]{spr-sola-addons} % For Solar Physics 
\usepackage{graphicx}                    % For eps figures, newer & more powerfull
\usepackage{color}                       % For color text: \color command
\usepackage{breakurl}                         % For breaking URLs easily trough lines
\newcommand{\kms}{$\mathrm{km}\,\mathrm{s}^{-1}$}
\newcommand{\cm}{$\mathrm{cm}^{-3}$}
\newcommand{\mathd}{\mathrm{d}}
\newcommand{\degr}{^\circ}

\begin{document}

\begin{article}

\begin{opening}

\title{Reconstruction of Helio-latitudinal Structure of the Solar Wind Proton Speed and Density}

%%%%%%%%%%%%%%%%%%%%%%%%%%%%%%%%%%%%%%%%%%%%%%%%%%%
%% Authors Names
%
\author[addressref={aff1},corref,email={jsokol@cbk.waw.pl}]{\inits{J.M.}\fnm{Justyna M.}~\lnm{Sok{\'o}{\l}}}
\author[addressref={aff1}]{\inits{P.}\fnm{Pawe{\l}}~\lnm{Swaczyna}}
\author[addressref={aff1}]{\inits{M.}\fnm{Maciej}~\lnm{Bzowski}}
\author[addressref={aff2}]{\inits{M.}\fnm{Munetoshi}~\lnm{Tokumaru}}

%%%%%%%%%%%%%%%%%%%%%%%%%%%%%%%%%%%%%%%%%%%%%%%%%%%
%% Runningheads
%
\runningauthor{Sok{\'o}{\l} et al.}
\runningtitle{Reconstruction of helio-latitudinal structure of the solar wind}

%%%%%%%%%%%%%%%%%%%%%%%%%%%%%%%%%%%%%%%%%%%%%%%%%%%
%% Affilations 
%% id should be the same with \author addressref value.
\address[id={aff1}]{Space Research Centre of the Polish Academy of Sciences, Warsaw, Poland}
\address[id={aff2}]{Solar-Terrestrial Environment Laboratory, Nagoya University, Nagoya, Japan}

%%%%%%%%%%%%%%%%%%%%%%%%%%%%%%%%%%%%%%%%%%%%%%%%%%%
%%% Abstract 
\begin{abstract}
The modeling of the heliosphere requires continuous three-dimensional solar wind data. The in-situ out-of-ecliptic measurements are very rare, so that other methods of solar wind detection are needed. We use the remote-sensing data of the solar wind speed from observations of interplanetary scintillation (IPS) to reconstruct spatial and temporal structures of the solar wind proton speed from 1985 to 2013. We developed a method of filling the data gaps in the IPS observations to obtain continuous and homogeneous solar wind speed records. We also present a method to retrieve the solar wind density from the solar wind speed, utilizing the invariance of the solar wind dynamic pressure and energy flux with latitude. To construct the synoptic maps of the solar wind speed we use the decomposition into spherical harmonics of each of the Carrington rotation map. To fill the gaps in time we apply the singular spectrum analysis to the time series of the coefficients of spherical harmonics. We obtained helio-latitudinal profiles of the solar wind proton speed and density over almost three recent solar cycles. The accuracy in the reconstruction is, due to computational limitations, about $20\%$. The proposed methods allow us to improve the spatial and temporal resolution of the model of the solar wind parameters presented in our previous paper (Sok\'o\l \,{\it et al.}, {\it Solar Phys.} {\bf 285}, 167, 2013) and give a better insight into the time variations of the solar wind structure. Additionally, the solar wind density is reconstructed more accurately and it fits better to the in-situ measurements from \textit{Ulysses}. 
\end{abstract}

%%%%%%%%%%%%%%%%%%%%%%%%%%%%%%%%%%%%%%%%%%%%%%%%%%%
%% Keywords
%
\keywords{Radio scintillation, interplanetary $\cdot$ Solar cycle $\cdot$ Solar wind $\cdot$ Sun: heliosphere}

\end{opening}
%-------------------------------------------------

%%%%%%%%%%%%%%%%%%%%%%%%%%%%%%%%%%%%%%%%%%%%%%%%%%%
%% Sections
%
\section{Introduction}
\label{secIntroduction}
The heliosphere is a cavity in the local interstellar medium created by the solar wind (SW thereafter), the plasma that originates from the Sun. For modeling the structure of the heliosphere and ionization conditions of the interstellar neutral gas inside the heliosphere, a three-dimensional (3D) structure of the SW parameters has to be known. The required dimensions are: time, latitude, and distance to the Sun. In-situ measurements of the SW are available from the mid-1960s, but mostly in the ecliptic plane and at 1~AU \citep[for recent review, see][]{bzowski_etal:13a}. In-situ measurements of the SW out of the ecliptic plane were only carried out by \textit{Ulysses} from 1990 to 2009 \citep[\textit{e.g.}][]{bame_etal:92a, mccomas_etal:00a, mccomas_etal:00b}. However, they were point measurements, and a time series of these observations is in reality a convolution of variabilities in time, distance, and helio-latitude. The information on continuous variations in the global (particularly spatial) structure of the SW is still missing in the in-situ measurements. On the other hand, the SW is also observed by remote-sensing methods, by interplanetary scintillation \citep[IPS thereafter; \textit{e.g.}][]{hewish_etal:64a, houminer:71a, coles_etal:80a, kojima_etal:04b} and by Lyman-$\alpha$ helioglow \citep[\textit{e.g.}][]{lallement_etal:85a, bzowski_etal:03a} observed by the \textit{Solar Wind ANisotropy} (SWAN) instrument onboard the \emph{SOlar and Heliospheric Observatory} (SOHO). 

\begin{figure}
	\centering
	\resizebox{\hsize}{!}{\includegraphics[angle=270]{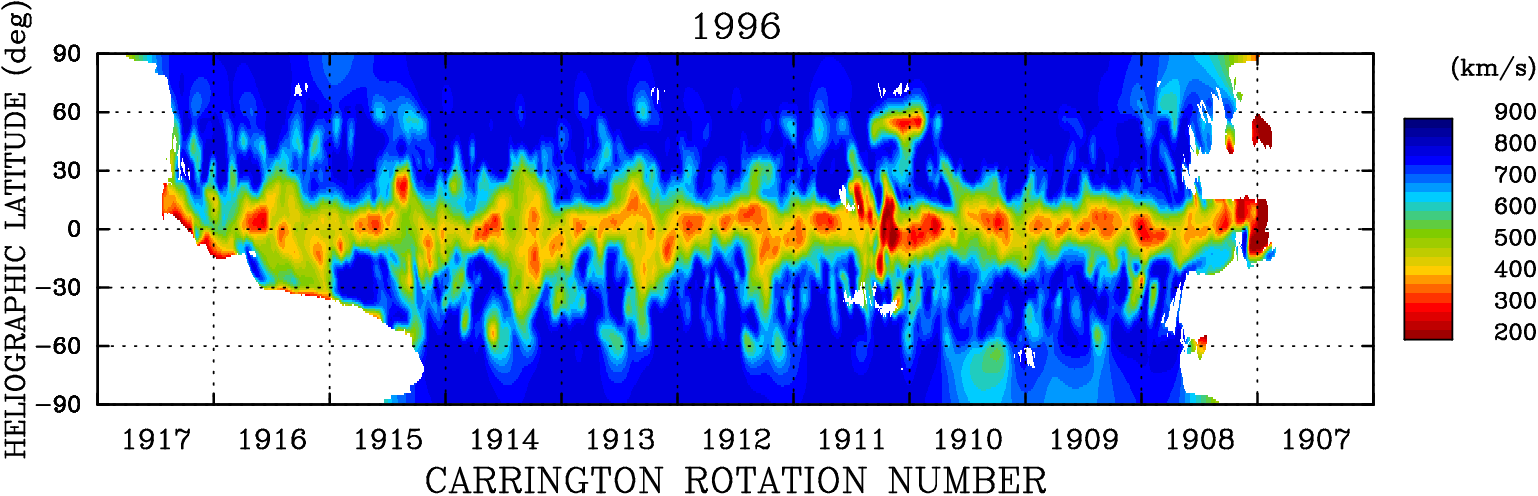}}
	\caption{Example annual map of the SW speed obtained from the IPS observations of STEL (\texttt{ftp://ftp.stelab.nagoya-u.ac.jp/pub/vlist/map/})}
	\label{figVmapAnnual1996}
	\end{figure}

\citet{sokol_etal:13a} used \textit{Ulysses} measurements, the OMNI in-ecliptic SW database \citep{king_papitashvili:05}, and the SW speed derived from IPS observations conducted at the Solar-Terrestrial Environment Laboratory (STEL), Nagoya University \citep{tokumaru_etal:10a} to reconstruct the SW speed and density evolution from 1990 to 2011. They constructed SW structures in a grid of 1 year spacing in time and $10\degr$ in helio-latitude. The choice of the grid steps was based on the available time coverage of the SW speed obtained from the IPS data, which have systematic yearly breaks, typically from December to March each year \citep{tokumaru_etal:10a}. Moreover, all helio-latitudes are not continuously observed; see the example SW speed map from IPS data in Figure~\ref{figVmapAnnual1996}. A more detailed description of the SW speed derived from IPS observations is given in Section~\ref{swSpeedIPS}. 

In this paper we eliminate the limits in the analysis of \citet{sokol_etal:13a} and fill the gaps and breaks in the SW speed data from IPS observations to obtain maps of the SW speed that are continuous in time and helio-latitude. We propose a method of filling the temporal and spatial gaps by adopting the techniques used with success in the analysis of geophysical, solar, and in-ecliptic SW data \citep[\textit{e.g.}][]{kondrashov_ghil:06a, kondrashov_etal:10a, dudokdewit:11b, kondrashov_etal:14a}. We decompose the SW speed maps from IPS observations into a set of spherical harmonics and perform a singular spectrum analysis (SSA; based on \opencite{ghil_etal:02a}) of the time series of the coefficients of spherical harmonics to fill the data missing in time.
	
\citet{sokol_etal:13a} proposed a method to calculate the SW proton density based on the SW speed using an empirical relation between the proton speed and density retrieved from \textit{Ulysses} measurements. We revise this method by comparing its results with the SW proton density calculated using the SW helio-latitudinal invariants, which are the SW dynamic pressure \citep{mccomas_etal:08a} and the SW energy flux \citep{leChat_etal:12a}.

\section{Solar Wind Speed from IPS}
\label{swSpeedIPS}
IPS is the phenomenon due to diffraction patterns on an observer's plane produced by interference of radio waves coming from a remote compact radio source (\textit{e.g.} quasar), scattered on electron density irregularities (fluctuations) in the SW (\textit{e.g.}~\opencite{hewish_etal:64a, houminer:71a, coles_maagoe:72a, kakinuma:77a, coles_kaufman:78a, coles_etal:80a, kojima_kakinuma:90a};
\citealp{jackson_etal:97a}, \citeyear{jackson_etal:98a}, \citeyear{jackson_etal:03a}; \opencite{tokumaru_etal:10a}). IPS observations are ground-based, line-of-sight (LOS) integrated measurements, which limit the availability of the data in the case of adverse weather conditions or insufficient elevation of the Sun above the horizon. 

In our analysis we use the SW speed data derived from the IPS observations of STEL \citep{tokumaru_etal:10a}. The IPS data were deconvolved using the computer-assisted tomography method (CAT; \opencite{asai_etal:98a, jackson_etal:98a, kojima_etal:98a}) to provide the reconstructed SW speed. This analysis provides Carrington rotation\footnote{The Carrington rotation period is the period of one sidereal solar rotation, 27.2753 days.} (CR) maps of the SW speed, which are available since the 1980s. In this method, spatial and temporal gaps are inevitable in the derived SW maps. Periodic gaps appear as a break in the SW speed because all or a portion of the STEL IPS system is closed during the winter months. Additionally, spatial gaps are present due to the reduction in valid observations at higher helio-latitudes, particularly in southern polar regions in winter. Figure~\ref{figMapSpeedStart} shows the time breaks from 1985 to 2013, and the left-hand panels in Figure~\ref{figMapsRecSphHarm} show spatial gaps in the CR maps. The SW speed retrieved from IPS observations does not exceed 800~\kms, because the SW model used in the CAT analysis has an upper bound at 800~\kms established based on the \textit{Ulysses} measurements of the fast SW. 

The retrieval of the SW speed from the IPS observations is possible owing to a relation between the electron density fluctuation level and the SW speed as a function of heliocentric distance \citep{asai_etal:98a} as
		\begin{equation}
		\Delta N_{\mathrm{e}}\left(r\right) \varpropto r^{-2}V^{-\gamma \left( r \right)},
		\label{eqSpeedRetrieve}
		\end{equation}
where $r$ is the heliocentric distance, $N_{\mathrm{e}}$ is the electron density, $V$ is the SW speed, and index $\gamma \left( r \right)$ varies with solar distance and must be determined experimentally \citep[\textit{e.g.}][]{coles_etal:95a, manoharan:93b, tokumaru_etal:12b}. The relation used to connect the fluctuation level with the SW speed had been established before the secular changes in the SW data were observed both in the ecliptic plane and out of the ecliptic by \textit{Ulysses} (see the discussion in \opencite{sokol_etal:13a}). This relation may be a possible source of systematic bias in the reconstruction of the SW speed.

The relation between $\Delta N_{\mathrm{e}}$ and $V$ was used to analyze the IPS data before 1997 using the CAT method. After 1997 the quantity called the $g$-value \citep{gapper_etal:82a} was introduced to the IPS analysis performed by STEL \citep{tokumaru_etal:00a}. It represents the relative variation of the scintillation strength with respect to the mean level: $g={\Delta I}/{\overline{\Delta I\left( R \right)}}$, where $\Delta I$ and $\overline{\Delta I\left( R \right)}$ are the observed instantaneous and average scintillation levels, respectively. $\Delta I$ is computed from the observed power spectrum of the fluctuations $P\left( f \right)$ as 
	\begin{displaymath}
	\Delta I = \int\limits_{0}^{\infty} P\left( f \right) \mathrm{d}f.
	\end{displaymath}
$\overline{\Delta I\left( R \right)}=aR^{-1.5}$ is a function of the solar offset distance $R$, by assuming a spherically symmetric distribution of the SW density fluctuations. The value of $a$ is derived by fitting the observations of $\Delta I$ at low latitudes \citep[see more details in][]{tokumaru_etal:12b}. Since 1997, therefore, two alternative reconstructions of the SW speed have been available from STEL: one obtained using the $g$-value and $\Delta N_{\mathrm{e}}$, and the other one using only the relation of Equation~(\ref{eqSpeedRetrieve}). Some revision of the $\Delta N_{\mathrm{e}} \sim V$ relation for the inner heliosphere was published \textit{e.g.} by \citet{hick_jackson:04a} and \citet{jackson_etal:10a}; they also found little variation in the level of $g$-values with solar distance beyond that imposed by an $R^{-2.0}$ radial $\Delta N_{\mathrm{e}}$ fall-off.

As shown in \citet{asai_etal:98a} and \citet{tokumaru_etal:12b}, the SW speed derived from the relation $\Delta N_{\mathrm{e}} \sim V^{-0.5}$ fits well the results of analysis with additional $g$-value data, except for the years of solar maximum. Therefore, the results from the CAT analysis using only the $\Delta N_{\mathrm{e}}$ data can be used to discuss long-term changes in the SW over solar cycles. 

In this paper we focus only on the SW speed data retrieved from the IPS observations of STEL. In Section~\ref{secSWDens} we present a simple method of calculation of the SW density based on the SW speed and helio-latitudinal invariants.

\section{Methods}
The IPS data are extensively used \citep[\textit{e.g.} recent works by][]{fujiki_etal:14a, kim_etal:14a, jackson_etal:15a} and widely incorporated into the MHD codes for the modeling of the near-Sun environment as well as the far regions of the heliosphere. In our analysis we want to improve the simple model proposed by \citet{sokol_etal:13a}, which reconstructed the structure of the SW proton speed and density at 1~AU averaged on a time scale of $\approx 1$ year, by increasing the temporal and spatial resolution.

As the starting point, we use the SW speed from the CAT analysis of STEL IPS observations, which are yearly grouped CR maps reduced to the distance of $2.5$ solar radii, with a $1\degr$ resolution in longitude and latitude, from 1985 to 2013. There was a one-year break in 2010, when the number of observations was insufficient for a complete analysis. For the period of 1985 to 2007 we adopt the maps prepared using the CAT analysis based on $\Delta N_{\mathrm{e}} \sim V$ only, and for 2008-2013 we use the data prepared with the additional information of the $g$-value. 

There were 389 CRs from the beginning of 1985 to the end of 2013. During this period we have 298 CR maps of the SW speed from IPS observations, among which 88 have more than 50$\%$ of surface empty. This means that we need to fill the spatial gaps in 210 maps and must fully reconstruct 179 CR maps. All calculations were done using \textit{Mathematica}\footnote{\texttt{http://www.wolfram.com/mathematica/}} (Wolfram Research). The details of the algorithms of the decomposition into spherical harmonics and singular spectrum analysis are described in Appendices~\ref{secAppendixWlm} and \ref{secAppendixSSA}, respectively.

\subsection{Reconstruction of the Helio-latitudinal Structure by Decomposition into Spherical Harmonics}
\label{secMethodReconstruction}
The IPS observations give information about the SW speed which originates from the surface of the Sun. Therefore the spherical frame is the most obvious coordinate system to use in our analysis. In the first step of the process of filling the SW speed structure we decompose the original maps into spherical harmonics which, due to their characteristics of orthogonality on a sphere, are a common tool used in similar studies. Because of the computational limitation (the analysis was performed on a personal computer) we reduce the spatial resolution to $3\degr \times 3\degr$ in longitude and latitude. Next we calculate the fraction of empty cells ($3\degr \times 3\degr$ in size) per map, and pick up the CR maps with less than $50\%$ data missing.

The SW speed maps obtained from the IPS observations were reconstructed by the decomposition into spherical harmonics,
	\begin{equation}
	V\left(\theta,\phi\right) \approx \sum_{\ell}\sum_{m}W_{\ell m} Y_{\ell m}\left(\theta,\phi\right),
	\label{eqDefDecomposition}
	\end{equation}
where the spherical harmonics $Y_{\ell m}(\theta,\phi)$ are defined in Equations~(\ref{eqDefWlm}) and (\ref{eqDefYlm}) as real functions in Appendix~\ref{secAppendixWlm}, $W_{\ell m}$ are the expansion coefficients, $\theta \in \left( 0,\pi\right)$ is the colatitude, and $\phi \in \left( 0, 2\pi \right)$ is the longitude. 

We set the limit $\ell \leq 12$, which allows us to reconstruct only structures larger than approximately $15\degr$. Each CR map of the SW speed was decomposed separately. The reconstruction of all maps for a given year was done with the information from the adjacent CR maps, but without information from the adjacent years. This approach is justified by the distribution of a few month breaks in the IPS data at the end and beginning of subsequent calendar years. Also, it is less probable that the SW source structures in the solar corona last longer than a few full rotations. 
	\begin{figure}
	\centering
	\resizebox{\hsize}{!}{\includegraphics{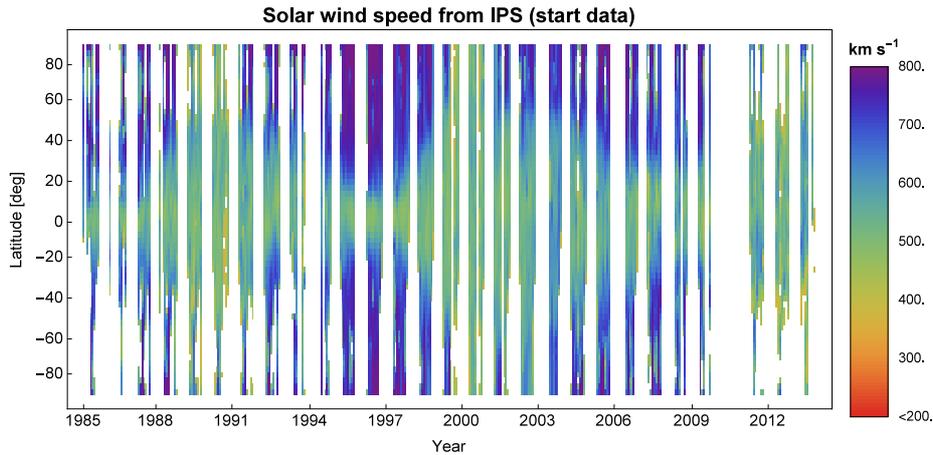}}
	\caption{A visualization of the distribution of spatial and temporal breaks in the SW speed in the original (but monthly averaged) IPS observations.}
	\label{figMapSpeedStart}
	\end{figure}
	
\begin{figure}
		\begin{tabular}{cc}
			\includegraphics[width=.45\textwidth]{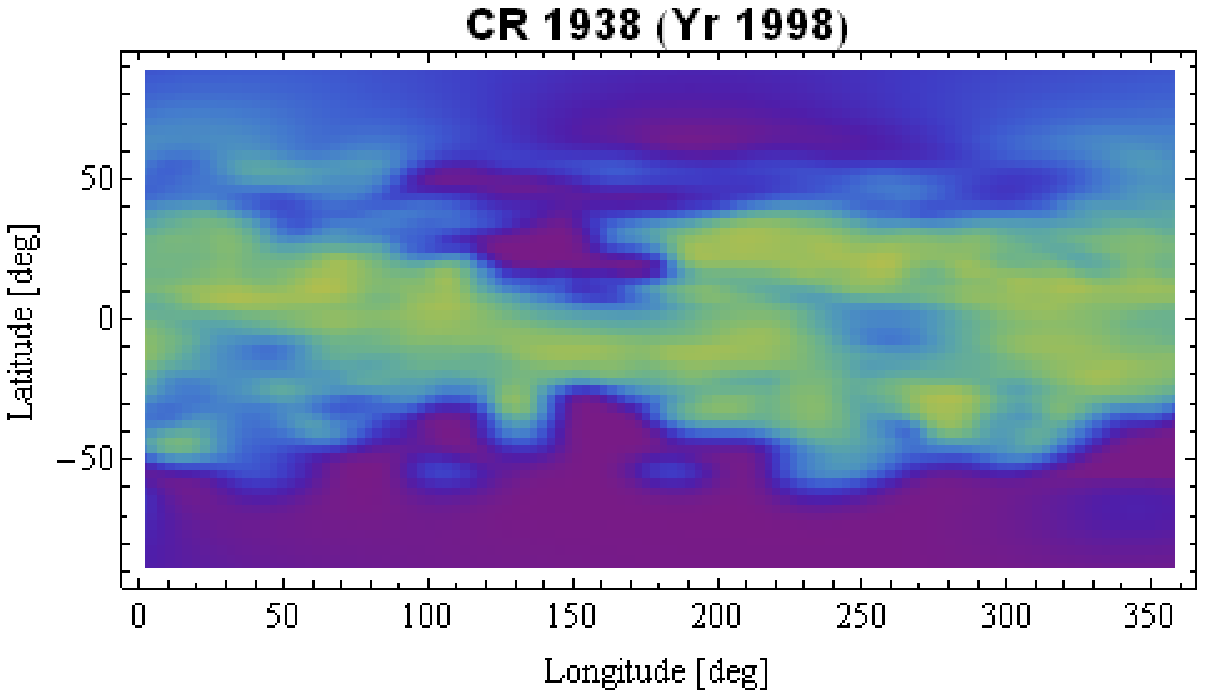} & \includegraphics[width=.45\textwidth]{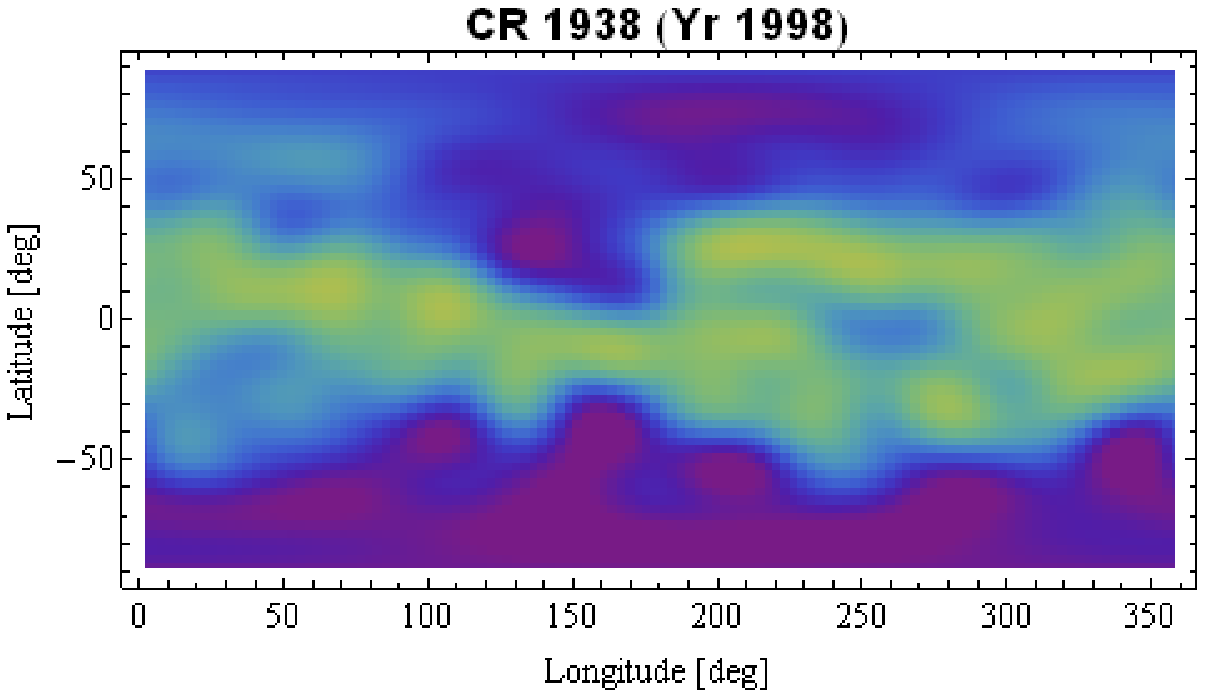}\\
			\includegraphics[width=.45\textwidth]{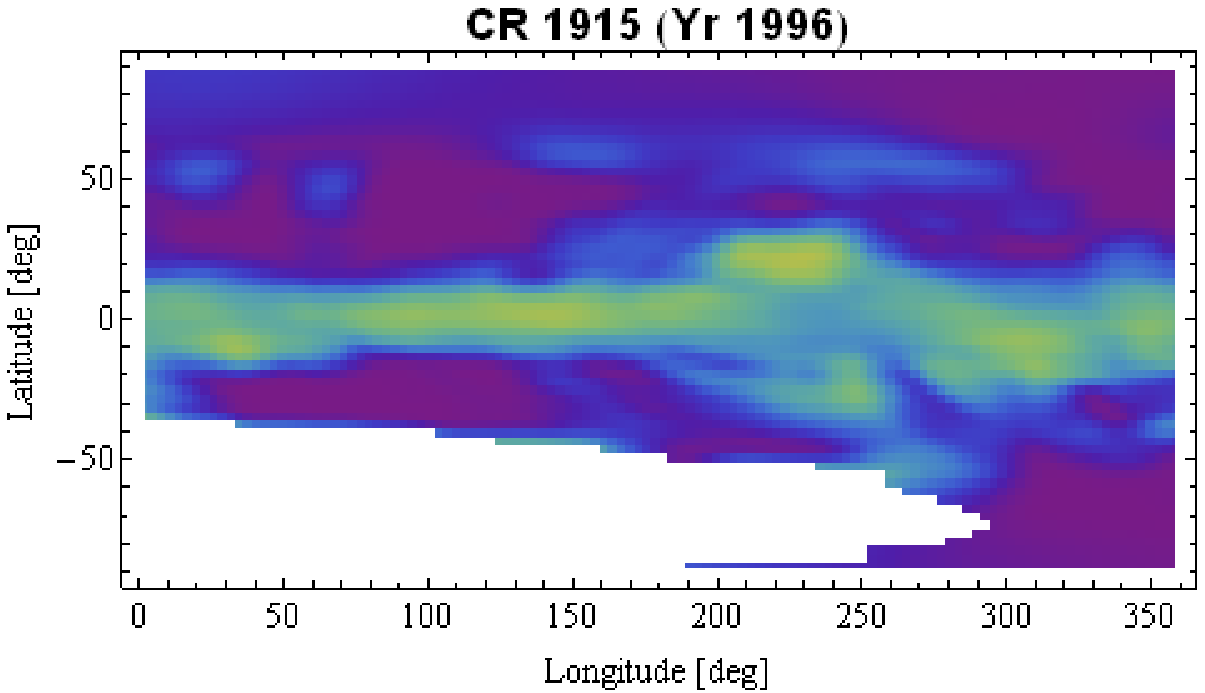} & \includegraphics[width=.45\textwidth]{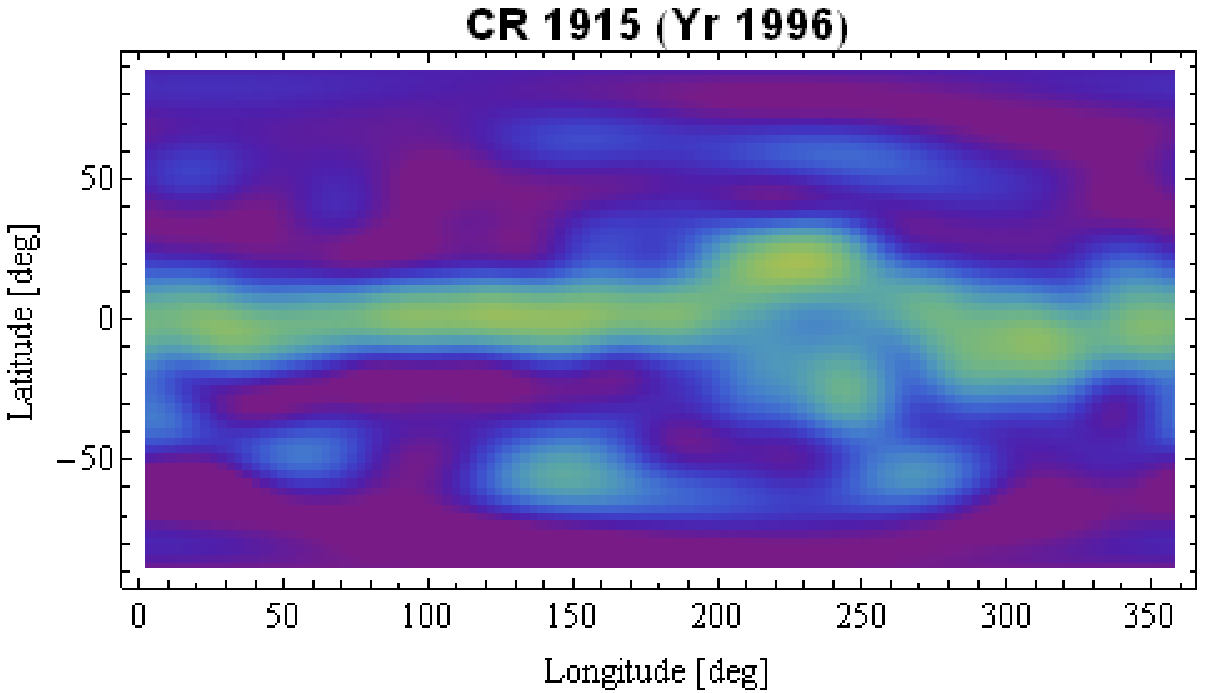}\\
			\includegraphics[width=.45\textwidth]{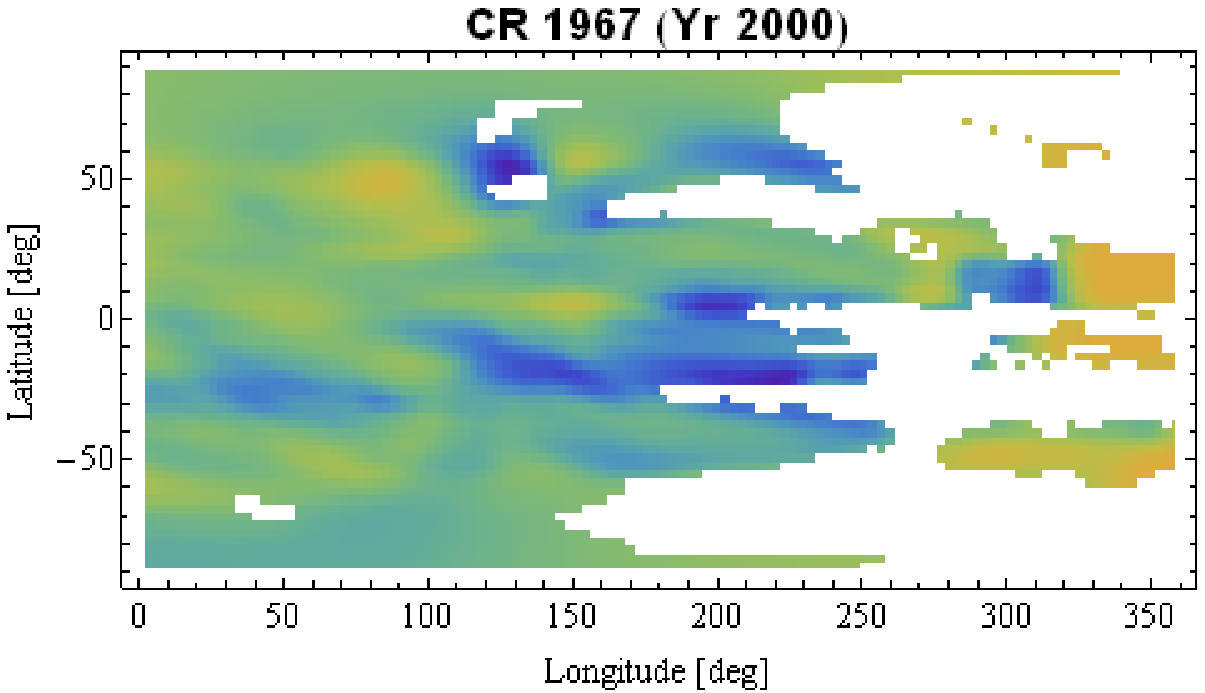} & \includegraphics[width=.45\textwidth]{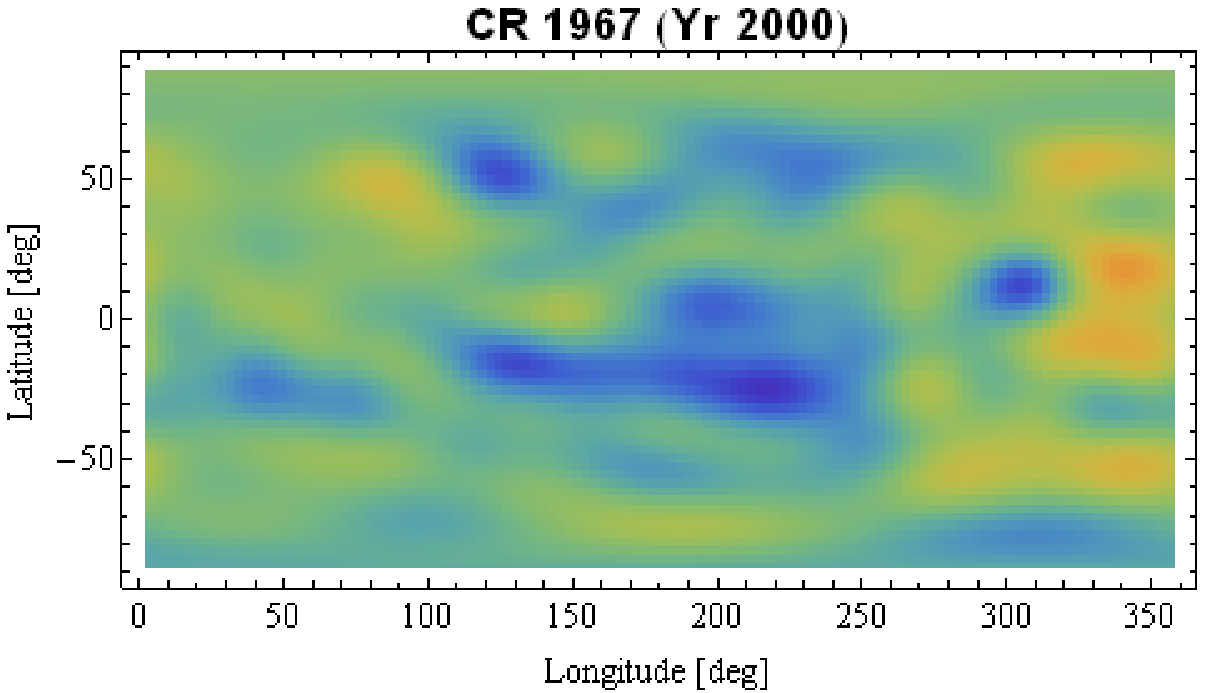}\\
			\includegraphics[width=.45\textwidth]{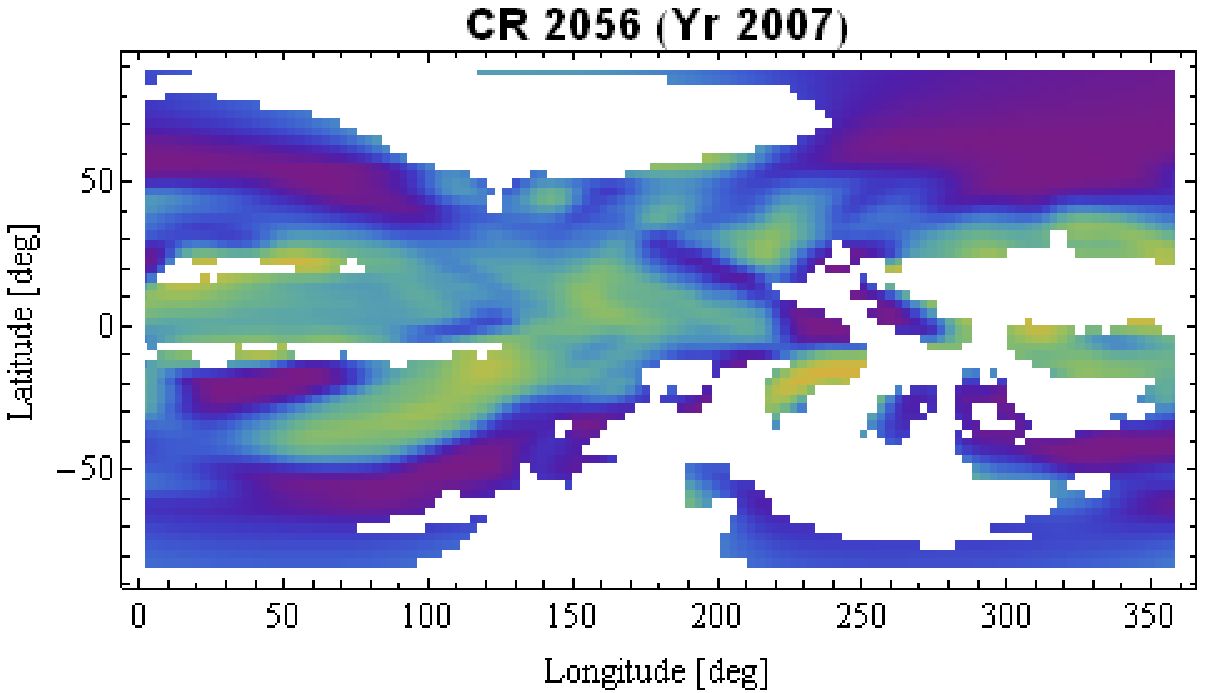} & \includegraphics[width=.45\textwidth]{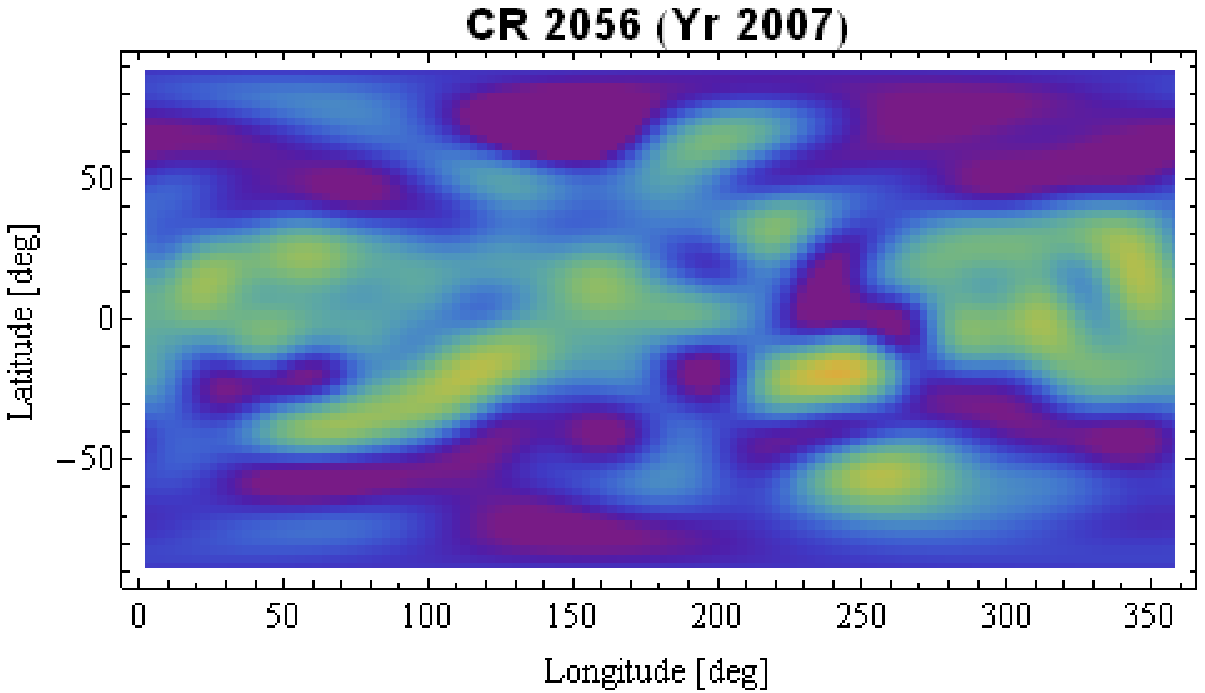}\\
			\multicolumn{2}{c}{\includegraphics[scale=0.45]{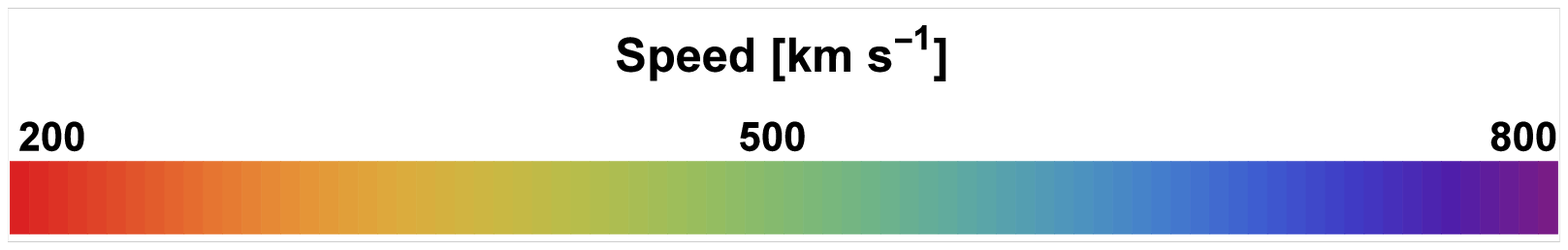}}\\
		\end{tabular}
		\caption{Example results of the reconstruction of CR maps of the SW speed by spherical harmonics. Left panel: original data of the SW speed retrieved by the CAT tomography of STEL IPS observations (resolution of $3\degr$ in longitude and latitude). Right panel: the same resolution maps reconstructed by spherical harmonics with the spatial gaps filled.}
		\label{figMapsRecSphHarm}
	\end{figure}

For each year we select the CR map with the smallest fraction of gaps (CR$_{0}^{\ast}$). The best possible case is the period without gaps in longitude and latitude, \textit{e.g.}, CR~1938 in 1998, shown in the top-left panel of Figure~\ref{figMapsRecSphHarm}. If there is a single map with the minimal fraction of gaps for the given year, then we use it as the start map for the analysis of this year. But if there are multiple maps with high coverage and they are not the consecutive CRs, and additionally each map has at least one non-empty cell in longitude for a given latitudinal $3\degr$-bin, then we set up the initial conditions for the neighboring maps using the $W_{\ell m}$ coefficients retrieved from the nearest start map. For most of the analyzed years, there are two distinct CR maps that can be used to set the initial conditions for the decomposition (\textit{e.g.} CR$_{0}^{1}$ and CR$_{0}^{2}$). When we start the decomposition of the CR maps for a given year, we fill the empty $3\degr \times 3\degr$ cells in the first step. For the start maps with the smallest fraction of gaps, we fill the gaps by an average value for each $3\degr$ latitudinal bin using the information from all longitudes. The gaps in the next neighboring map are filled at the first step of the iterations by using the spherical harmonic decomposition of either the start map for a given year (CR$_{0}^{\ast}$), if it is the nearest, or by the nearest completed map. The same process is applied to all CR-maps for the given year. This method makes effective use of long-lived, large and unambiguous structures at the source surface of the SW. 

We calculate the spherical harmonic coefficients $W_{\ell m}$ as the scalar product of the filled map with spherical harmonics with appropriate weight function on the sphere given by Equation~(\ref{eqDefWlm}) in Appendix~\ref{secAppendixWlm}. For each map the procedure is repeated until the mean relative difference of speed values for the initially empty cells (which were the gaps) stop to change by more than 0.01 or start to increase (see Appendix~\ref{secAppendixWlm} for more details). 

The decomposition into spherical harmonics was carried out to fill the spatial gaps for all the CR maps analyzed in this study. As shown in Figure~\ref{figMapsRecSphHarm}, this method works very well compared to the original data for all kinds of spatial distribution of gaps both during the solar minimum and solar maximum. The left-hand column in Figure~\ref{figMapsRecSphHarm} presents various completeness levels of the original IPS SW speed CR maps (\textit{i.e.}, complete, with gaps only on one hemisphere, with gaps distributed only on one half of the map in longitude, and with a random gap distribution), and the right-hand column illustrates the final reconstruction by decomposition into spherical harmonics.
	
\subsection{Filling in the Temporal Breaks by Singular Spectrum Analysis}
\label{sec:fillingSSA}
Having the $W_{\ell m}$ coefficients for each CR map, we have filled the empty cells in maps, but we still have the breaks in time. As shown in Figure~\ref{figMapSpeedStart}, the time breaks occur mostly at the beginning and end of the year. We have also one longer gap in 2010. We treat the coefficients $W_{\ell m}$ of spherical harmonic decomposition as a time series and we use the singular spectrum analysis (SSA) to fill the time breaks. We only focus on the monthly average \footnote{By monthly we actually mean the average over one CR period.} of the helio-latitudinal profiles of the SW. Therefore, we reproduce the maps that are averaged over longitude, namely we only retain coefficients $W_{\ell m}$ with $m=0$\footnote{Note that for $m\neq 0$, the longitudinal average of the spherical harmonics vanishes.}. 

\begin{figure}
		\centering
		\includegraphics[scale=0.65]{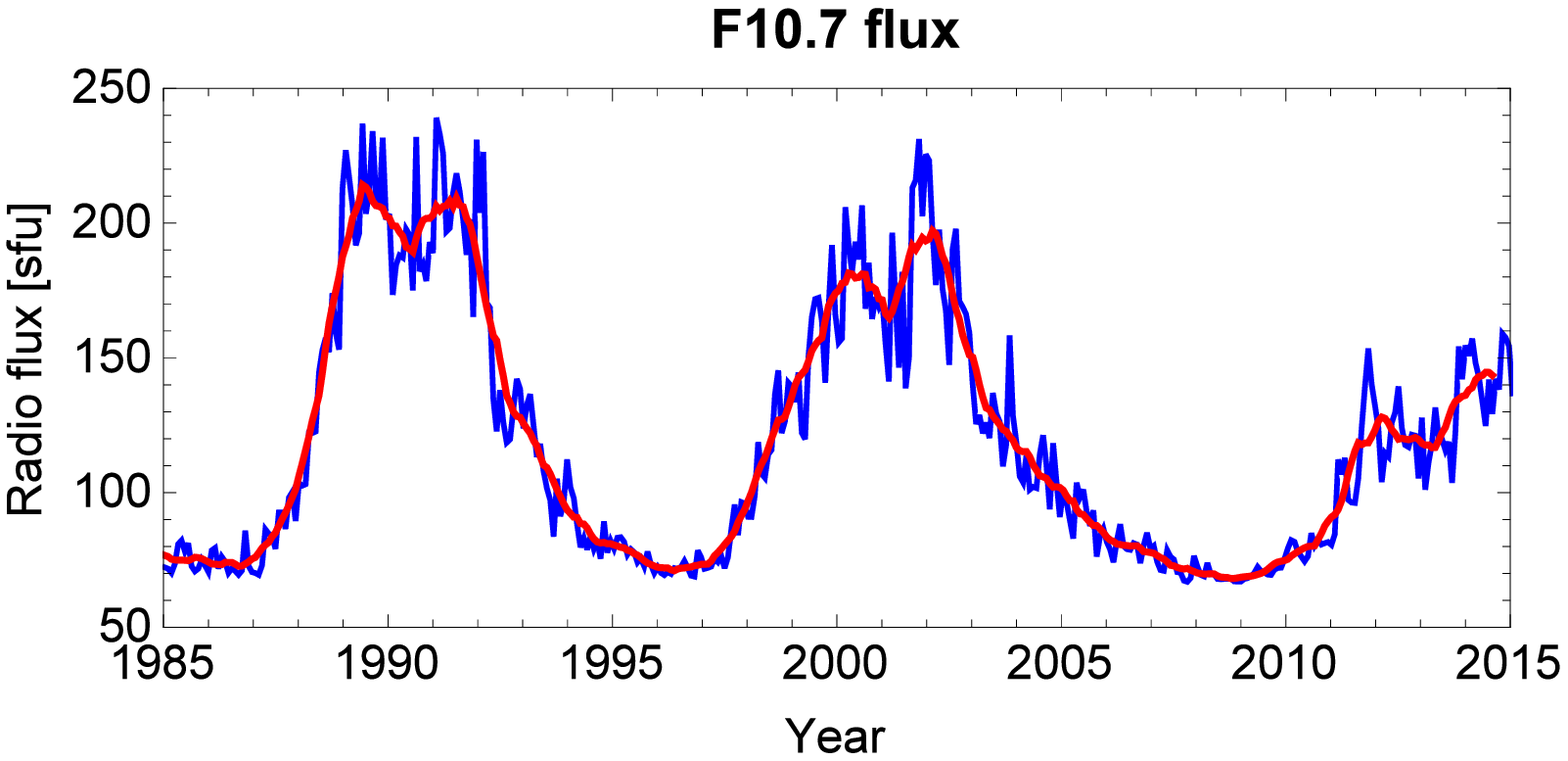}
		\caption{The F10.7 radio flux at 1~AU. Blue: CR-averaged measurements; red: the CR averages smoothed by a 13 CR moving average.}
		\label{figF107}
		\end{figure}

For the SSA to reconstruct the solar cycle (SC) variations of the SW, 2.5~SCs may not be sufficient to provide a valid reconstruction of this periodicity. Thus, we decided to use the solar F10.7 radio flux\footnote{We use the daily noon time series of 10.7~cm radio flux measured by DRAO Penticton and collected by NOAA: \texttt{ftp://ftp.ngdc.noaa.gov/STP/space-weather/solar-data/solar-features/solar-radio\\/noontime-flux/penticton/penticton\_adjusted/listings/}.} (\opencite{tapping:13a}; see Figure~\ref{figF107}) as an indicator of the SC variations because this record is being measured longer than three SCs. We correlate and normalize the $W_{\ell m}$ coefficients with the F10.7 flux, smoothed by a 13-CR running average as illustrated in Figure~\ref{figF107} (see Appendix~\ref{secAppendixSSA} for more details).
		
\begin{figure}
	\centering
		\begin{tabular}{cc}
		\includegraphics[width=.4\textwidth]{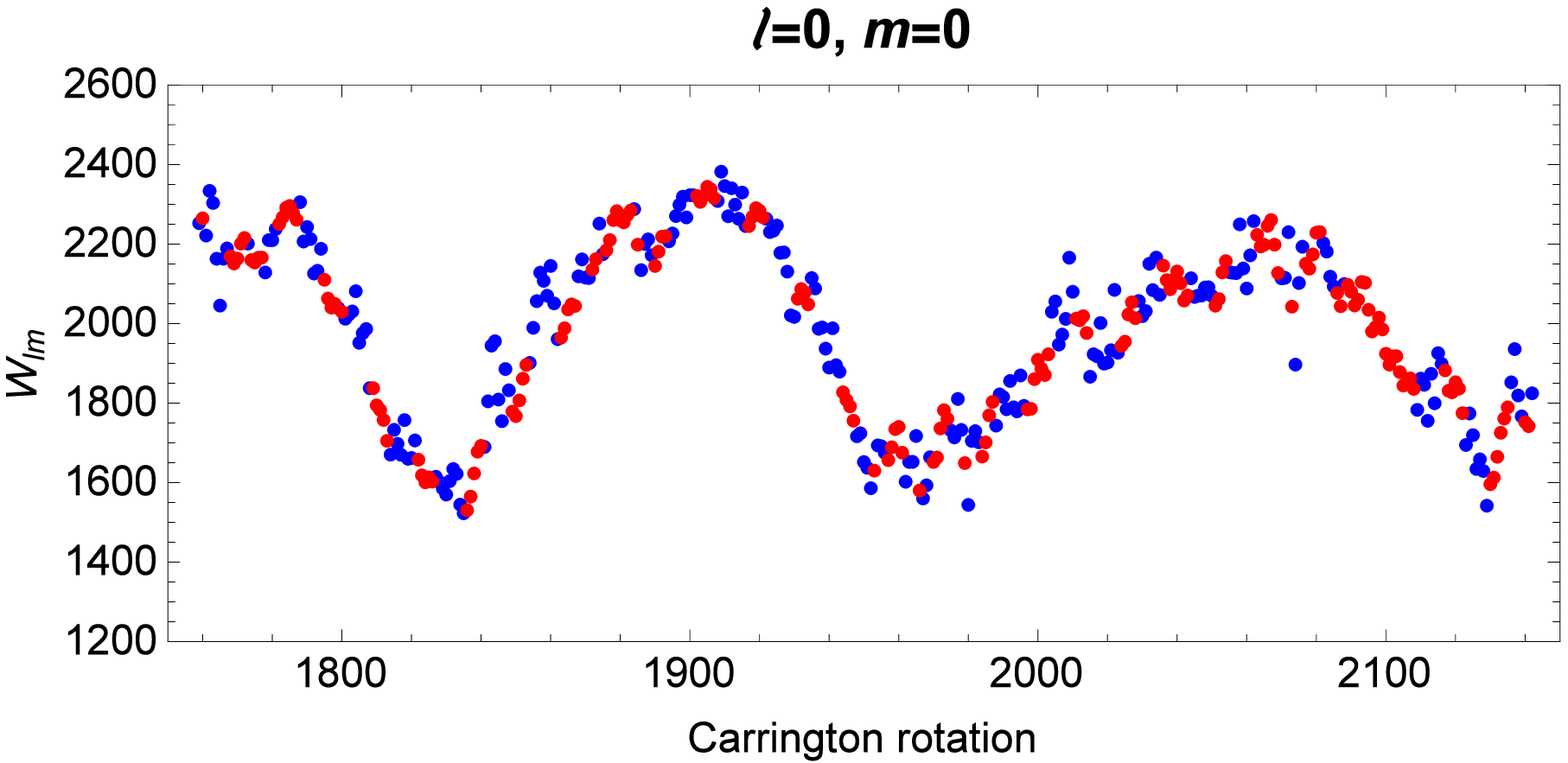} & \includegraphics[width=.4\textwidth]{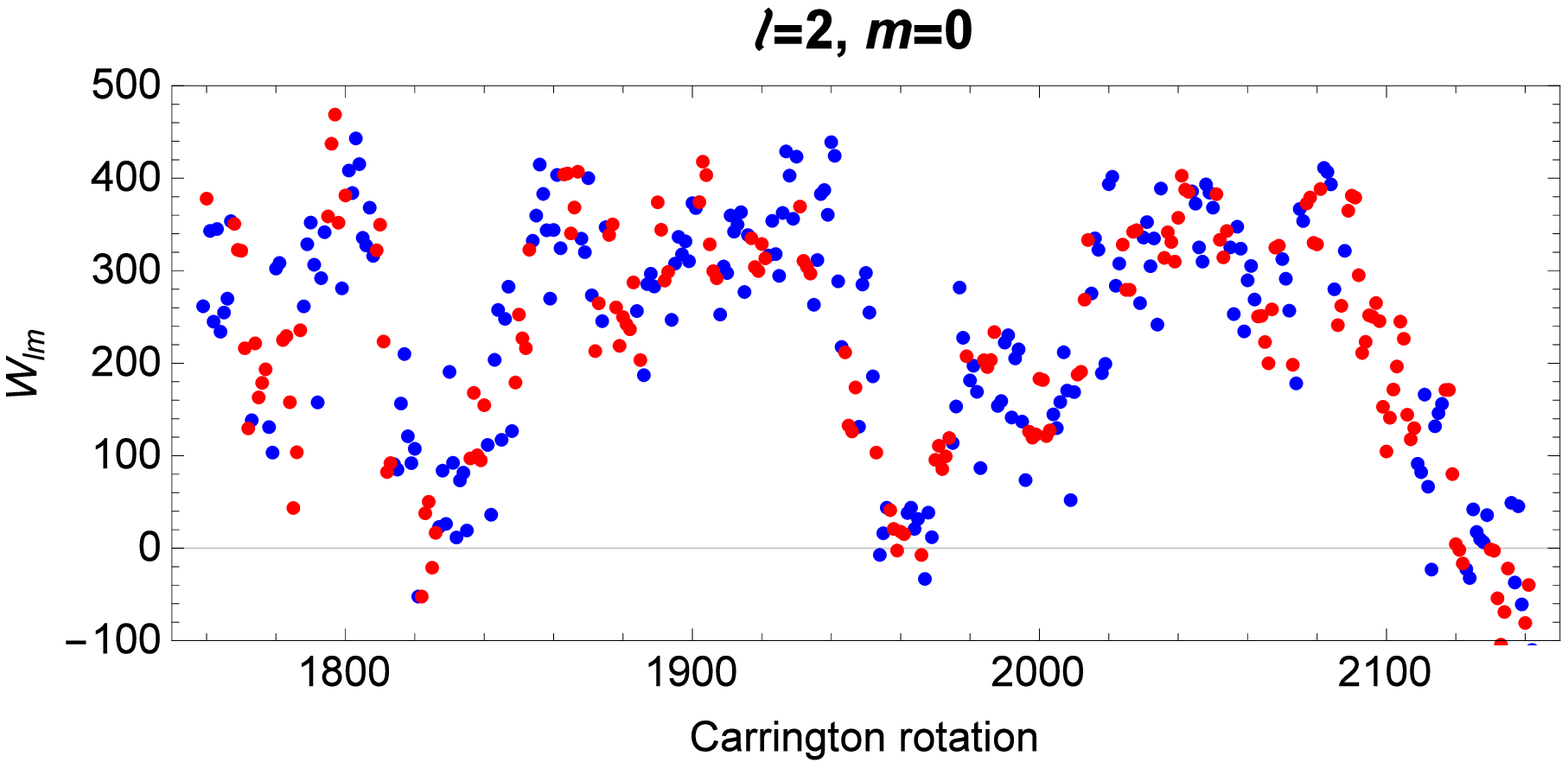}\\
		\includegraphics[width=.4\textwidth]{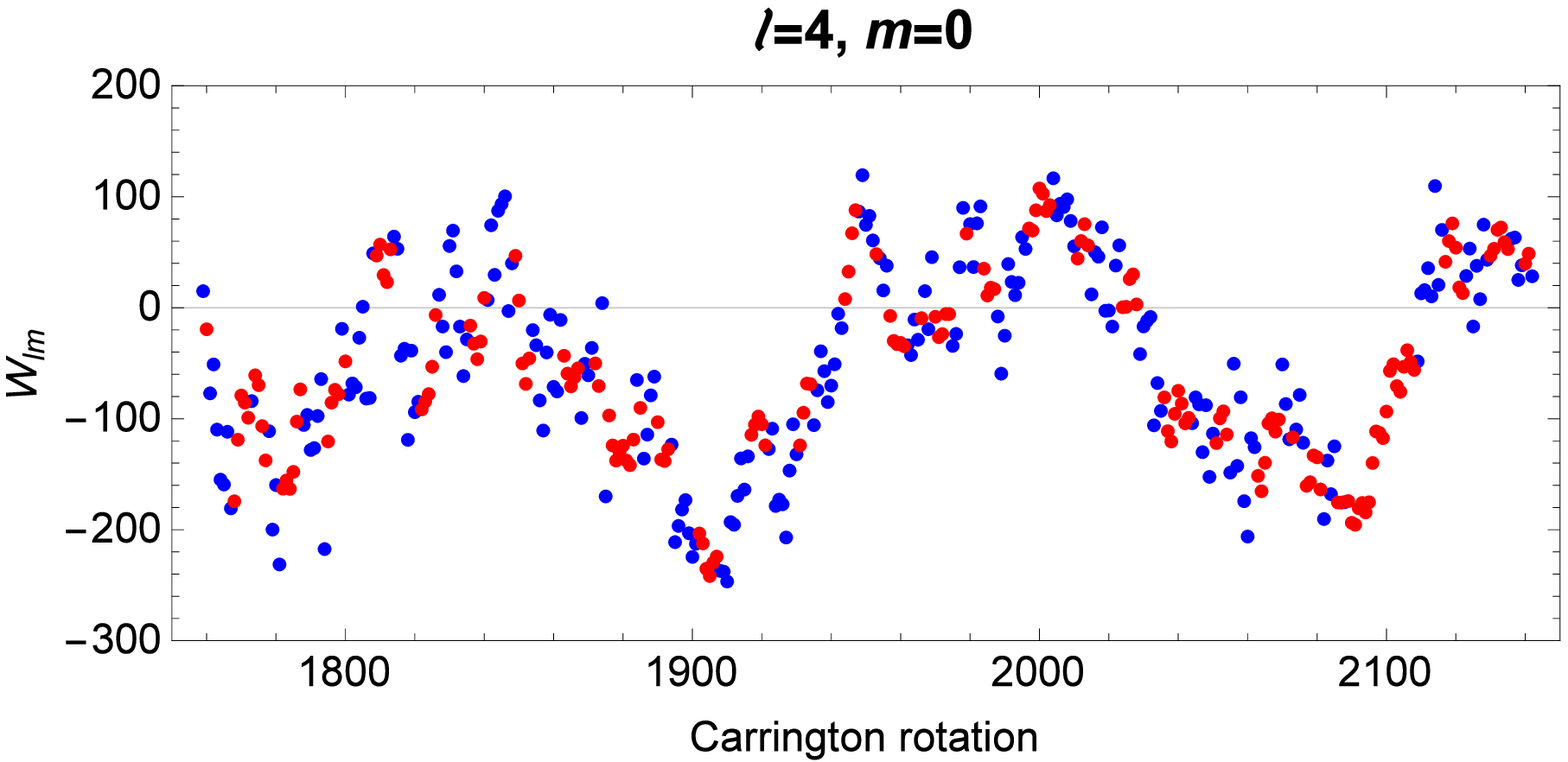} & \includegraphics[width=.4\textwidth]{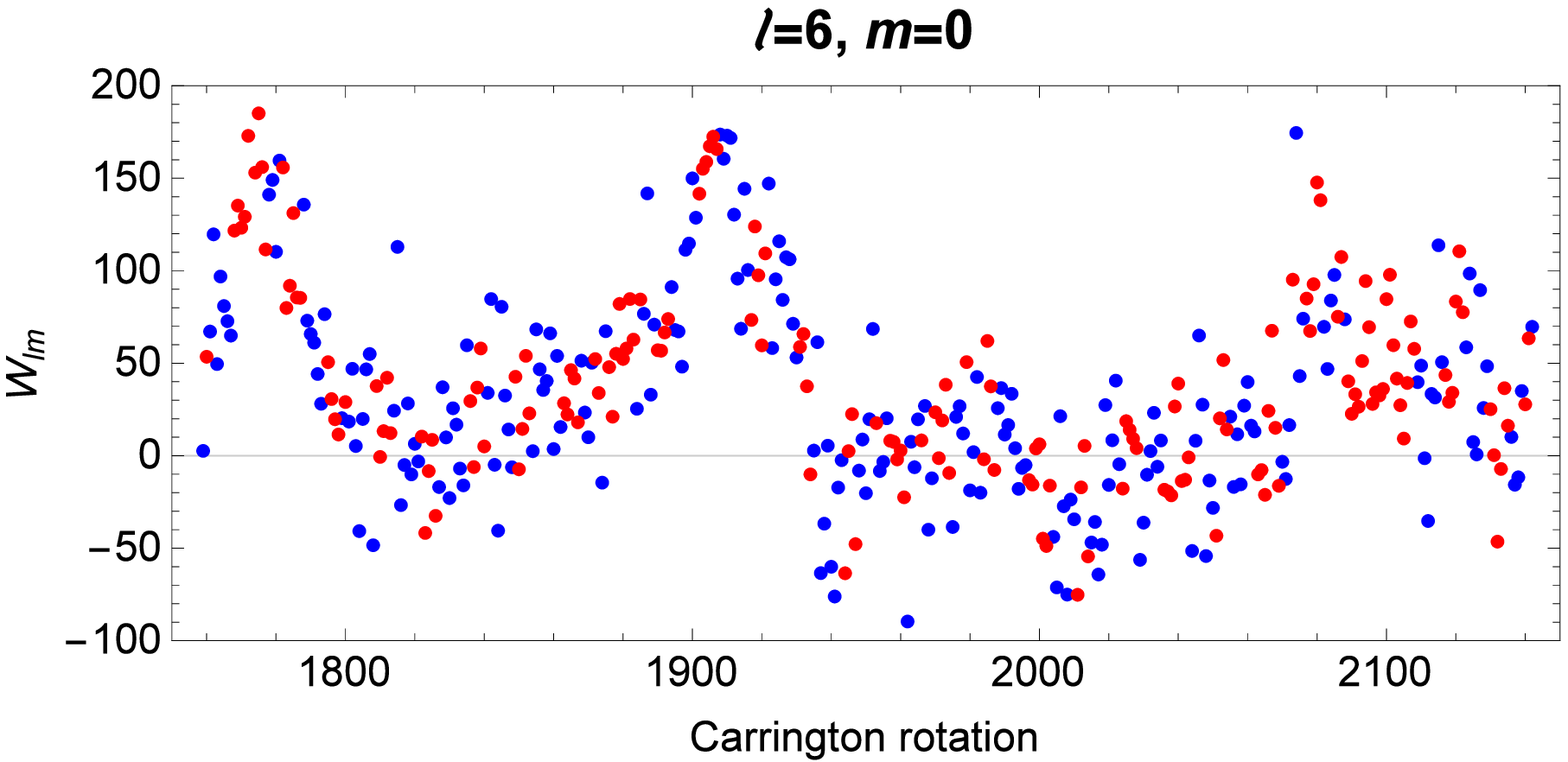}\\
		\includegraphics[width=.4\textwidth]{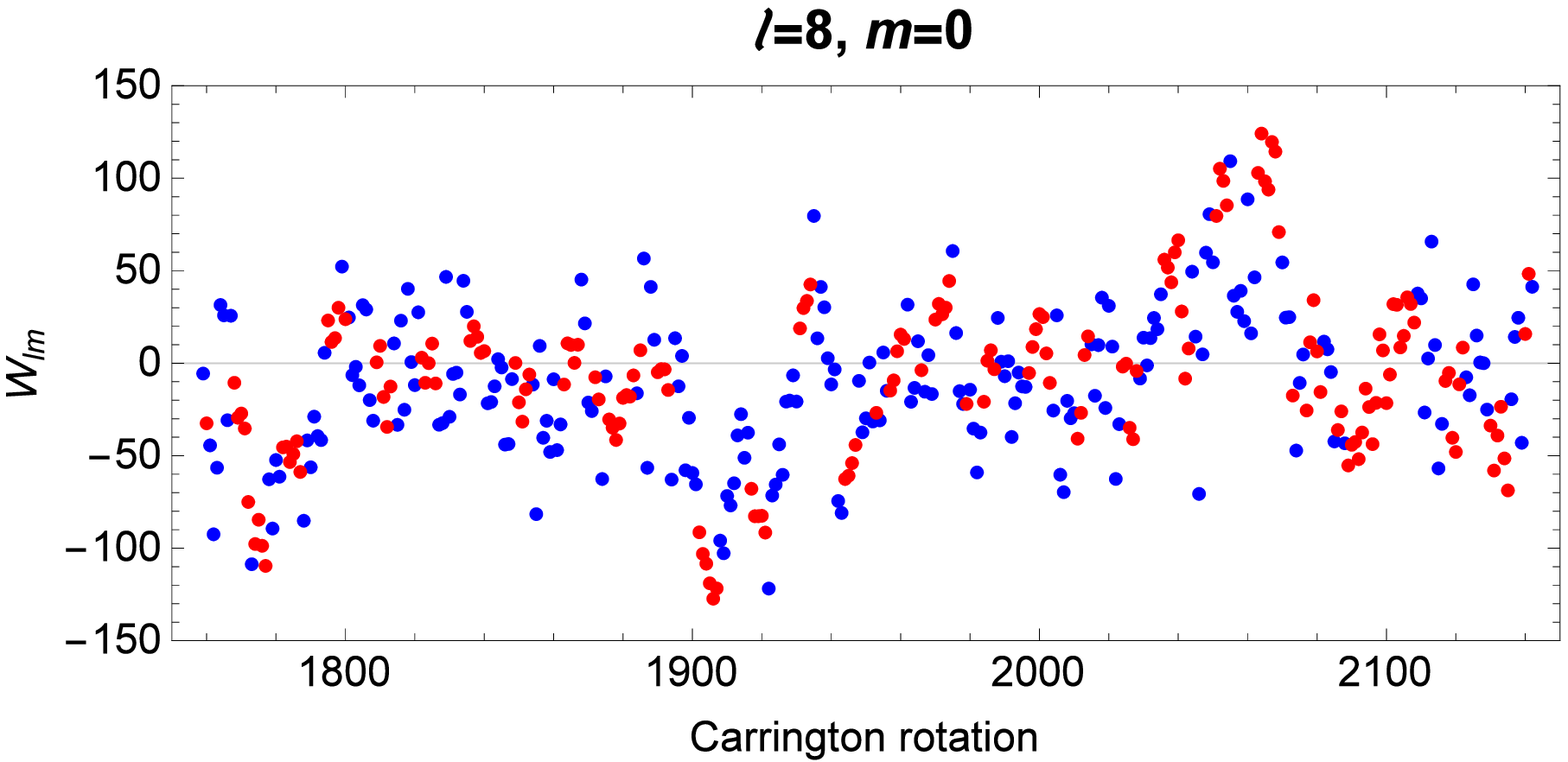} & \includegraphics[width=.4\textwidth]{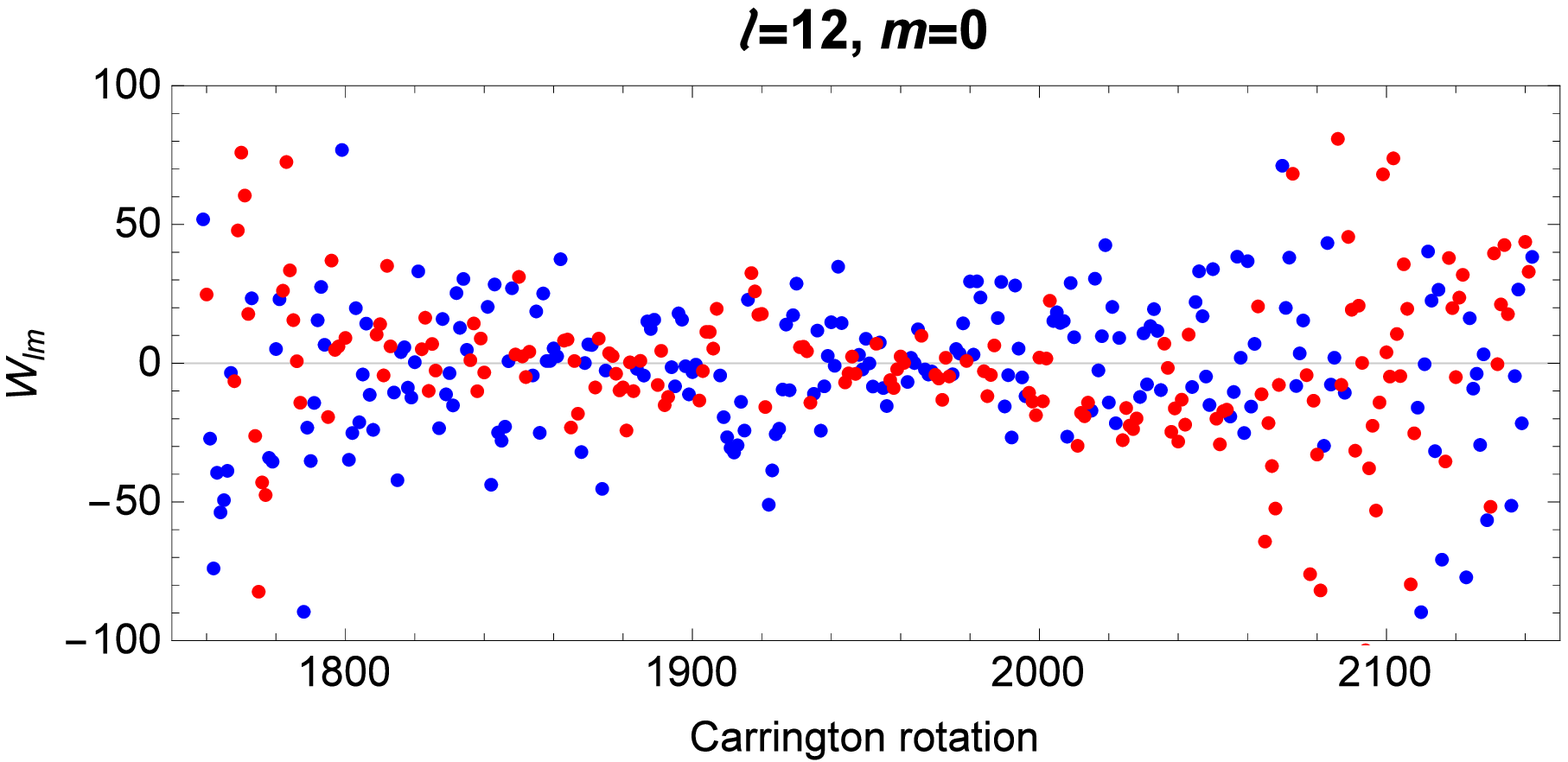}\\
		\end{tabular}
	\caption{Time series of $W_{\ell m}$ for $\ell=$0, 2, 4, 6, 8, and 12 (blue dots) with time breaks filled (red dots).}
	\label{figWlmFill}
	\end{figure}

We use the SSA algorithm proposed by \citet{ghil_etal:02a}. The window width $M$, which determines the longest periodicity captured by SSA, is in our analysis adopted as constant (more details in Appendix~\ref{secAppendixSSA}). We search for the best representation iteratively up to the moment when the root-mean-square difference between the computed coefficients and the known ones is smaller than 0.01. The procedure is repeated for each $\ell$ of $W_{\ell m}$ coefficients separately. The results are shown in Figure~\ref{figWlmFill}, which presents the time series for some example numbers $\ell$. The values found for the missing elements in the time series of coefficients fit very well the general shapes of the time series and trace their variations in time. 

With the gaps in the time series of the spherical harmonics coefficients filled, we are able to calculate the helio-latitudinal profiles of the SW speed over the entire period of our analysis.

\section{Results of Reconstruction of Gaps in the Solar Wind IPS Data}
\begin{figure}
	\centering
		\begin{tabular}{ccc}
		\includegraphics[width=.28\textwidth]{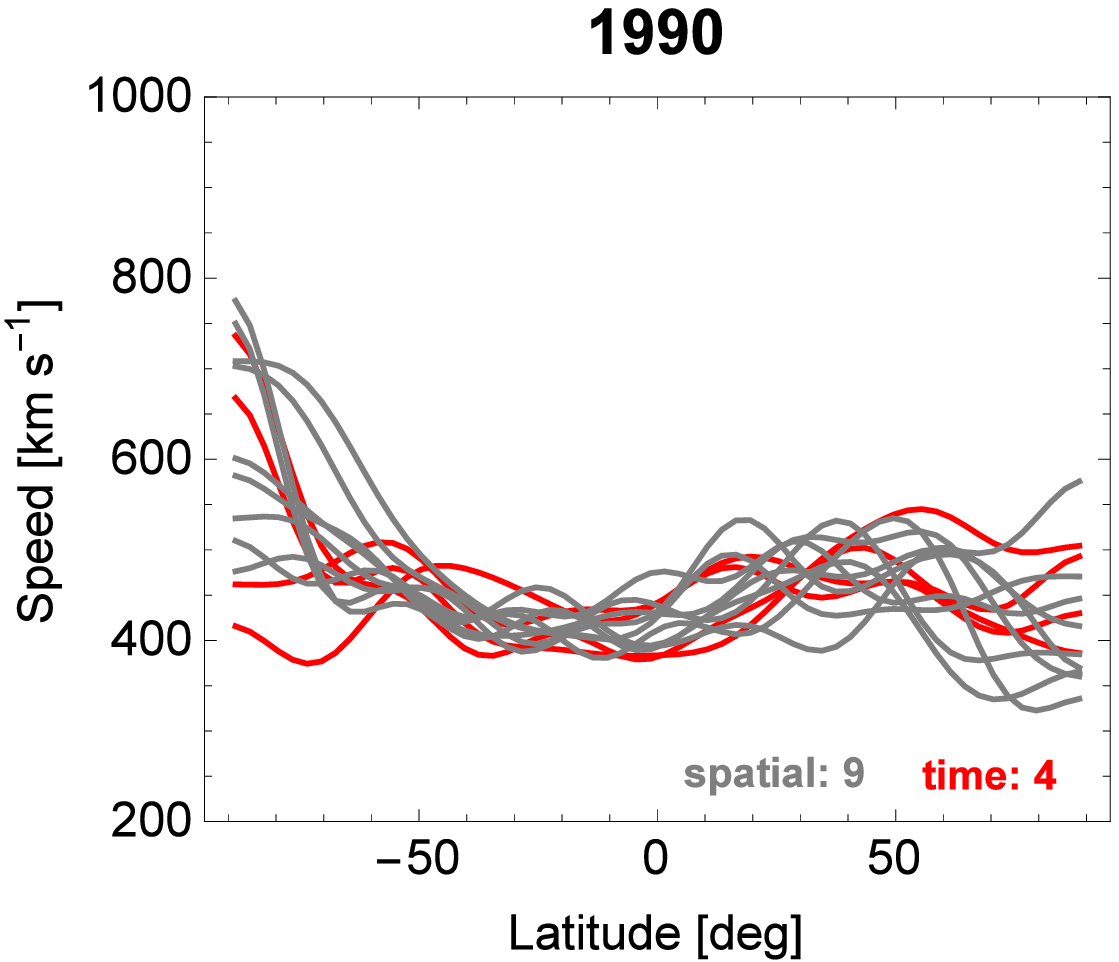} & \includegraphics[width=.28\textwidth]{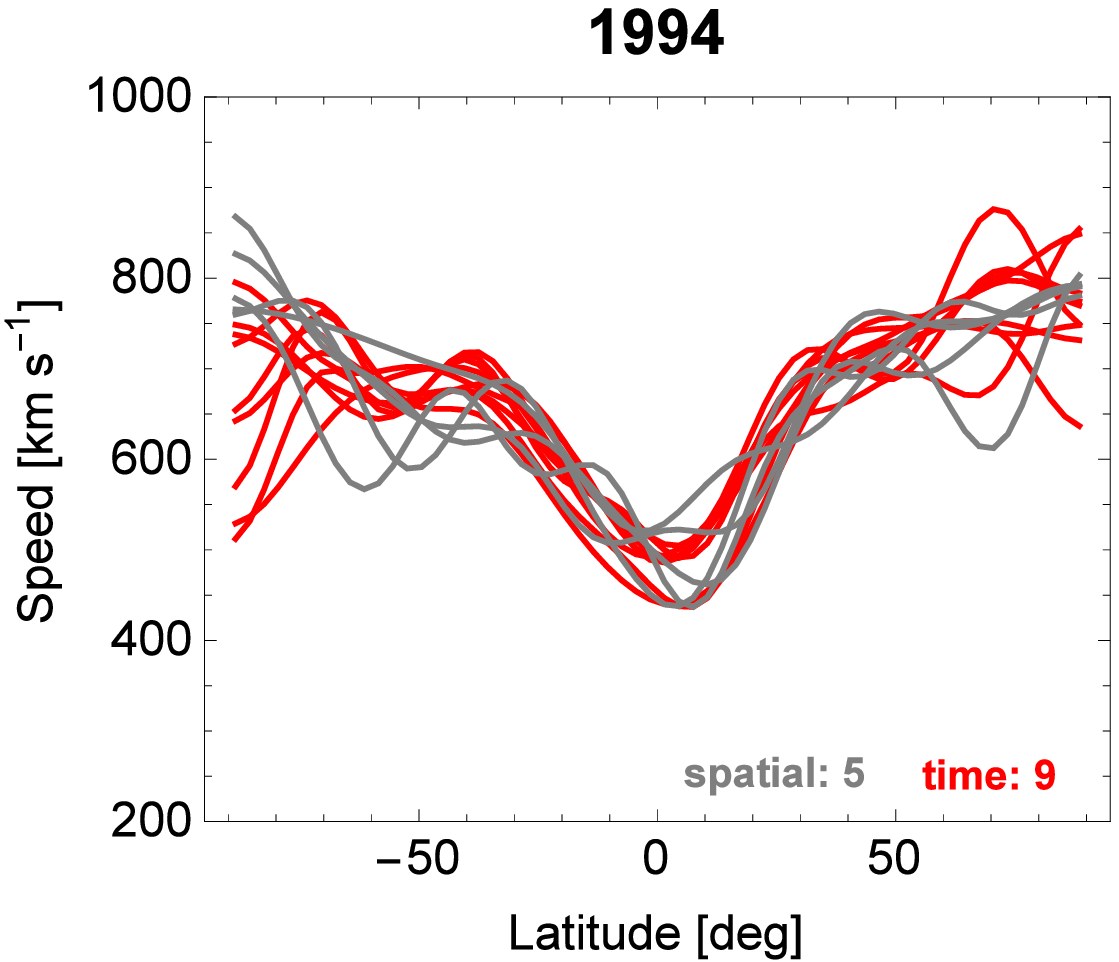} & 	\includegraphics[width=.28\textwidth]{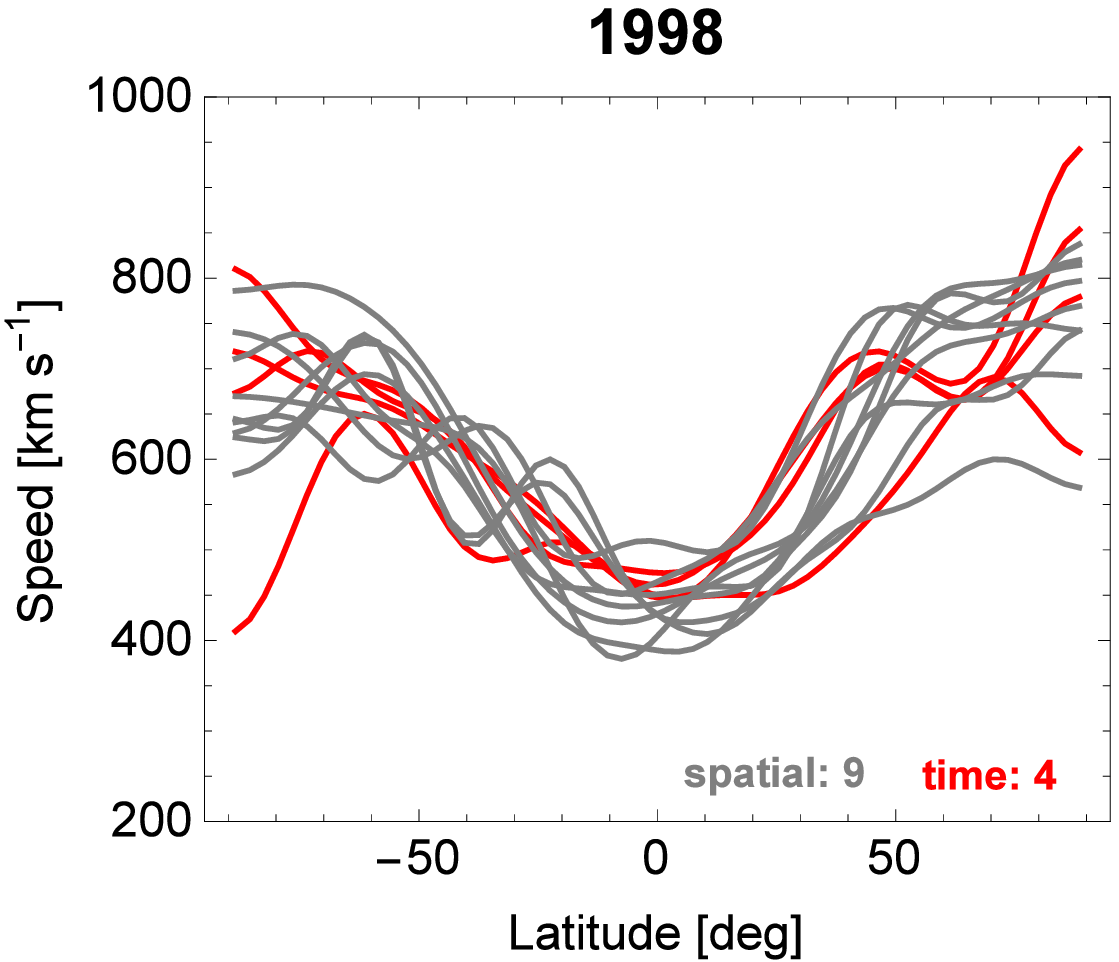}\\
		\includegraphics[width=.28\textwidth]{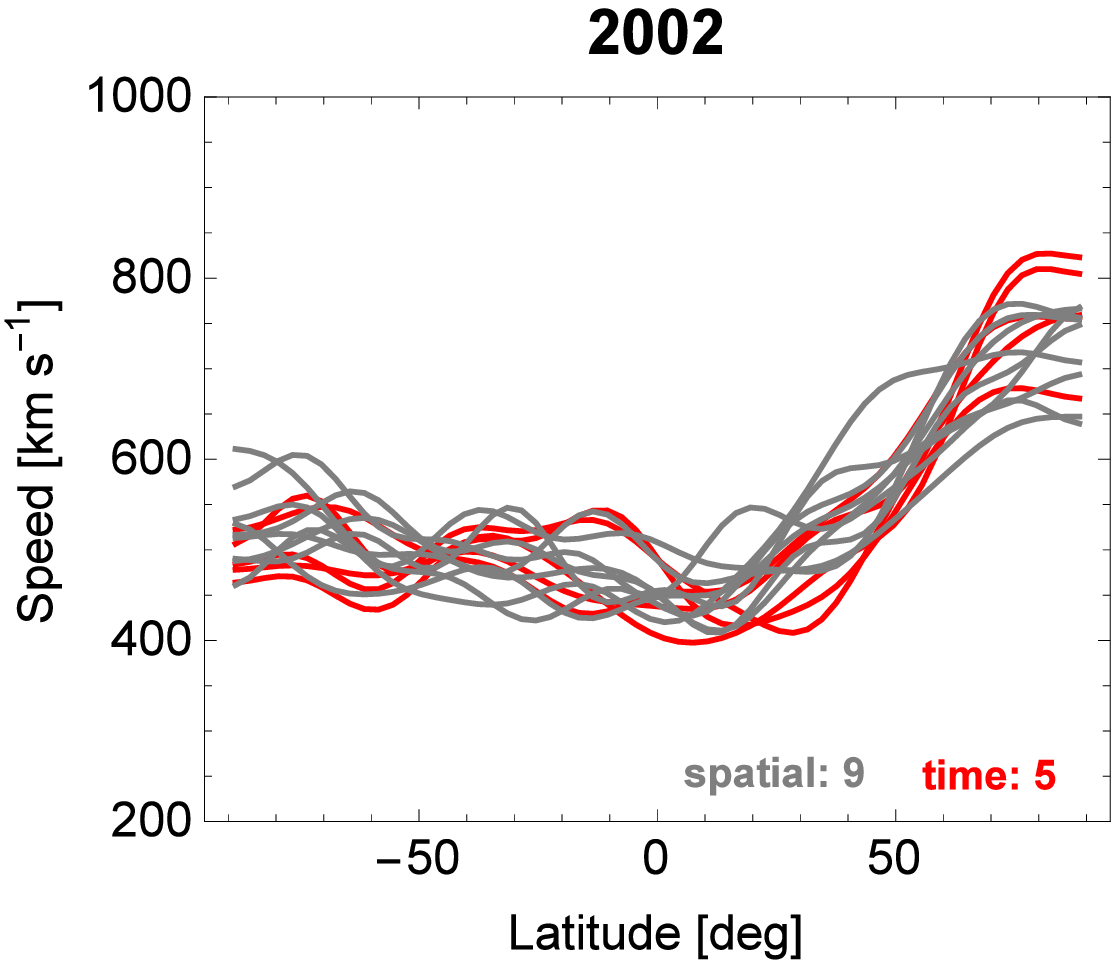} & \includegraphics[width=.28\textwidth]{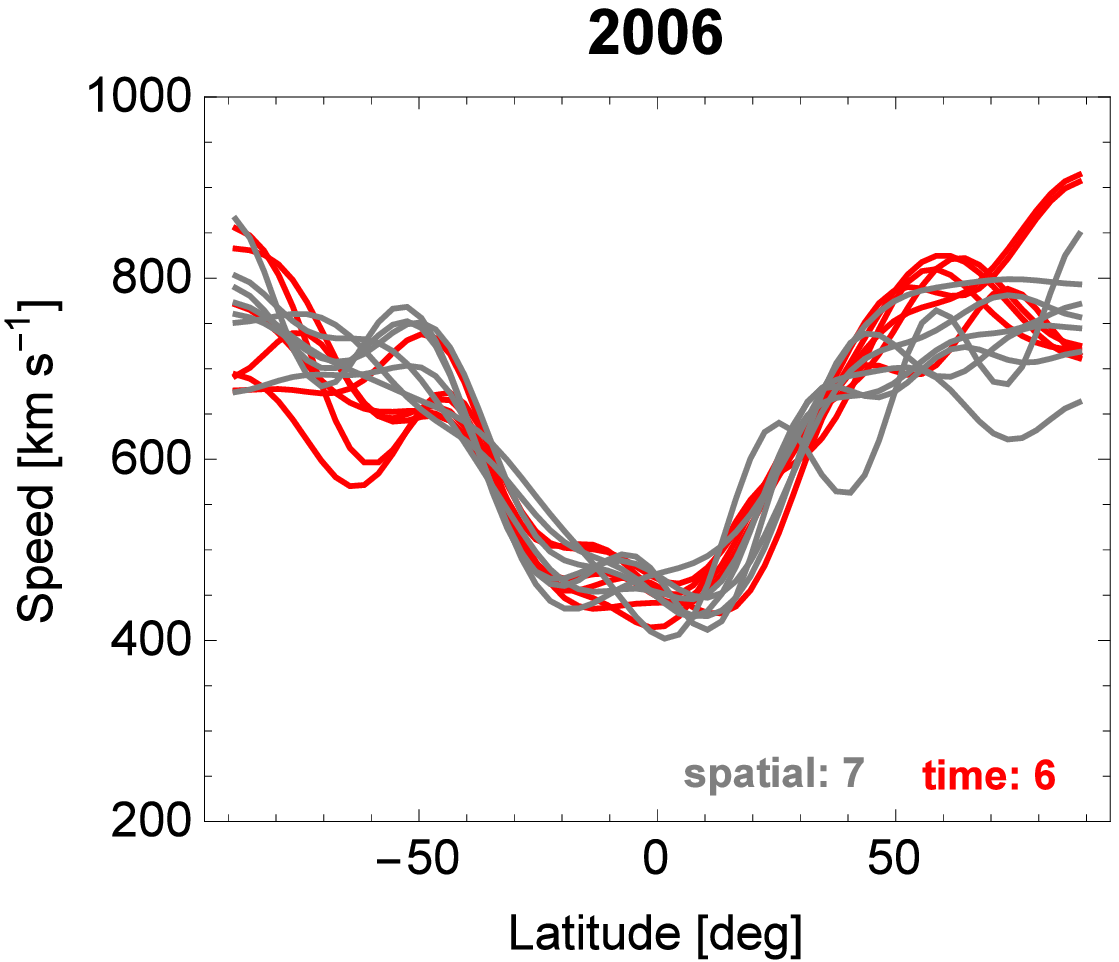} & \includegraphics[width=.28\textwidth]{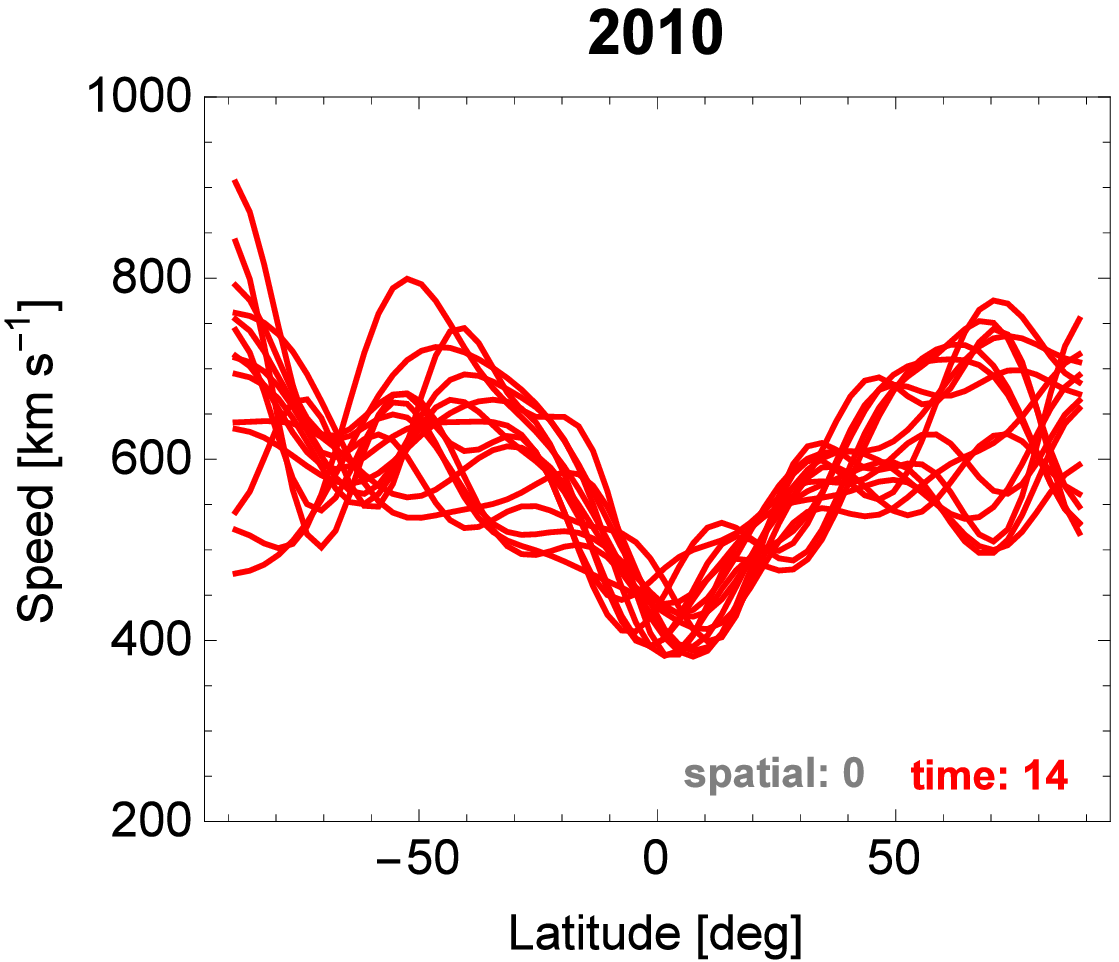}\\		
		\end{tabular}
	\caption{Helio-latitudinal profiles of the reconstructed SW speed. Gray: SW speed approximated using spherical harmonics. Red: reconstructions of the SW speed by filling the missing CR obtained by SSA. The numbers in the lower right corners indicate how many profiles for the given year come from the reconstruction of spatial gaps (in gray) and missing CR dataset (in red).}
	\label{figSpeedProfiles}
	\end{figure}

\begin{figure}
	\centering
	\includegraphics[scale=0.55]{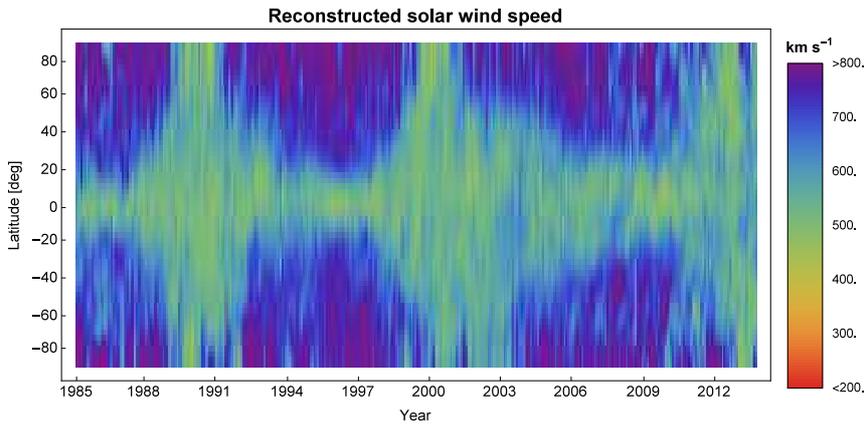}\\
	\caption{Helio-latitude vs. time map of the SW speed reconstructed by the present analysis.}
	\label{figMapFinalSpeed}
	\end{figure}

The procedures described in the previous sections allow us to fill the spatial and temporal gaps in the SW speed maps from 1985 to 2013. Figure~\ref{figSpeedProfiles} presents the helio-latitudinal profiles of the SW speed for all the CR periods under study. Shown in gray are the profiles by filling the spatial gaps, and in red are shown the profiles after filling the time gaps. The reconstructed profiles trace the general shape of the latitudinal dependence of the SW speed very well for all phases of solar activity, from almost flat profiles during the solar maximum (\textit{e.g.}, 1990 and 2002) to a bi-modal structure close to the solar minimum (\textit{e.g.}, 1998, 2006). The complete map of the 3D structure of the SW proton speed at 1~AU is presented in Figure~\ref{figMapFinalSpeed}.
	
\section{Solar Wind Density}
\label{secSWDens}
The heliosphere is shaped by the SW ram pressure, which is proportional to the product of the SW density and squared speed. With the time- and latitudinal variations of the structure of the SW speed available, one needs information on these variations in the SW density for complete modeling. The SW density in the ecliptic plane has been measured in-situ by a fleet of spacecraft since mid-1960s, and the data are now available in a compiled and meticulously curated form; the OMNI database\footnote{Source: \texttt{ftp://spdf.gsfc.nasa.gov/pub/data/omni/low\_res\_omni/}} \citep{king_papitashvili:05}. The out-of-ecliptic in-situ observations are solely available from \textit{Ulysses}, but only for years 1990--2009. The out-of-ecliptic remote-sensing observations of the heliospheric Lyman-$\alpha$ glow from SOHO/SWAN can provide the required information about the structure of the SW density after complex modeling (see a sketch of the algorithm of reconstruction of the SW density from the SWAN and SW speed from IPS data in \citet{bzowski_etal:13a}), but this procedure is not free from yet unanswered questions \citep{katushkina_etal:13a}. Also the $g$-level values from IPS observations can serve as a proxy to the out-of-ecliptic SW density \citep[\textit{e.g.}][]{houminer_hewish:72a, houminer_hewish:74a, tappin:86a, hick_jackson:04a, jackson_hick:04a}. There exists a nonlinear relationship between the bulk density and the IPS scintillation level that has been widely and successfully used to study the corotating structures and coronal mass ejections (\textit{e.g.} \opencite{jackson_etal:98a, tokumaru_etal:07a, bisi_etal:09a, bisi_etal:10d, fujiki_etal:14a}; a comprehensive review by \opencite{jackson_etal:11b}), but in this study we focus only on the use of the SW speed determined from IPS observations.

\begin{figure}
		\centering
		\includegraphics[scale=0.65]{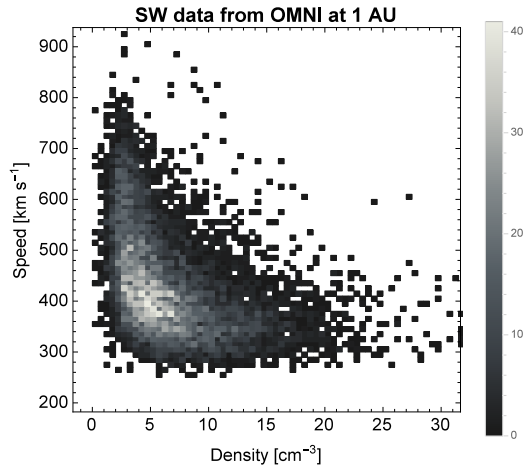}
		\caption{Relation between daily SW proton density at 1~AU and speed in the ecliptic plane (source: OMNI dataset) shown as a 2D histogram (100$\times$100 bins) with the count scale illustrated by the side bar.}
		\label{figSWSpeedDensRelation}
		\end{figure}

As is already known, there is no clear correlation between the SW proton speed and density in the ecliptic plane at 1~AU based on the OMNI dataset (Figure~\ref{figSWSpeedDensRelation}). It seems that the crescent-like relation may be expected, but the shape of the crescent is too wide to provide a clearly defined correlation function. 
		
We propose a simple approach with the use of the SW invariants to determine the SW density from the SW speed. Based on the \textit{Ulysses} measurements two alternative SW quasi-invariants were inferred: the SW dynamic pressure \citep{mccomas_etal:08a}
			\begin{equation}
  		 \rho = \frac{1}{2} n_{\mathrm{p}} \left( m_{\mathrm{p}} + \xi_{\alpha} m_{\alpha}\right) V^2
  		 \label{eqSWDynPress}
		  \end{equation}
and the SW energy flux \citep{leChat_etal:12a}
			\begin{equation}
			W = n_{\mathrm{p}} \left( m_{\mathrm{p}} + \xi_{\alpha} m_{\alpha}\right) V \left( \frac{1}{2} V^2 + G \frac{M_{\odot}}{R_{\odot}} \right)
			\label{eqSWEnergyFlux}
			\end{equation}
where $n_{\mathrm{p}}$ is the SW proton density, $V$ is the SW proton speed, $m_{\mathrm{p}}$ is the proton mass, $\xi_{\alpha}$ is the abundance of $\alpha$-particles ($\approx 4\%$ after \citet{kasper_etal:12a}), $m_{\alpha}$ is the mass of the $\alpha$-particles, $M_{\odot}$ is the mass of the Sun, $R_{\odot}$ is the solar radius, and $G$ is the gravitational constant.
		 \begin{figure}
		 \resizebox{\hsize}{!}{\includegraphics{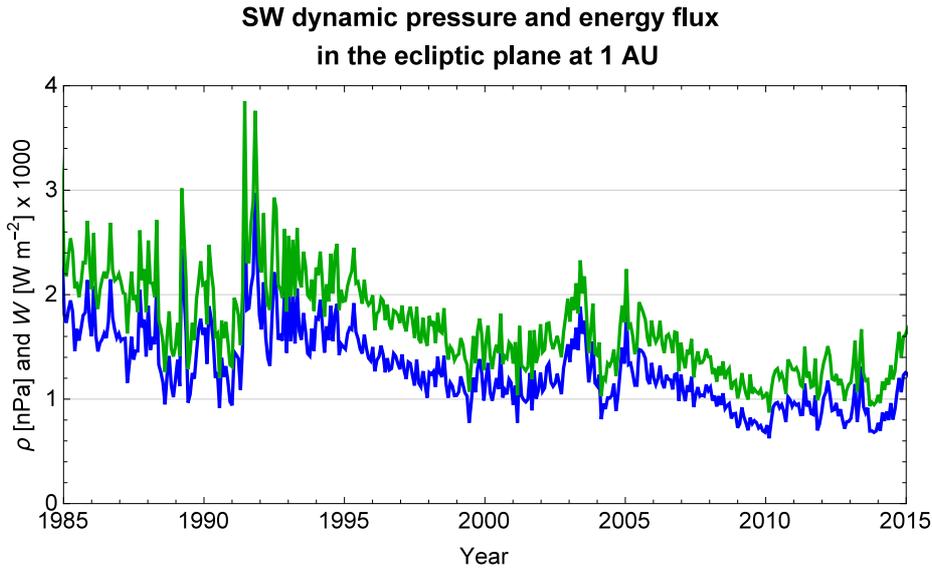}}
     \caption{The CR-averaged SW dynamic pressure (blue; Equation~(\ref{eqSWDynPress})) and energy flux (green; Equation~(\ref{eqSWEnergyFlux})) in the in-ecliptic plane at 1~AU based on OMNI. The latter was multiplied by 1000 to roughly match the two scales.}
		\label{figSWDynPressEnergyFlux}
		\end{figure}
As shown in Figure~1 of \citet{leChat_etal:12a} and in Figure~3 of \citet{mccomas_etal:08a} for the \textit{Ulysses} measurements, both the SW energy flux and the dynamic pressure are almost constant, independent of latitude. Figure~\ref{figSWDynPressEnergyFlux} presents two quantities as a function of time, calculated for the in-ecliptic SW data at 1~AU from the OMNI database. Both quantities show very similar variations as a function of time and they can serve as the SW helio-latitudinal invariants to calculate the out-of-ecliptic structure of the SW density from the latitudinal variations of the SW speed as is discussed in Appendix~B of \citet{mccomas_etal:14b}.
		
\begin{figure}
		\centering
			\begin{tabular}{ccc}
			\includegraphics[width=.28\textwidth]{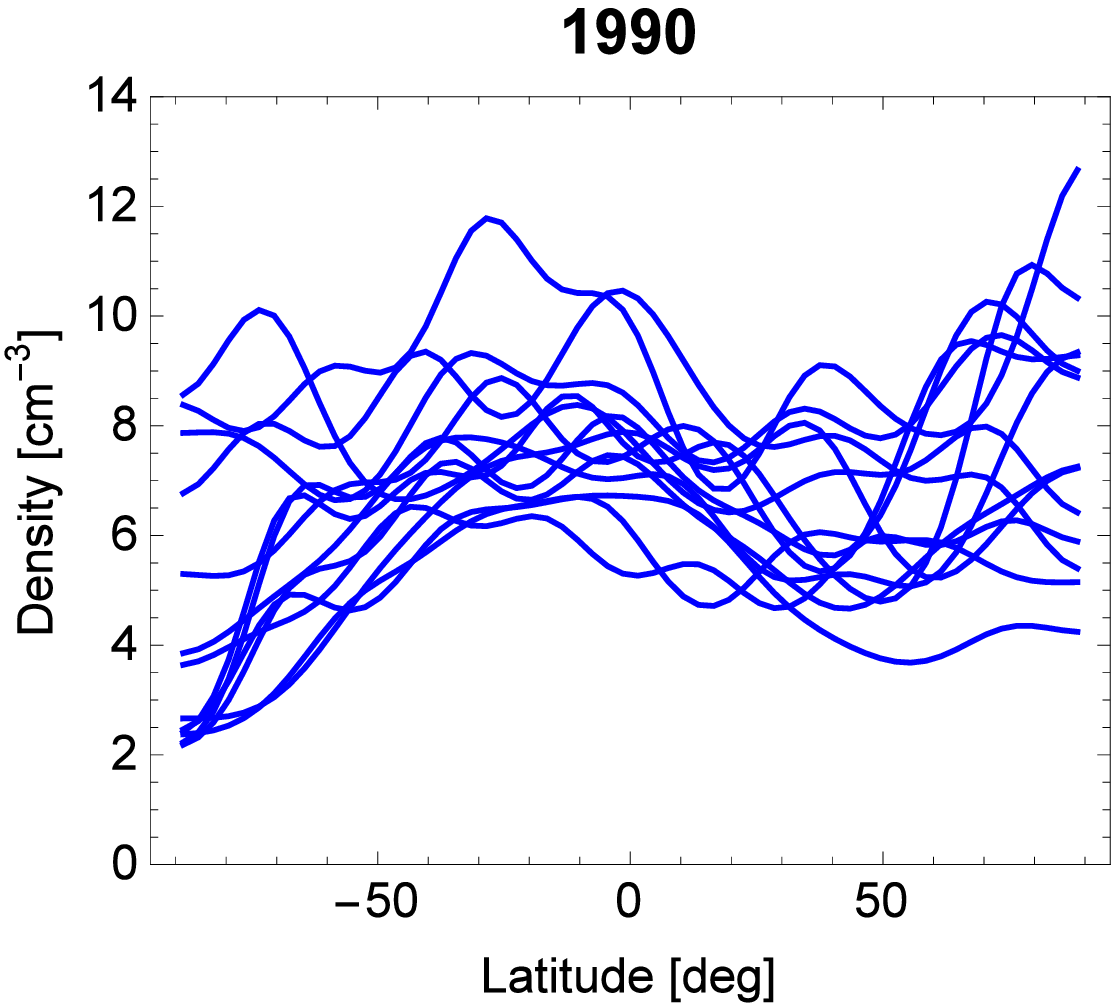} & \includegraphics[width=.28\textwidth]{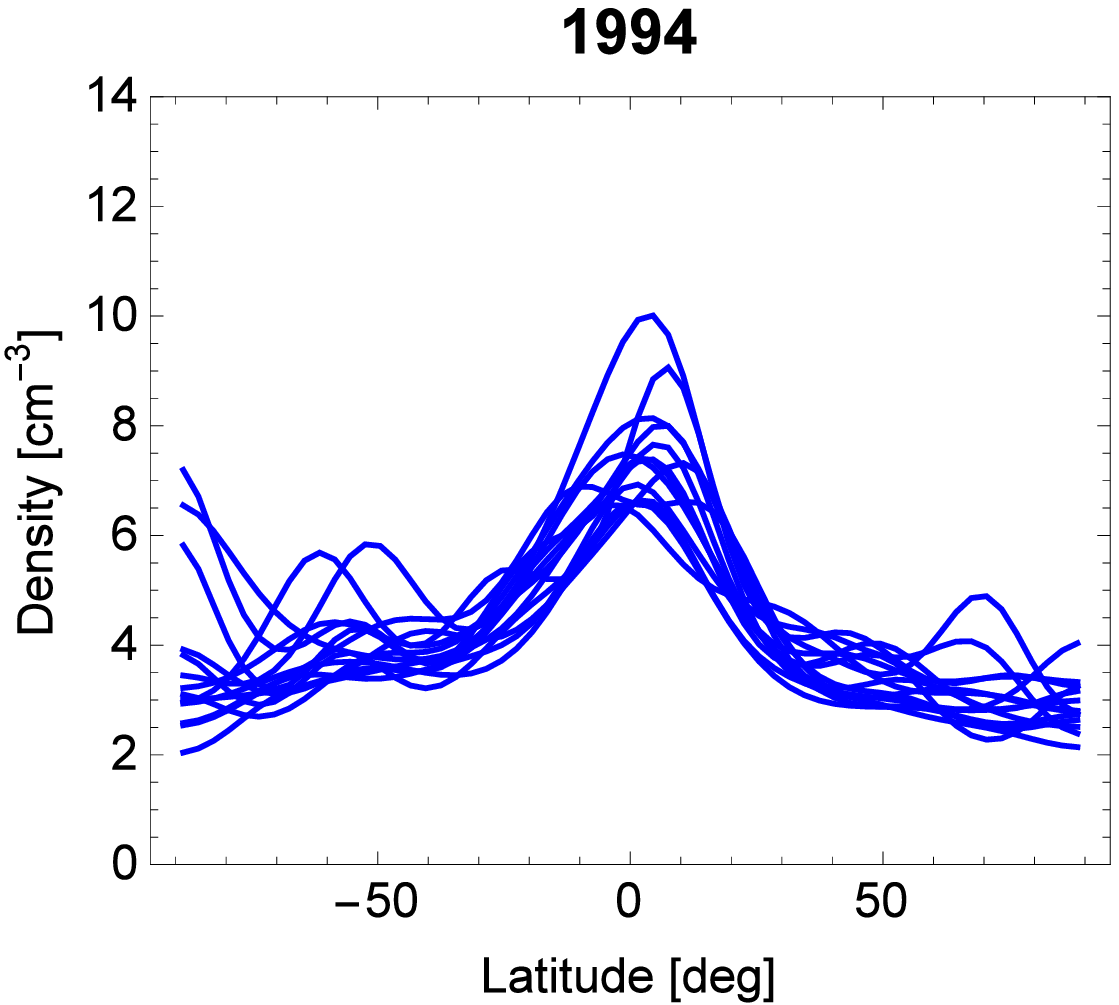} & \includegraphics[width=.28\textwidth]{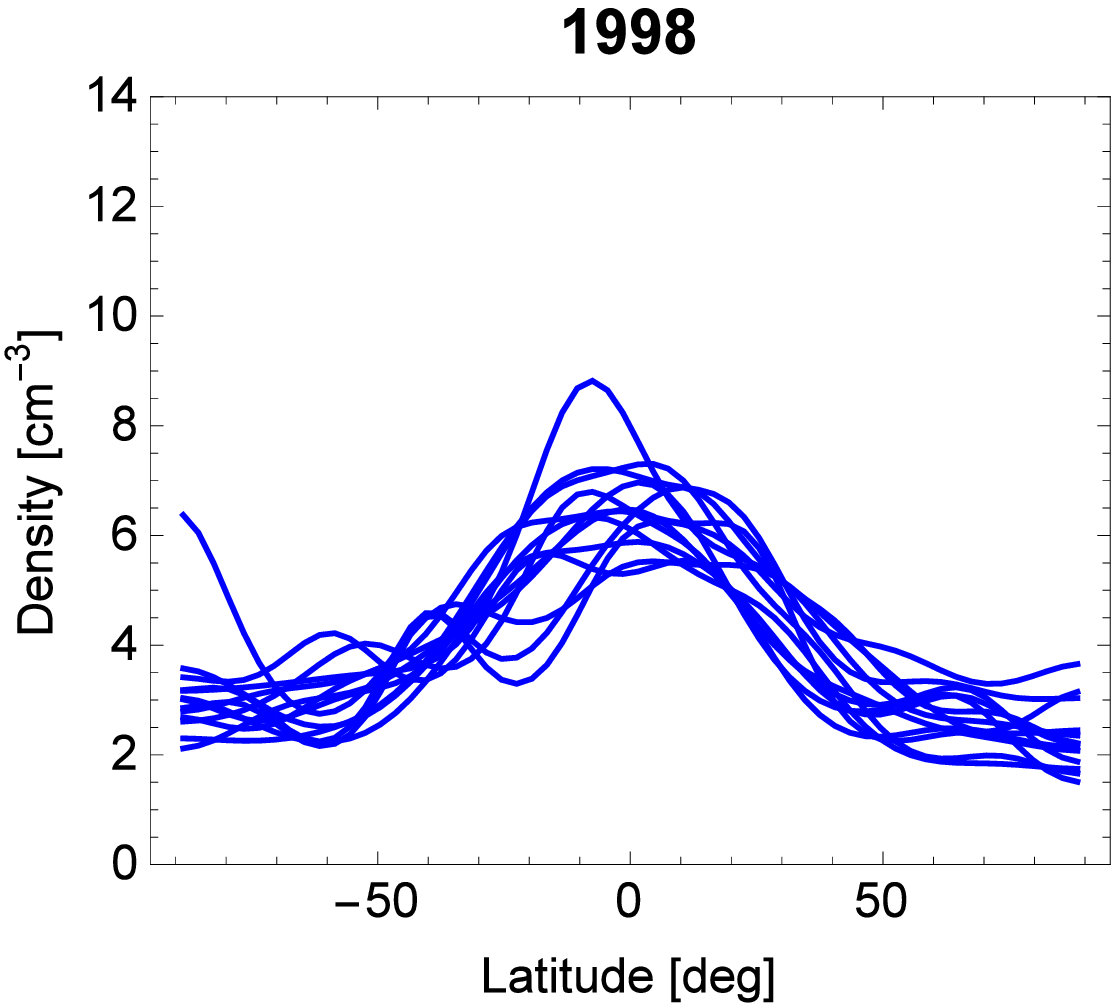}\\
			\includegraphics[width=.28\textwidth]{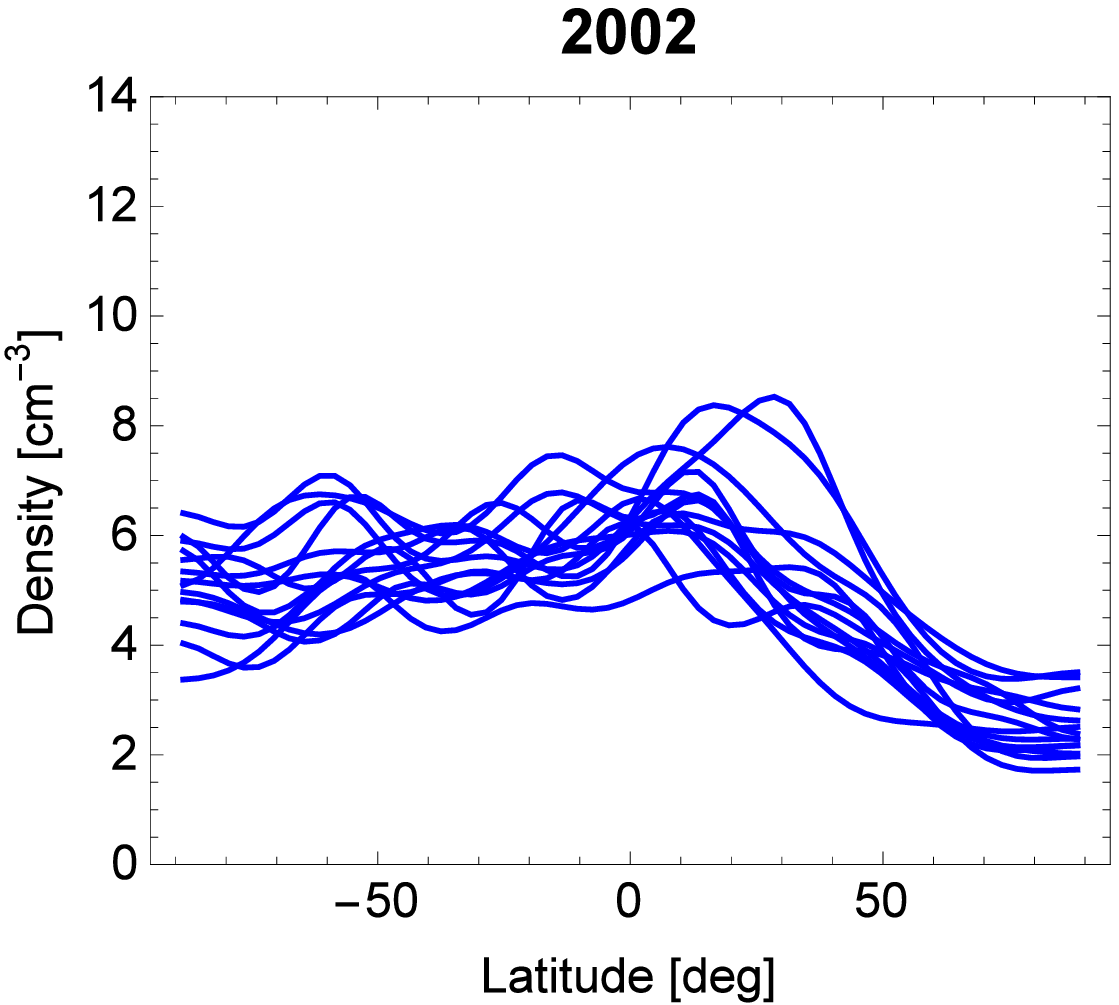} & \includegraphics[width=.28\textwidth]{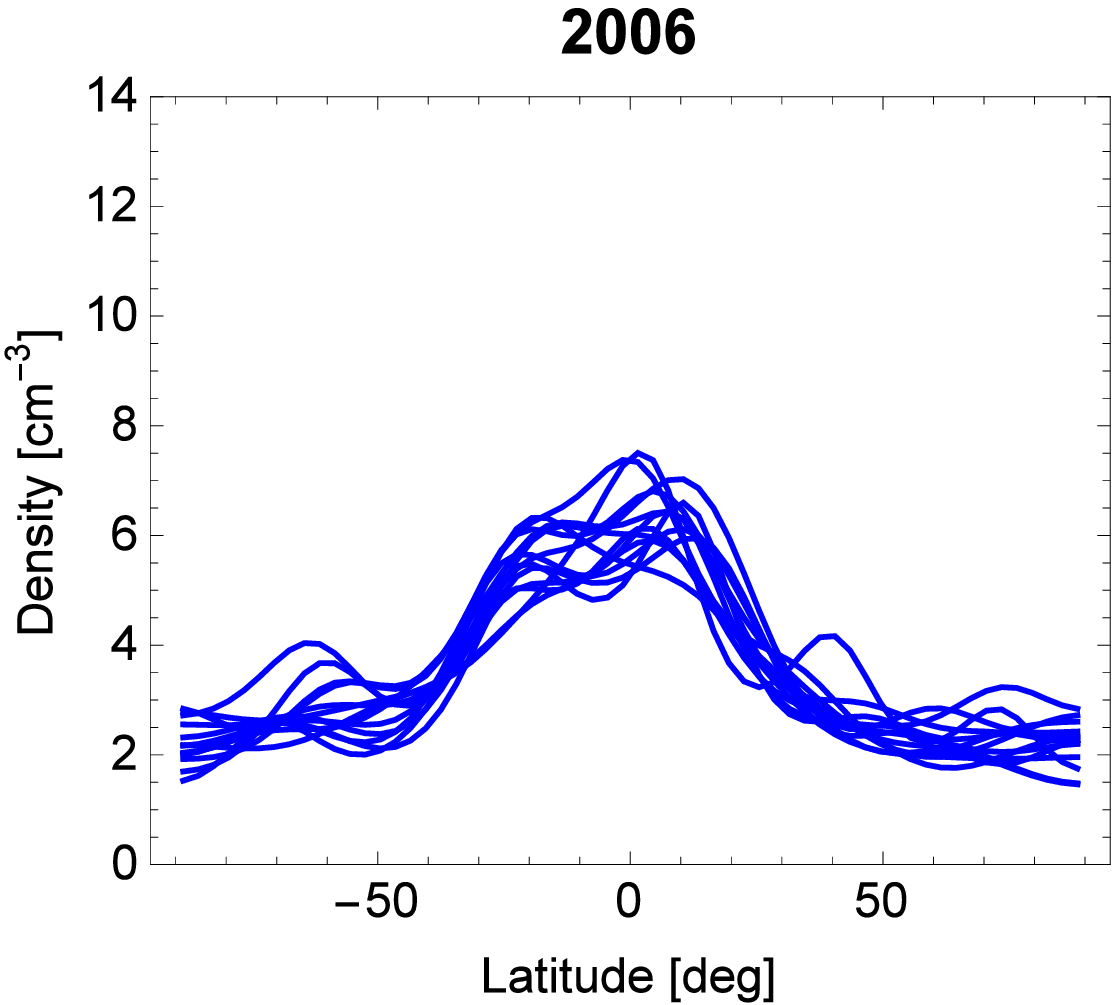} & \includegraphics[width=.28\textwidth]{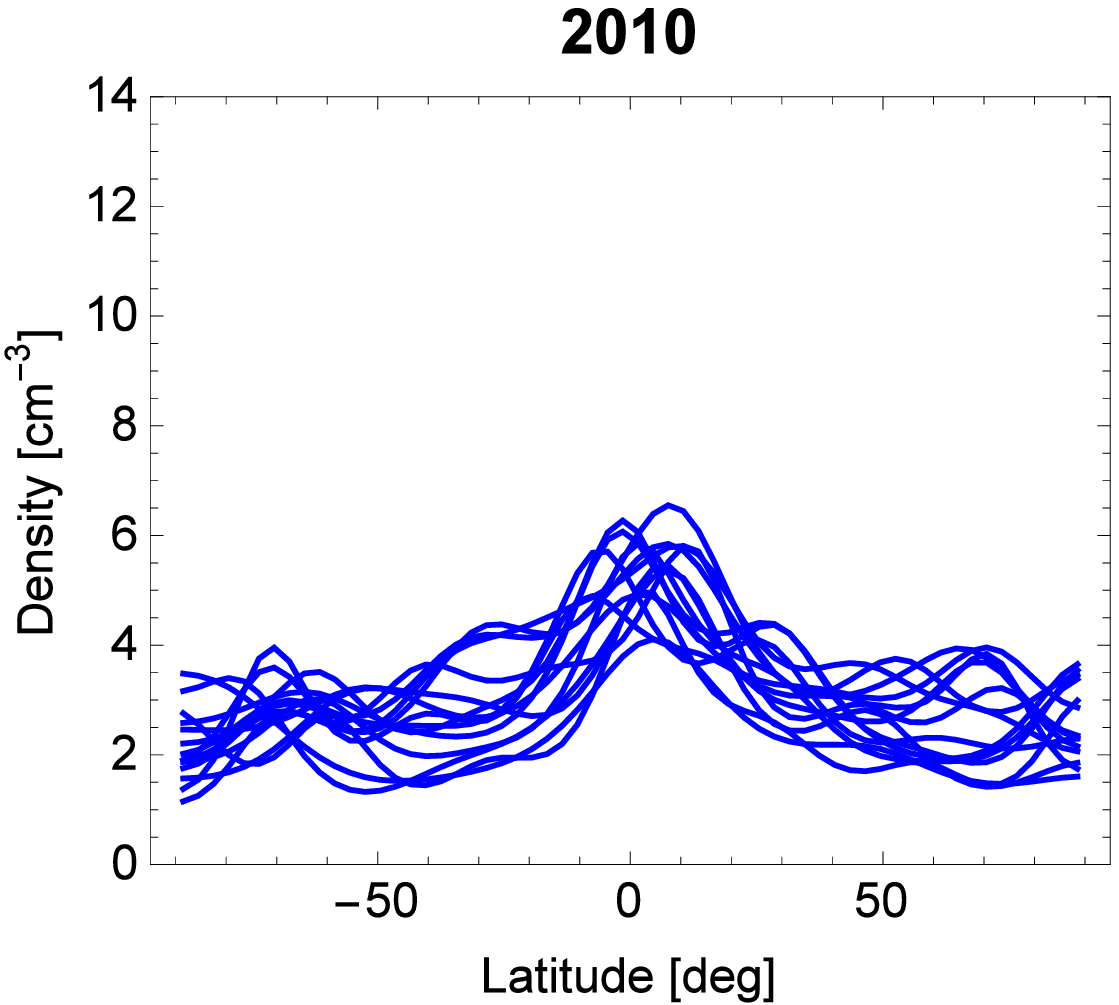}\\		
			\end{tabular}
		\caption{Helio-latitudinal profiles of the SW density at 1~AU calculated from the SW energy flux obtained from in-ecliptic measurements and the reconstructed SW speed.}
		\label{figDensProfiles}
		\end{figure}

In the analysis below we calculate the SW proton density as a function of helio-latitude based on the SW proton speed and the SW invariants in latitude. We calculate the SW dynamic pressure from Equation~(\ref{eqSWDynPress}) and the SW energy flux from Equation~(\ref{eqSWEnergyFlux}) for the in-ecliptic SW at 1~AU from the OMNI database as averages of the daily data over each CR period for the time ranges in question. Next, we calculate the density profiles ($n_{\mathrm{p}}$) as a function of helio-latitude based on the SW speed profiles presented in Figure~\ref{figSpeedProfiles} and the invariant calculated from the in-ecliptic measurements (Figure~\ref{figSWDynPressEnergyFlux}). Results for the corresponding years are presented in Figure~\ref{figDensProfiles}. Because the SW energy flux and dynamic pressure are invariant in latitude, the difference between the ecliptic plane and the solar equatorial plane does not matter. The invariance of the SW dynamic pressure and energy flux derived from \textit{Ulysses} is also seen in the in-ecliptic values measured by spacecraft collected by OMNI, as both datasets agree quite well for the ecliptic values, as shown by \citet{sokol_etal:13a} in their Figures~11 and 12, over the yearly and hourly time resolution. 

\begin{figure}
	\centering
	\includegraphics[scale=0.55]{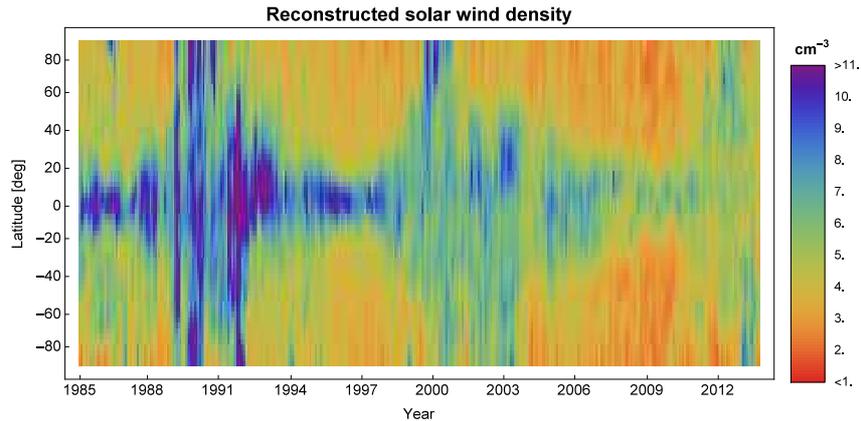}\\
	\caption{Helio-latitude vs. time maps of the SW density at 1~AU calculated from the SW speed and SW energy flux (Equation~(\ref{eqSWEnergyFlux})).}
	\label{figMapFinalDens}
	\end{figure}

The complete map of the 3D structure of the SW proton density at 1~AU is presented in Figure~\ref{figMapFinalDens}. The approach we adopted has led to the confirmation of a general dichotomy of the SW, \textit{i.e.}, dense and slow wind vs. low density and fast-wind, as is seen in Figures~\ref{figMapFinalSpeed} and \ref{figMapFinalDens}. 

\section{Comparison with OMNI and \textit{Ulysses}}
\label{secComparison}
SW measurements made in-situ give the reference for all studies of the SW structure. The in-ecliptic measurements at 1~AU have been gathered by various spacecraft since 1960s and are compiled in the OMNI database. The only available in-situ measurements out of the ecliptic plane come from the unique \textit{Ulysses} mission, but as it samples a single point at a time in a wide range of radial distance (from $\approx$1.2 to $\approx$5 AU), latitude ($\approx \pm 80\degr$), and time (1990-2009), they cannot answer all questions about the latitudinal structure of the SW and its evolution over the SC. Nevertheless, to be credible, all attempts of reconstruction of the SW must give results in agreement with these two datasets.

\begin{figure}
		\centering
			\begin{tabular}{cc}
			\includegraphics[scale=0.28]{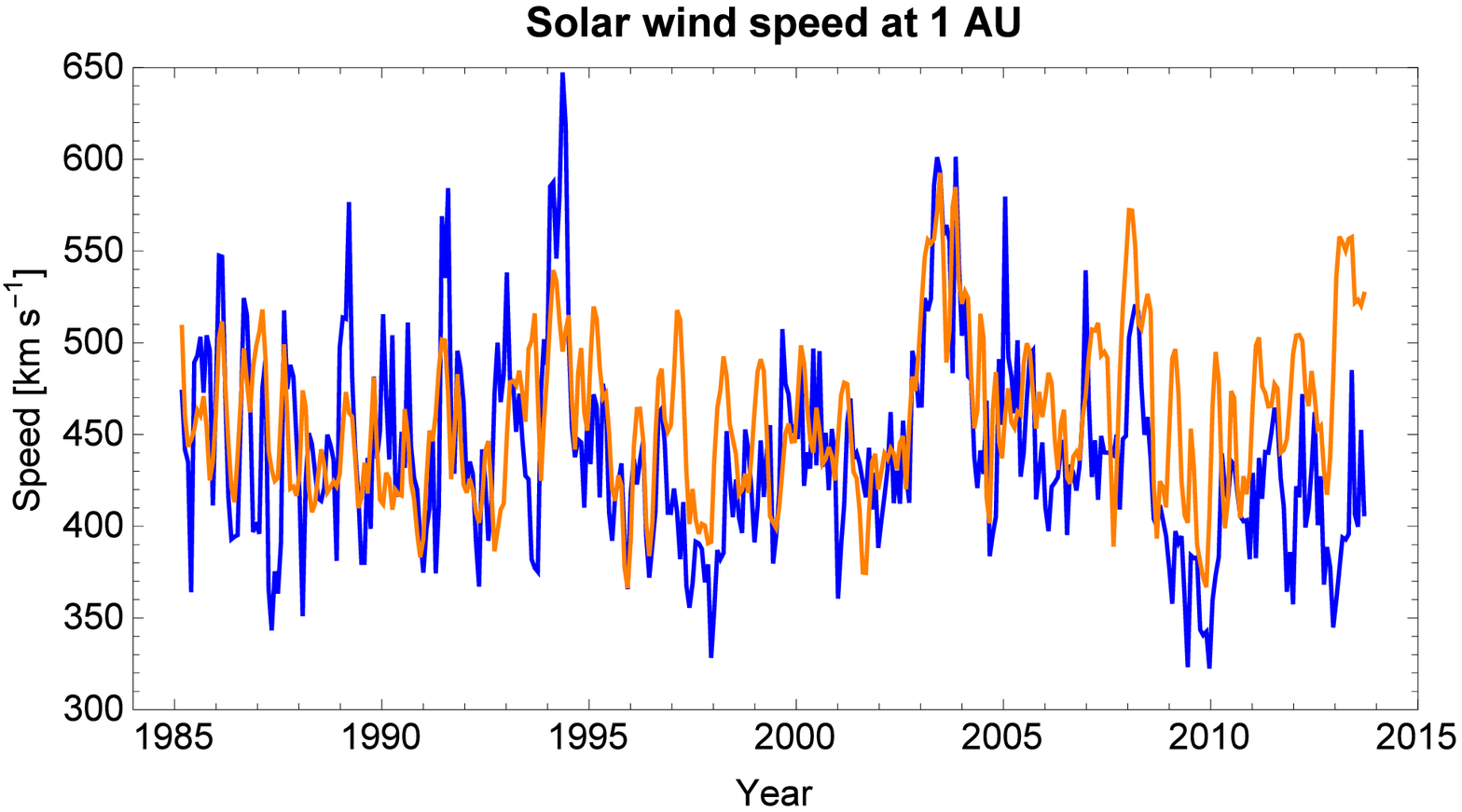} & \includegraphics[scale=0.28]{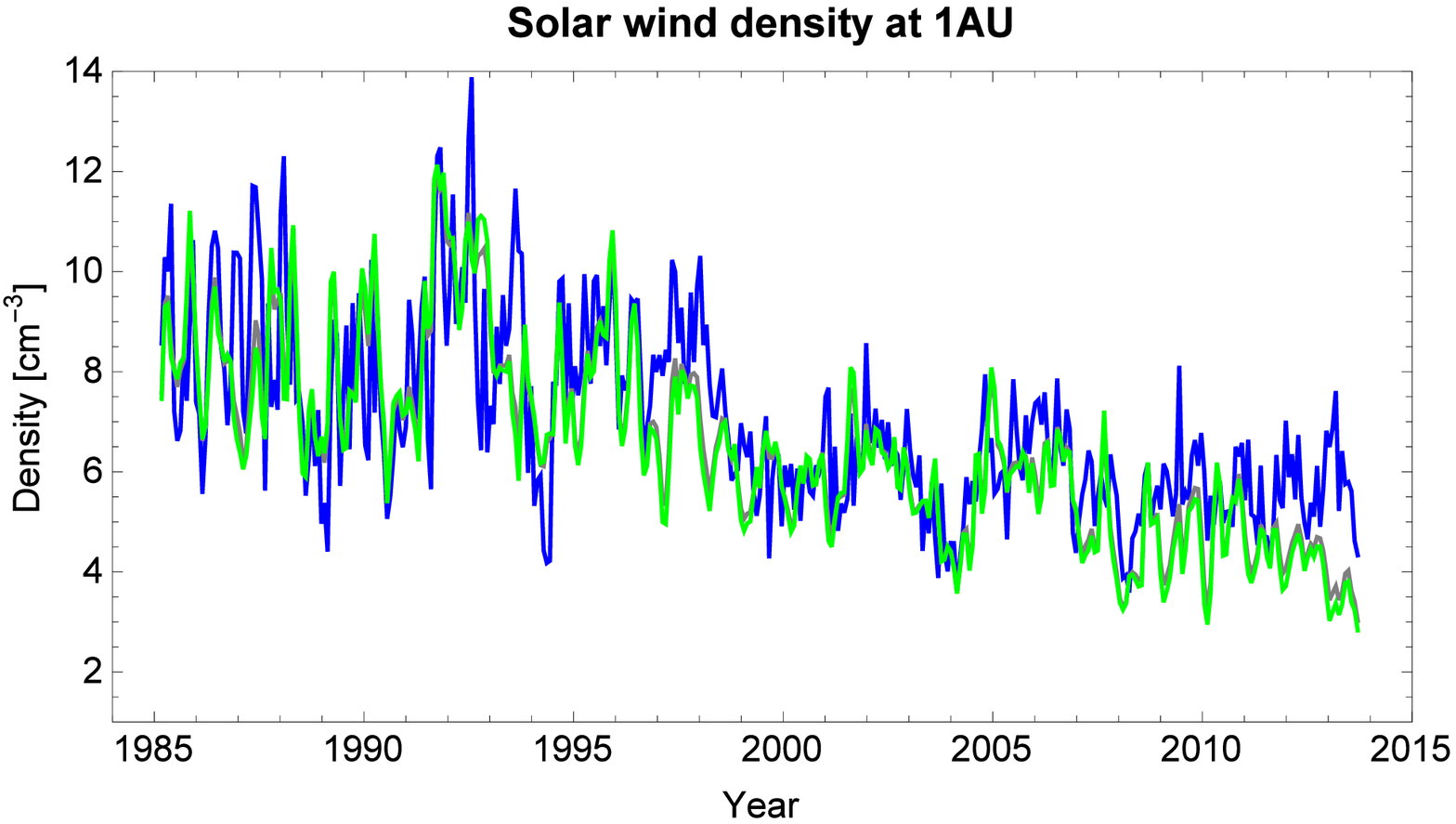}\\
			\includegraphics[scale=0.28]{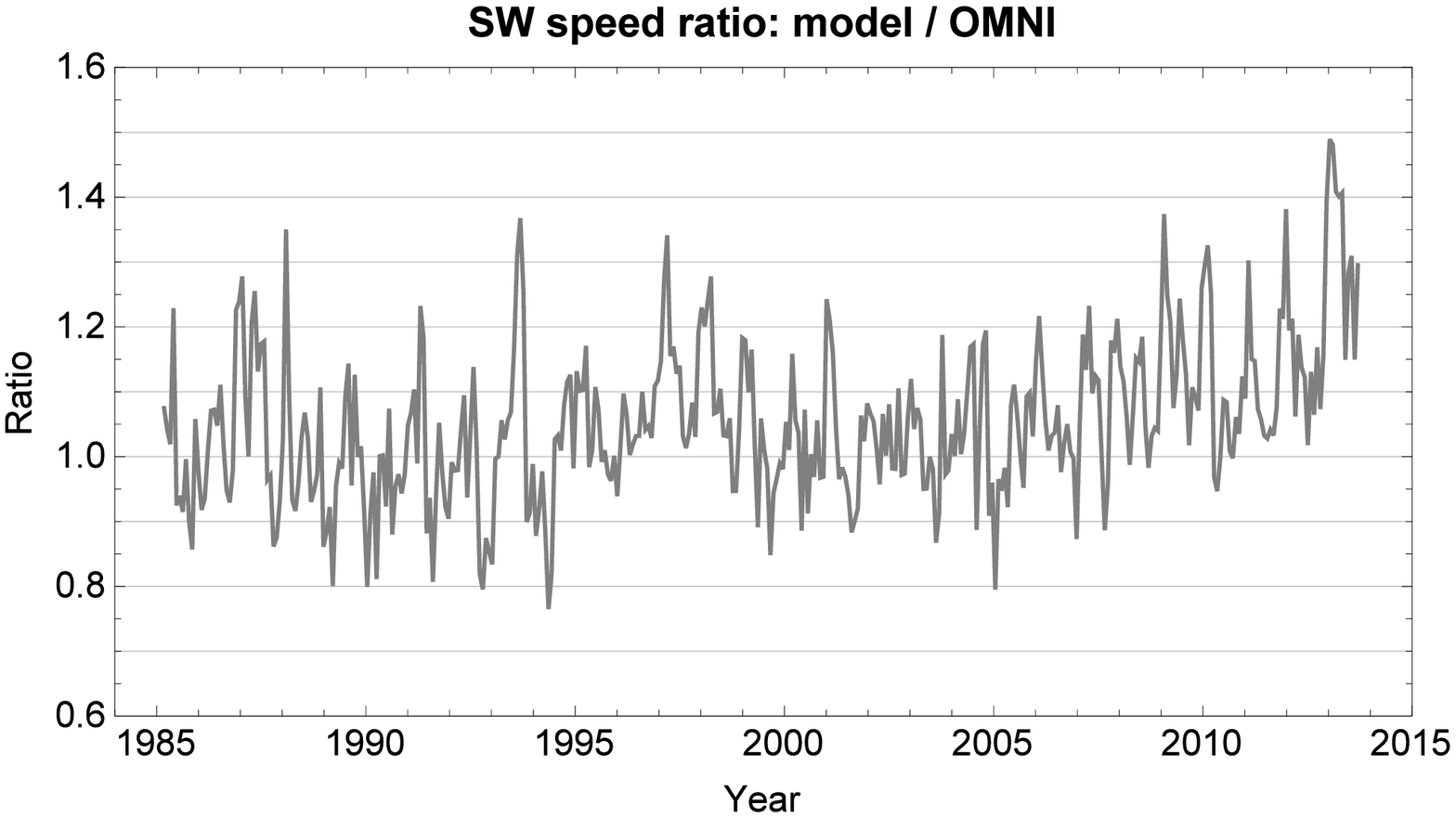} & \includegraphics[scale=0.28]{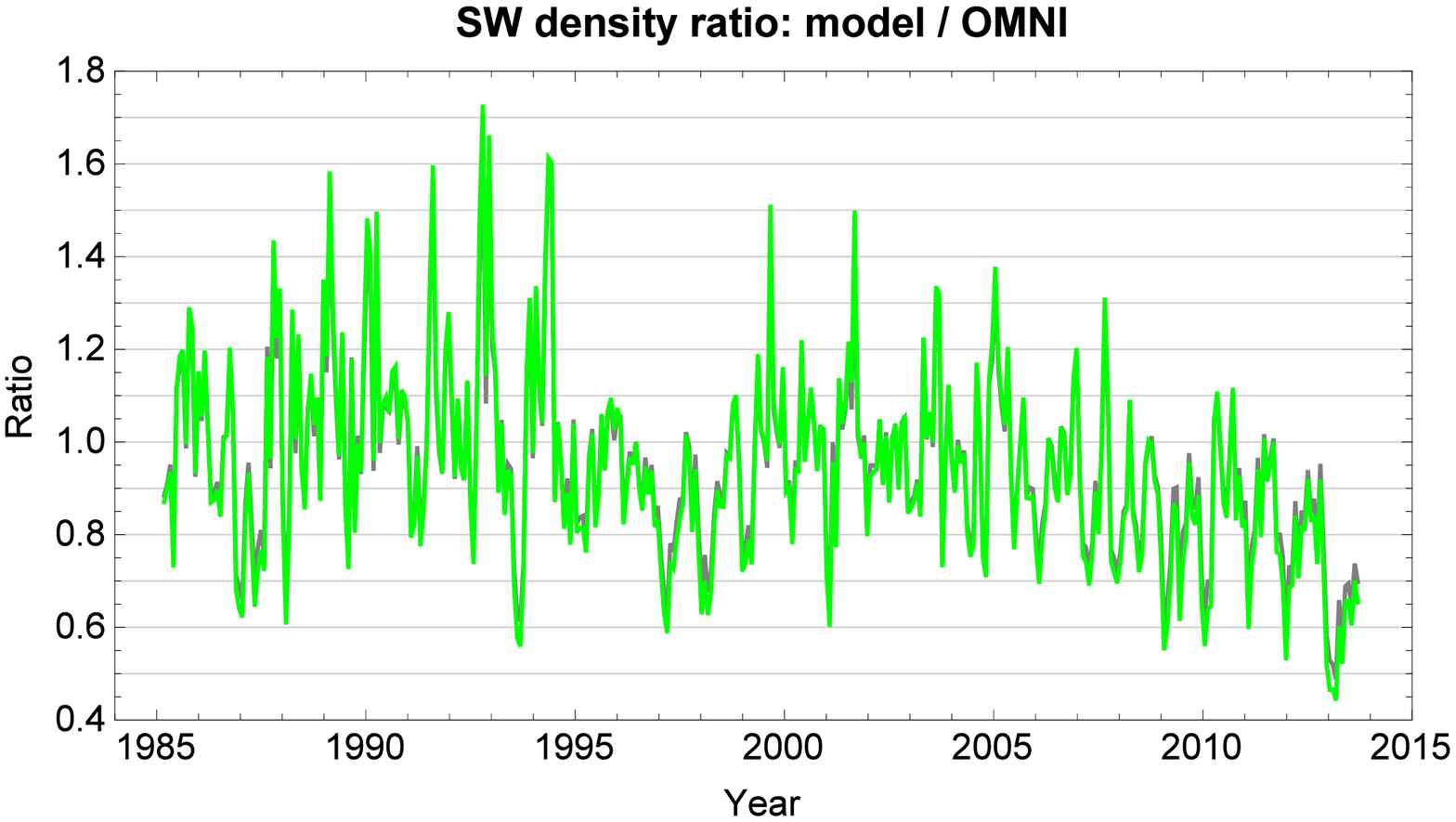}\\
			\includegraphics[scale=0.3]{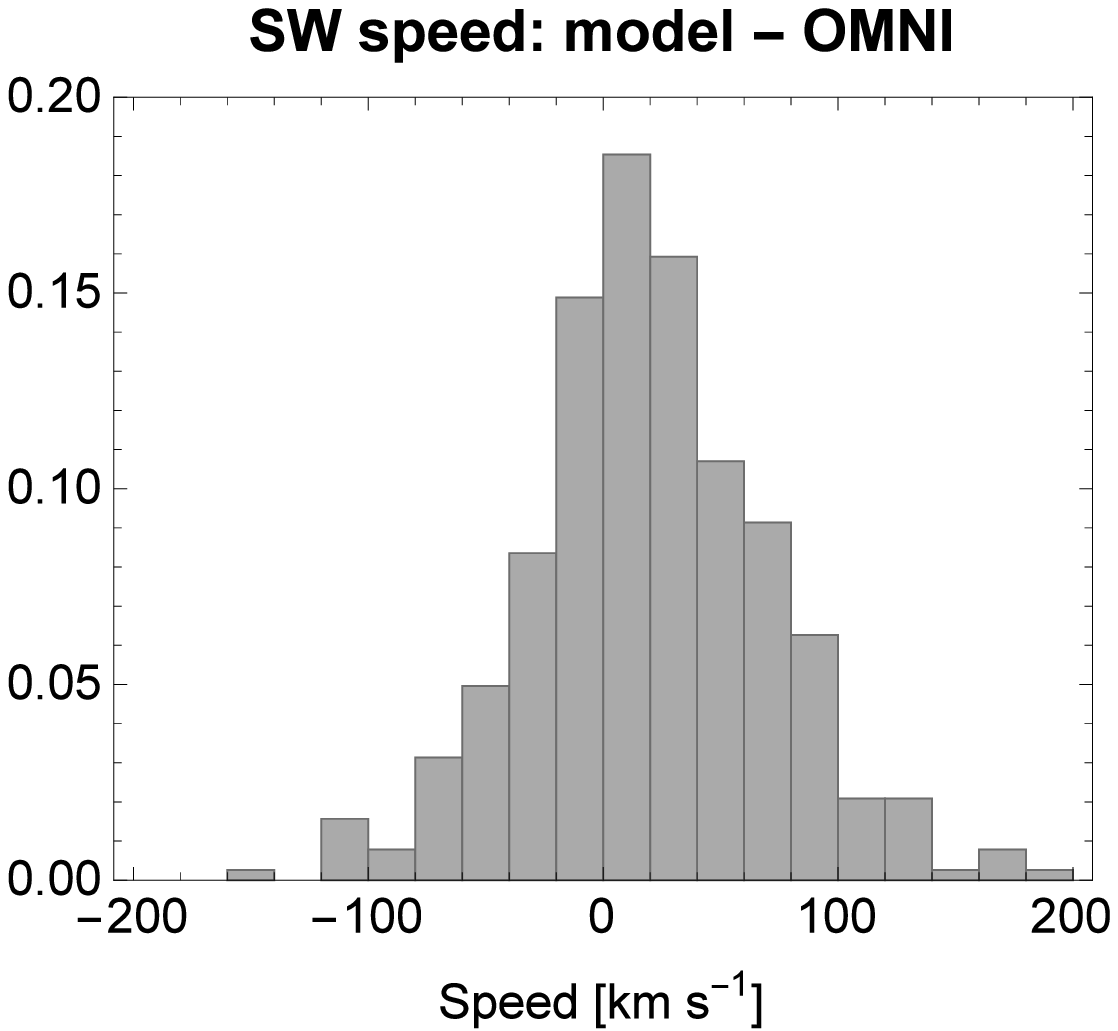} & \includegraphics[scale=0.3]{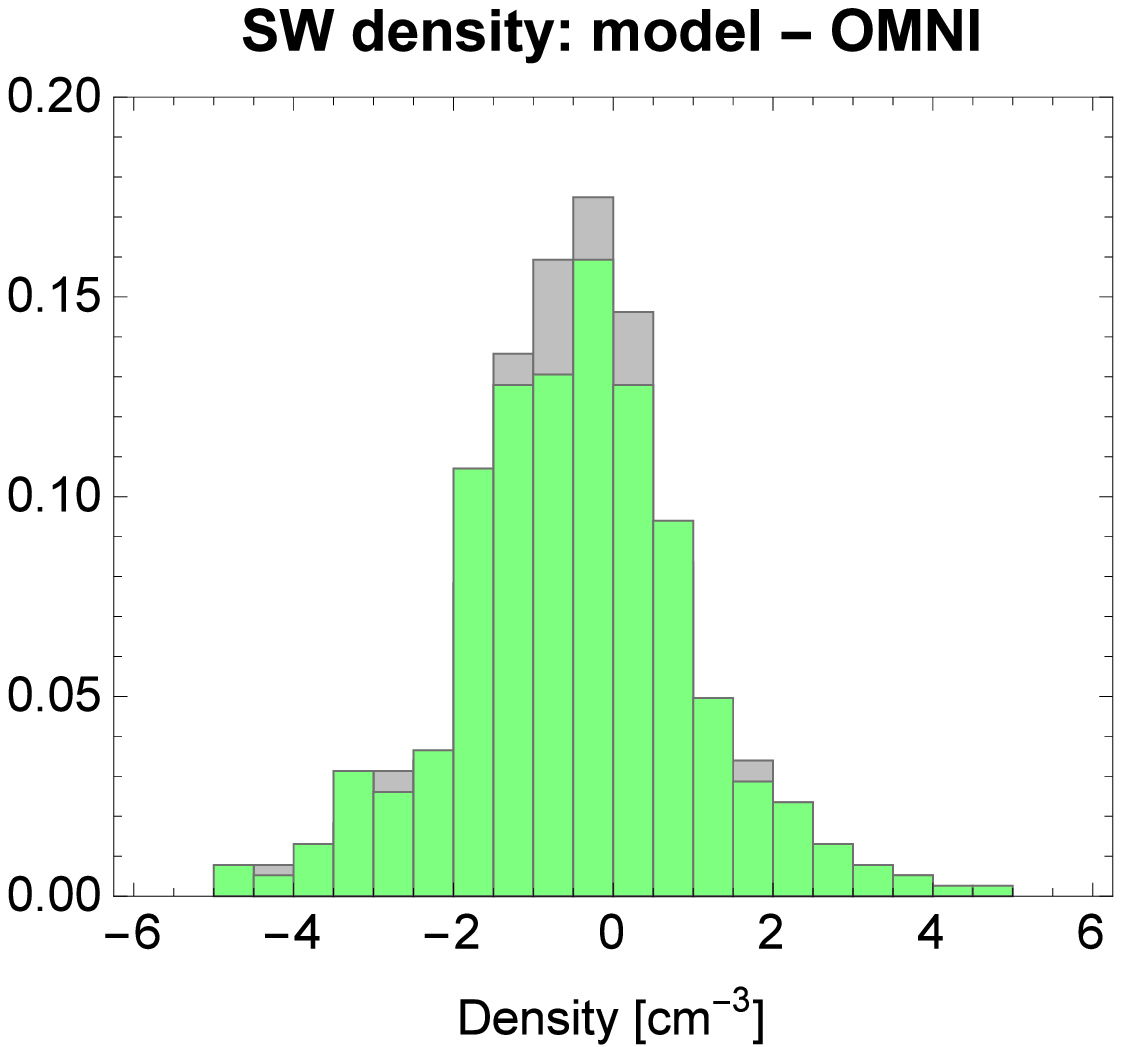}\\				
		\end{tabular}
		\caption{CR-averaged SW proton speed (left-hand column) and density (right-hand column) derived in our model are compared with the values from the OMNI database (blue lines in the top-row panels) in the ecliptic plane at 1~AU. In the top-left panel the orange line shows the SW speed reconstructed in this study; in the top-right panel the gray and green lines show the SW density calculated from the SW energy flux or dynamic pressure, respectively. The middle-row panels show the ratios of the SW quantities obtained in our analysis with respect to the OMNI data. The bottom panels show the histograms of the differences between our model and the OMNI values; the color coding is the same as in the middle-row panels.}
		\label{figCompareOMNI}
		\end{figure}

In the first step, we compared the results of our analysis with the in-ecliptic SW parameters at 1~AU from the OMNI database. We compared the CR-averaged values of the SW proton speed and density at 1~AU with those obtained from our model. The comparison for the speed is shown in the top-left panel of Figure~\ref{figCompareOMNI}; the blue line is for the time series from OMNI and the orange line is for the time series from the SW reconstructed from the IPS data. The agreement between the two is very good until the end of 2008; afterwards a systematic deviation appears, with the lower values from OMNI. Before this time, which coincides with the deepest solar minimum since the beginning of the space-age, the SW reconstruction traces the variations in the in-ecliptic SW very well. The middle left panel of Figure~\ref{figCompareOMNI} presents the ratio between the SW speed from OMNI to the SW speed reconstructed in this analysis. For almost all years, the ratio is constant with a mild systematic upward trend. The source of this trend may be in the assumed relation between the density fluctuations and the SW speed from Equation~(\ref{eqSpeedRetrieve}) and may not be related to the adopted version of analysis of the IPS data (either $\Delta N_{\mathrm{e}}$ or $\Delta N_{\mathrm{e}}$ and $g$-value relation). This difference after 2008 suggests that either the $\Delta N_{\mathrm{e}} \sim V$ relation needs verification considering the secular changes in the SW or, although highly unlikely, the OMNI data are in error.

The right-hand column of Figure~\ref{figCompareOMNI} presents the results for the SW proton density adjusted to 1~AU from the Sun. The OMNI values (blue line) are compared with the density calculated using the SW latitudinal invariants; the SW dynamic pressure (green line, Equation~(\ref{eqSWDynPress})) and the SW energy flux (gray line, Equation~(\ref{eqSWEnergyFlux})). Similar to the case of the SW speed, the agreement between the two is very good up to 2009. Then the discrepancy increases, with the density from our analysis underestimated. This difference is understandable, because in our method of calculation of the SW density, any inaccuracy in the SW speed affects the derived density. As shown by the histograms of differences of the absolute values in the bottom panels in Figure~\ref{figCompareOMNI}, the mean accuracy of reconstruction is about $50$~\kms\,for the SW speed and $1.5$~\cm\,for the SW density.
		
		\begin{figure}
		\centering
			\begin{tabular}{cc}
			\includegraphics[scale=0.28]{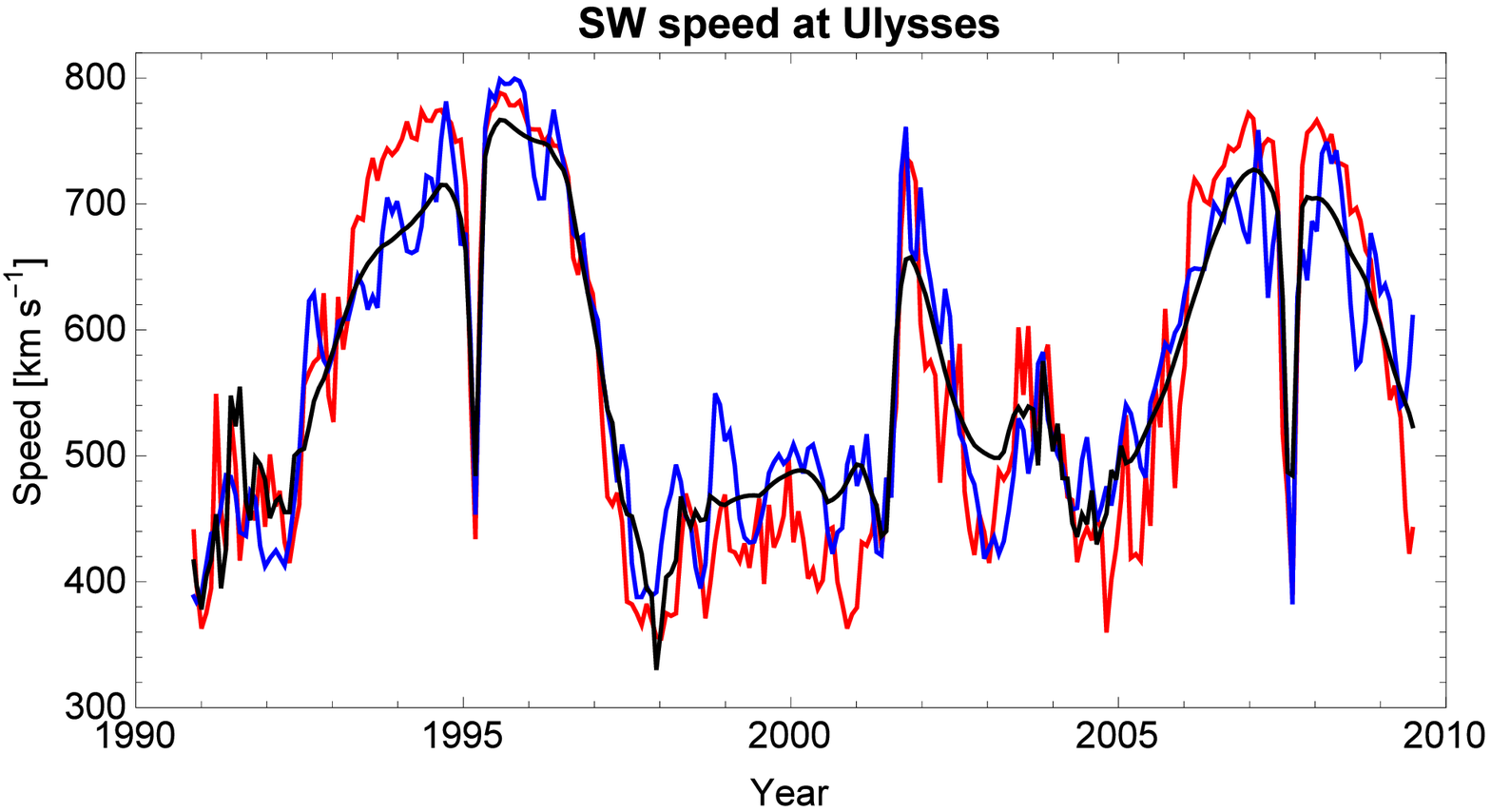} & \includegraphics[scale=0.28]{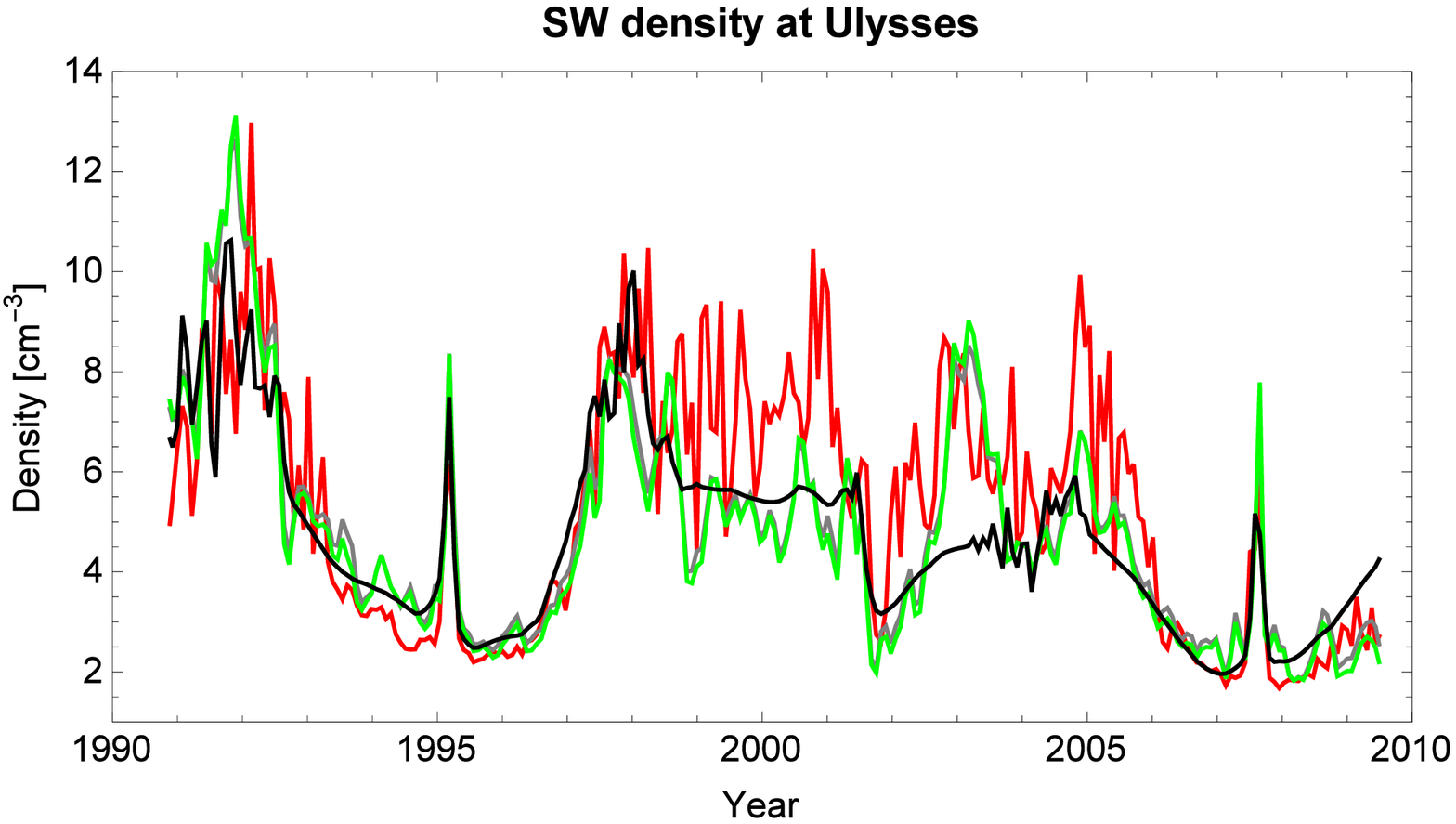}\\
			\includegraphics[scale=0.28]{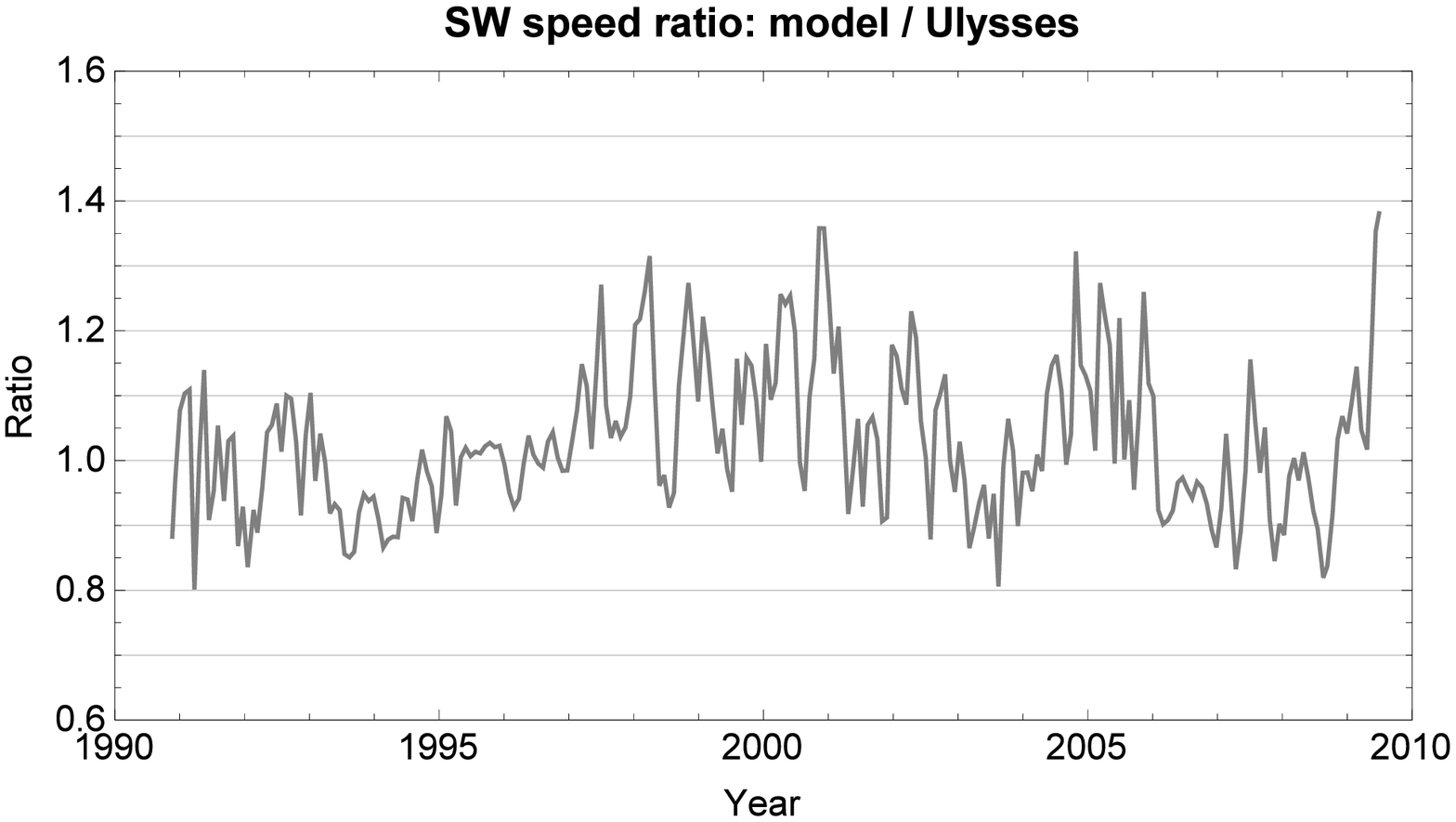} & \includegraphics[scale=0.28]{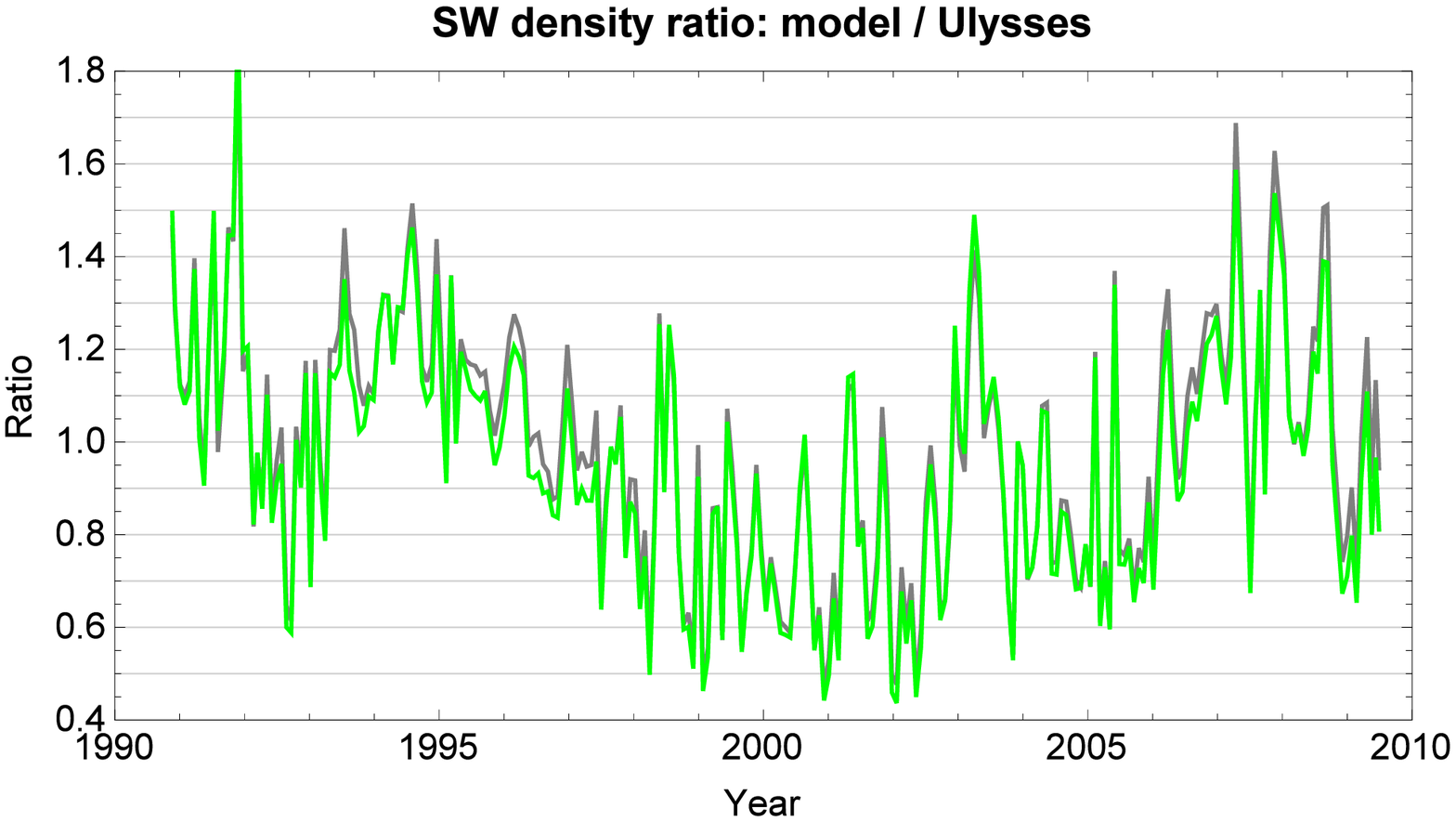}\\
			\includegraphics[scale=0.3]{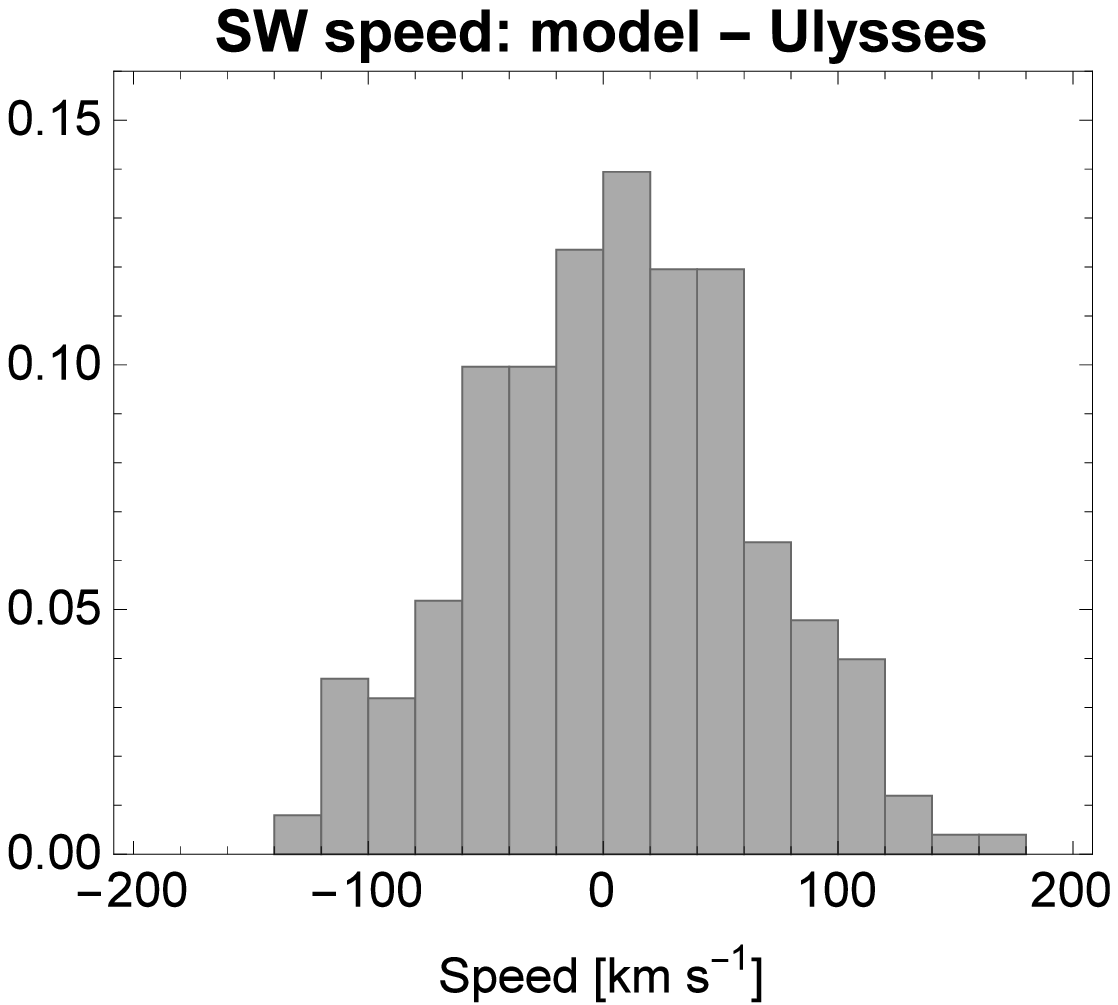} & \includegraphics[scale=0.3]{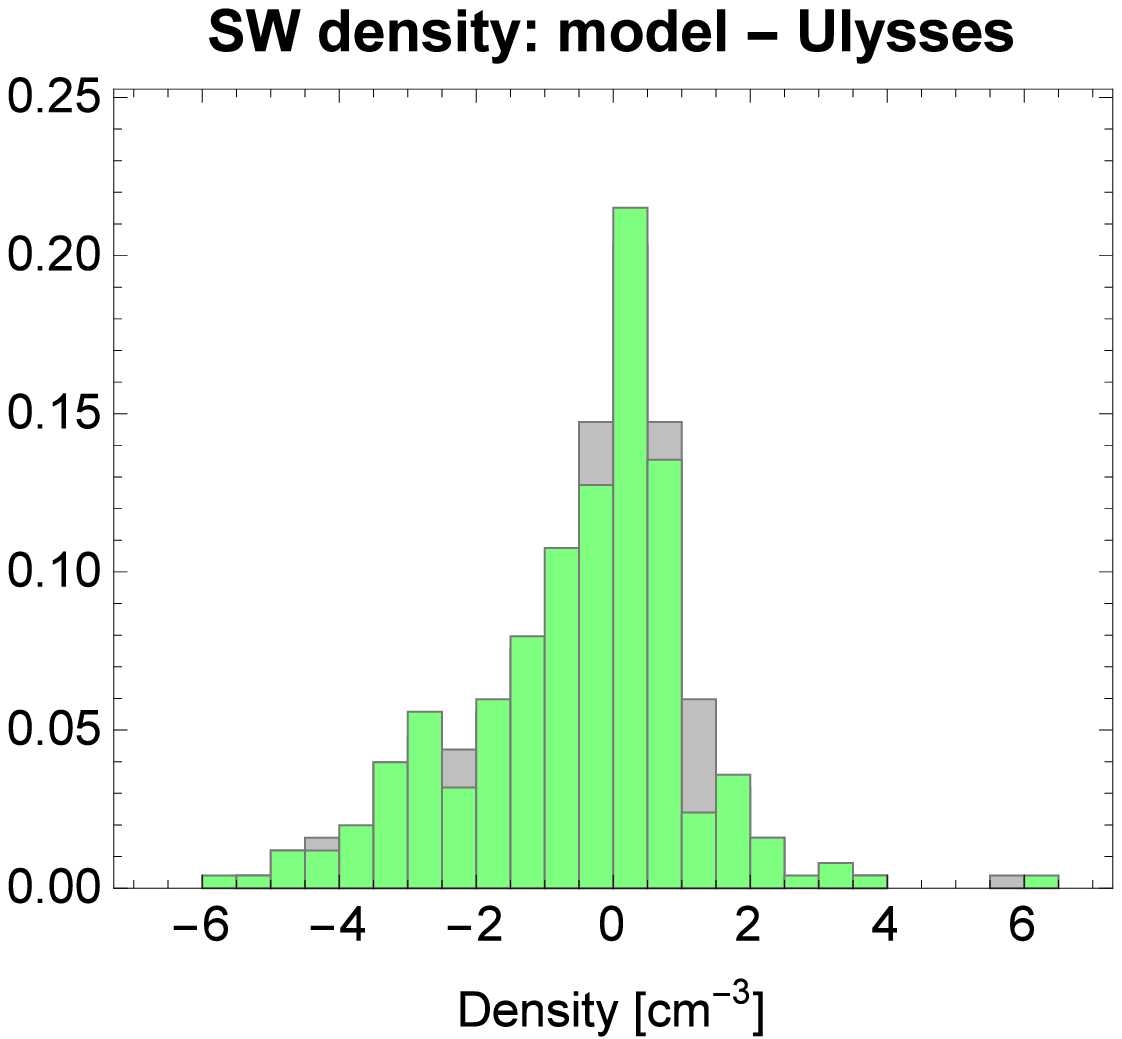}\\				
		\end{tabular}
		\caption{CR-averaged SW proton speed (left-hand column) and density (right-hand column) derived in our model are compared with the \textit{Ulysses} data (red lines in the top-row panels) at the positions of the spacecraft, scaled to 1~AU. In the top-left panel the blue line shows the SW speed reconstructed in this study; in the top-right panel the gray and green lines show the SW density calculated from the SW energy flux or dynamic pressure, respectively. The black lines in the top-row panels show the SW parameters given in \citet{sokol_etal:13a}. The middle-row panels show the ratios of the SW quantities in our model with respect to the \textit{Ulysses} data. The  bottom panels show the histograms of the differences between our model and the \textit{Ulysses} values; the color coding is the same as in the middle-row panels.}
		\label{figCompareUlysses}
		\end{figure}

We also compared the results of the present method with the out-of-ecliptic measurements from \textit{Ulysses}\footnote{\textit{Ulysses} \emph{Solar Wind Observations Over the Poles of the Sun} (SWOOPS) experiment, data source: \texttt{ftp://nssdcftp.gsfc.nasa.gov/spacecraft\_data/ulysses/plasma/swoops/ion/}.} and our previous model \citep{sokol_etal:13a} (Figure~\ref{figCompareUlysses}). The \textit{Ulysses} data are scaled to the distance of 1~AU. The present model traces the short time-scale variations of the SW speed much better than the previous model by \citet{sokol_etal:13a}. However, for some years it does not precisely reproduce the SW speeds measured by \textit{Ulysses}, where the speeds faster by more than 50~\kms (1993--1994) or slower (1999--2001) are measured. The comparison of SW density looks much better; the present model reproduces the SW density measured by \textit{Ulysses} with higher precision except for the solar maximum of SC~23, when \textit{Ulysses} sampled higher values.

The results of the reconstruction of the SW structure proposed by \citet{sokol_etal:13a} and in this paper are, for both speed and density, similar in general, but the present one is much better in detail. The blue line in the top-left panel of Figure~\ref{figCompareUlysses} oscillates around the black line, which means that the present model gives more precise information on time variations of the SW speed. Furthermore, the new method based on the SW helio-latitudinal invariants reproduces the SW density better than the model proposed by \citet{sokol_etal:13a} which was based on the relation between the SW speed and density measured by \textit{Ulysses} during its fast latitudinal scans (\textit{i.e.}, part of the orbit closest to the Sun). The higher resolution in latitude and in time in the present model has provided us with a better reconstruction of the monthly variations of the SW data, compared to the smooth time series in the yearly-averaged series from \citet{sokol_etal:13a}; see the comparison of density time series from 2002 to 2005 in Figure~\ref{figCompareUlysses} where the present model fits much better the in-situ measurements from \textit{Ulysses}. 

The previous \citep{sokol_etal:13a} and the present models of reconstruction of the SW structure, based on the IPS observations, indicate deviations from the measured \textit{Ulysses} values at the same places (in time and latitude). Since both models are based on the same SW speed data from IPS observations, these differences may be real; the \textit{Ulysses} data are only point measurements while the SW IPS data analysis gives information from a broader region of the sky. On the other hand, these differences may also point out that the IPS observations underestimate the SW speed at higher latitudes during the years with high solar activity.

Our final results for the SW speed and density, when compared to the in-situ values from in-ecliptic and out-of-ecliptic spacecraft, show that typically the reconstructed values differ by about $20\%$ from the measured values, except for solar maximum years of SC 23, when the discrepancies in the SW density increase to $40\%$. 

\section{Discussion}
The time vs. latitude maps of the SW proton speed and density at 1~AU (Figures~\ref{figMapFinalSpeed} and \ref{figMapFinalDens}) illustrate the variation of the SW structure over almost three SCs. Our analysis begins in 1985, during the low activity level of SC~22 and ends in 2013 in the middle of the maximum of SC~24. During the activity maximum the SW is almost homogeneous in helio-latitude, with slow and dense streams distributed from the equator to both poles. Inspection of the density map (Figure~\ref{figMapFinalDens}) shows a clear difference in the SW density from the maximum of SC~23. The SW is less dense during the maximum close to the solar equator than it was at the end of 1980s; the density is reduced from about 10~\cm~around 1991 to $\approx$5~\cm~in 2013. This confirms the secular changes in the SW observed in the in-ecliptic measurements (see Figures~1--3 in \citet{sokol_etal:13a} and also \citet{mccomas_etal:13b}). The maps show also very distinct solar minima ($\approx$1985, $\approx$1996, and $\approx$2009) with a slow (about 500~\kms) and dense (about 8~\cm) SW at low latitudes around the solar equator, and a fast (about 800~\kms) and dilute (about 3~\cm) flow at higher latitudes (\textit{i.e.}, outside the $\pm 20\degr-30\degr$ equatorial band). But the structure of the SW during these two solar minima is quite different; the slow SW extended much higher in helio-latitude (to about $\pm30\degr$) during the last SC than during the previous ones, when it reached to about $\pm20\degr$. These differences were observed also by \textit{Ulysses} (see Figure~6 in \citet{sokol_etal:13a} and also \citet{mccomas_etal:06a, mccomas_etal:08a}) and Figure~\ref{figCompareUlysses} in this article. The slow SW prevails in mid-latitudes much longer after the maximum of SC~23 than after the maximum of SC~22, as can be seen from the latitudinal structure of the SW speed and density for the years 2002--2005 in comparison with those for the years 1991--1993 (Figures~\ref{figMapFinalSpeed}, \ref{figMapFinalDens}, and \ref{figCompareUlysses}).

Our method also correctly reproduced the asymmetry between the north and south hemispheres, manifested mainly during the maximum of SC in the form of phase shift between the hemispheres. This aspect cannot be retrieved from the point measurements from in-situ measurements. The phase shift, noticed already by \citet{sokol_etal:13a}, has been observed, \textit{e.g.}, in the polar coronal holes by \citet{hess-webber_etal:14a} and in the analysis by \citet{tokumaru_etal:15a}.

\section{Summary and Conclusions}
The SW is one of the main factors that create the heliospheric environment from distances within a few solar radii of the Sun to beyond the edge of the solar system, where it interacts with the surrounding interstellar medium. To correctly account for its role in the heliospheric models, one needs to include time variations of its latitudinal structure. The distribution of the SW speed has a significant imprint on the production of energetic neutral atoms (ENA) of various energies \citep{mccomas_etal:12c, mccomas_etal:14b}. The \textit{Interstellar Boundary Explorer} (IBEX; \opencite{mccomas_etal:09a, mccomas_etal:09c}) probed the structure of the boundary layers of the heliosphere and its interaction with the surrounding interstellar medium through ENA observations \citep[\textit{e.g.}][]{zirnstein_etal:13a, heerikhuisen_etal:14a, czechowski_etal:15a}. Additionally, the 3D structures of the SW speed and density are also needed for the modeling of He ENA \citep{grzedzielski_etal:13a, grzedzielski_etal:14a, swaczyna_etal:14a}, to understand the measurements from the \textit{Voyager} spacecraft carried out in the inner heliosheath (\textit{e.g.} \opencite{stone_etal:05a, stone_etal:08a, richardson_etal:08b}; and recent works by \opencite{provornikova_etal:14a, grygorczuk_etal:15a}), and the origin and structure of the IBEX ribbon \citep[\textit{e.g.}][]{schwadron_mccomas:13a, funsten_etal:15a}. The continuous information about the latitudinal structure of the SW parameters is also essential for modeling of interstellar neutral (ISN) hydrogen (\textit{e.g.}~\citealp{katushkina_etal:13a}, \citeyear{katushkina_etal:14c}; \opencite{katushkina_etal:15a}) and explanation of the difference between the inflow direction derived from the observations of ISN He and H and its imprint on the asymmetry of the front structure of the heliosphere. Furthermore, knowledge of the variation of the global structure of the SW provides us with information about the long-term processes on the Sun, evolution of its cycle of activity, and differences between the north and south hemispheres.

In this paper we focused on variations of the helio-latitudinal profile of the SW proton speed and density from 1985 to 2013, \textit{i.e.}, during almost three SCs. We are interested in the long-term and large-scale variations of the SW for the purpose of modeling of the shape of the boundaries of the heliosphere and the ionization conditions in the heliosphere from the SW termination shock to 1~AU. For these studies we do not concentrate on the short time and small spatial scale details of SW variations that are necessary in the modeling of space climate or the MHD modeling of boundaries of the heliosphere \citep[\textit{e.g.}][]{vanderHolst_etal:10a, washimi_etal:11a, fujiki_etal:14a, kim_etal:14a}.  

Our aim was to increase the temporal and spatial resolution of the model developed by \citet{sokol_etal:13a}. We used the SW speed data retrieved from IPS observations from STEL \citep{tokumaru_etal:12b} and we focused on reconstruction of the data missing in time and helio-latitude in the CR maps. We adopted the methods commonly used in studies of geophysical data by decomposing them into spherical harmonics to reconstruct the missing regions in longitude and latitude, and the singular spectrum analysis to fill in the breaks in time. We left out the longitudinal variations of the SW speed and restricted the results only to the helio-latitudinal profiles. With this two-step reconstruction algorithm we were able to obtain a continuous time series of the SW speed from 1985 to 2013 with more detailed information about the variations at a given latitude over the whole CR. The details of the reconstruction are described in Appendices~\ref{secAppendixWlm} and \ref{secAppendixSSA}. The final time versus helio-latitude map of the SW speed is presented in Figure~\ref{figMapFinalSpeed}. 

After the SW speed was reconstructed, we calculated the helio-latitudinal variations in the SW density using the method that is based on latitudinal invariants of the SW: the SW dynamic pressure (Equation~(\ref{eqSWDynPress})) and the SW energy flux (Equation~(\ref{eqSWEnergyFlux})). The invariants were calculated from the in-ecliptic data taken from the OMNI database, and densities at various helio-latitudes were computed using the SW speeds that we obtained in the present analysis (see Figures~\ref{figSpeedProfiles} and \ref{figDensProfiles}). The resulting time versus helio-latitude map of the SW density at 1~AU, complementary to the analogous map of the SW speed, is presented in Figure~\ref{figMapFinalDens}. 

The SW speed and density profiles presented in Figures~\ref{figMapFinalSpeed} and \ref{figMapFinalDens} are available via the online supplementary material. 

We cross-validated our resulting SW speed and density data with other available observations, \textit{i.e.}, with the OMNI database for the ecliptic plane (Figure~\ref{figCompareOMNI}) and with \textit{Ulysses} measurements out of the ecliptic plane (Figure~\ref{figCompareUlysses}). The agreement in both cases is very good, except a small deviation in the ecliptic after 2008, which is probably related to the assumed $\Delta N_{\mathrm{e}} \varpropto V^{\gamma}$ relation used in the CAT analysis of IPS data (see Equation~(\ref{eqSpeedRetrieve})). The agreement with \textit{Ulysses} measurements is also very good. Additionally, the retrieval of the SW density based on the new method is better than in the model presented by \citet{sokol_etal:13a}. From the differences between our results and  the measured in-situ values, we conclude that the average goodness of the reconstruction is $\approx20\%$.

%%%%%%%%%%%%%%%%%%%%%%%%%%%%%%%%%%%%%%%%%%%%%%%%%%%%%%%%%%%%%%%%%%%%%%%%%%%
%% Acknowledgements
%
\begin{acks}
We acknowledge Space Physics Data Facility of NASA/GSFC for the \textit{Ulysses}/SWOOPS and the OMNI data, and DRAO and NOAA for F10.7 data. The IPS observations were carried out under the solar wind program of the Solar-Terrestrial Environment Laboratory of Nagoya University. The authors from SRC PAS were supported by grant 2012/06/M/ST9/00455 from the Polish National Science Center. 
\end{acks}

%%%%%%%%%%%%%%%%%%%%%%%%%%%%%%%%%%%%%%%%%%%%%%%%%%%%%%%%%%%%%%%%%%%%%%%%%%%
%% Appendix

\appendix 
\section{Algorithm of the Decomposition into the Spherical Harmonics of the Solar Wind Speed Data Retrieved from the IPS Observations}
\label{secAppendixWlm}
We use the spherical coordinate system with colatitude $\theta \in \left( 0,\pi\right)$, with $0$ for the north pole and $\pi$ for the south pole, and longitude $\phi \in \left( 0, 2\pi \right)$. On input, set $A$ includes all points on the sphere belonging to one CR map for which we have available SW speed values retrieved by the CAT from the STEL IPS observations (with the spatial resolution reduced to $3\degr \times 3\degr$ in $\theta$ and $\phi$), set $B$ is complementary to set $A$ and contains the coordinates of data gaps:
	\begin{equation}
	A=\left\{ \left( \theta_{i},\phi_{i} \right), \quad i=1,\ldots,N_{A}  \right\}, \qquad
	B=\left\{ \left( \theta_{j},\phi_{j} \right), \quad j=1,\ldots,N_{B}  \right\},
	\label{eqADefSetsAB}
	\end{equation}
where $N_{A}$ and $N_{B}$ are the numbers of elements in sets $A$ and $B$, respectively. By $f(\theta_i,\phi_i)$ we denote the SW speed values for the elements of set $A$, and by $\hat{f}(\theta_j,\phi_j)$ the estimates of the SW speed for the elements of set $B$. 

The coefficients $W_{\ell m}$ of the decomposition are given by
	\begin{equation}
	\begin{array}{lll}
	W_{\ell m} & = & \int\limits_{0}^{\pi}{\sin \theta \mathd{\theta}}\int\limits_{0}^{2\pi}{\mathd{\phi}Y_{\ell m}f(\theta,\phi)} \\
	& = & \left[ \sum_{\left( \theta_i,\phi_i\right)\in A}^{}{\sin{\theta_i}Y_{\ell m}\left( \theta_i, \phi_i \right) f \left( \theta_i, \phi_i  \right)} \right.\\
	& + & \left. \sum_{\left( \theta_j,\phi_j\right)\in B}^{}{\sin{\theta_j}Y_{\ell m}\left( \theta_j, \phi_j \right) \hat{f} \left( \theta_j, \phi_j \right)} \right] \Delta \theta \Delta \phi.
	\end{array}
	\label{eqDefWlm}
	\end{equation}
Here $Y_{\ell m}(\theta,\phi)$ is a real representation of the spherical harmonics. We have
	\begin{equation}
	Y_{\ell m}\left( \theta, \phi \right) = \left\{ 
	\begin{array}{ll}  
	\frac{i}{\sqrt{2}}\left[Y\left( \ell, -m, \theta, \phi \right)-\left( -1\right)^m Y\left( \ell,m,\theta,\phi\right) \right] & m < 0,\\
	Y\left( \ell,0,\theta,\phi\right) & m=0,\\
	\frac{1}{\sqrt{2}}\left[Y\left( \ell, -m, \theta, \phi \right)+\left( -1\right)^m Y\left( \ell,m,\theta,\phi\right) \right] & m > 0. 
	\end{array}
	\right.
	\label{eqDefYlm}
	\end{equation} 
The spherical harmonics $Y\left( \ell,m,\theta,\phi\right)$ were calculated using the built-in function \texttt{SphericalHarmonicY} in \emph{Mathematica~8} using the default precision and options. The integers $\ell$ and $m$ take the following values:
\begin{displaymath}
\ell = 0,1,2,\ldots,\ell_{\mathrm{max}}, \quad m=-\ell,-\ell+1,\ldots,0,1,\ldots,\ell-1,\ell .
\end{displaymath}

The assumed maximal order of $\ell$ gives the maximum spatial size $s$ of the structures that we are able to reconstruct by decomposition into spherical harmonics, according to the relation
\begin{displaymath}
\frac{180\degr}{\ell_{\mathrm{max}}} = s.
\end{displaymath}
This means that with a limiting value $\ell_{\mathrm{max}}=12$ we are able to reconstruct the spatial structures larger than $15\degr$. Additionally, the condition $\ell_{\mathrm{max}}^{2}<N_{A}$ needs to be fulfilled.

In the first step of analysis we select the maps with the minimal fraction of gaps in the CR maps for the given year, and we replace the elements of set $B$ with the average values over longitude for the given $3\degr$-latitude according to the description in Section~\ref{secMethodReconstruction}. The initial elements of set $B$ are replaced with values calculated using the coefficients of the decomposition into spherical harmonics,
	\begin{equation}
	\forall \left( \theta_j, \phi_j \right) \in B, \quad \hat{f}\left( \theta_j, \phi_j \right) = \sum_{l,m}^{}{W_{lm}Y_{lm}\left( \theta_j,\phi_j \right)}.
	\label{eqADefNewB}
	\end{equation}

In the first filling of elements in set $B$, the gaps in a given map for a selected year are approximately filled. In the next iteration step we search for the $W_{\ell m}$ coefficients with the elements of set $B$ filled with the values calculated from Equation~(\ref{eqADefNewB}). We iterate this process, calculating new sets of coefficients of the decomposition in spherical harmonics, until the average of $\delta_{k+1} = \left| \left( \hat{f}_k\left( \theta_j, \phi_j \right) - \hat{f}_{k+1}\left( \theta_j, \phi_j \right) \right) / \hat{f}_k\left( \theta_j, \phi_j \right) \right|$ for all elements in set $B$ is smaller than $0.01$, or if the ($\delta_{k+1}-\delta_{k}$) start to be greater than zero. 
 
\section{Algorithm of Singular Spectrum Analysis for Filling the Gaps in the Time Series of Coefficients of Spherical Harmonics}
\label{secAppendixSSA}
In this part of the analysis we introduce set $\mathcal{A}$ of the elements for which we have the coefficients of spherical harmonics, and set $\mathcal{B}$ of the elements for which we want to find the values (\textit{i.e.}, the gaps); $\mathcal{N}$ is the total number of elements in both sets $\mathcal{A}$ and $\mathcal{B}$. The window width $M$ of the maximal periodicity captured must fulfill the requirement of $\mathcal{N}/M \approx$~a few (for discussion see \opencite{ghil_etal:02a}). We take $M=60$, which means that the longest periodicity is determined at most by 60 consecutive points in the time series. The maximal order of iterations $J$ is set to $M/2$, and the condition for the convergence of the iteration adopted is $\epsilon = 0.01$.

In the SSA analysis we use only the coefficients of spherical harmonics for those CR maps of the SW speed for which the fraction of gaps is smaller than $50\%$. Additionally, we focus only on the helio-latitudinal profiles, so we limit the $\mathcal{A}$ set to the coefficients with $m=0$. 

Before we start the SSA we correlate the time series $W_{\ell m}(t)$ with a proxy $\eta(t)$ of the solar activity cycle for elements from set $\mathcal{A}$ (following the rationale given in Section~\ref{sec:fillingSSA}), with $t=1,\ldots,\mathcal{N_{\mathcal{A}}}$, where $\mathcal{N_{\mathcal{A}}}$ is the length of set $\mathcal{A}$. We use the CR averages of solar-radio 10.7~cm flux (F10.7) smoothed by a 13~CR running average (Figure~\ref{figF107}). We look for $\alpha_{\ell m}$ and $\beta_{\ell m}$ from the minimization of the expression
	\begin{equation}
	\sum_{t \in \mathcal{A}}^{}{|W_{\ell m}\left( t \right) - \alpha_{\ell m} \eta\left( t \right) - \beta_{\ell m}|^2}.
	\end{equation}
Next we normalize $\kappa_{\ell m}\left( t \right)$ dividing it by the standard deviation (stdDev),
	\begin{equation}
	\kappa_{\ell m}\left( t \right) = W_{\ell m}\left( t \right) - \alpha_{\ell m} \eta\left( t \right) - \beta_{\ell m},
	\label{eqAKappalm}
	\end{equation}
	\begin{equation}
	X_{\ell m}\left( t \right) = \frac{\kappa_{\ell m}\left( t \right)}{\mathrm{stdDev}\left( \kappa_{\ell m} \right)}.
	\label{eqAXlm}
	\end{equation}	
Then we apply SSA to the time series $X_{\ell m}\left( t \right)$.

The SSA analysis of the time series $X(t)$ of the normalized coefficients $W_{\ell m}$ of the elements $t \in \mathcal{A}$ starts with the initial conditions of the iterators $p=1$, $r=1$, a normalization parameter $S\left(p,r \right)=1$, and $\forall_{t \in \mathcal{A}} X^{(p=1,r=1)}(t)=X(t)$, $\forall_{t \in \mathcal{B}} X^{(p=1,r=1)}(t)=0$. In the first step, we calculate the matrix $C^{(p,r)}$ with $i = 1,\ldots,M$, $j=1,\ldots,M$,
	\begin{equation}
	C_{ij}^{(p,r)} = \frac{1}{\mathcal{N}-|i-j|}\sum_{t=1}^{\mathcal{N}-|i-j|}{X^{(p,r)}(t)X^{(p,r)}(t+|i-j|)},
	\label{eqACijp}
	\end{equation}
where $X^{(p,r)}$ is a set with a sum of elements from sets $\mathcal{A}$ and $\mathcal{B}$. Next, the eigenvectors and eigenvalues of the matrix $C_{ij}^{(p,r)}$ are found $\left( E_{k}^{(p,r)}, \lambda_{k}^{(p,r)}\right)$, respectively, decreasingly ranked with respect to the eigenvalues. In the following step, we calculate
	\begin{equation}
	R^{(p,r)}\left( t \right) = M_t \sum_{k=1}^{p}\sum_{j=L_t}^{U_t}{A_{k}^{(p,r)}\left( t-j+1 \right)E_{k}^{(p,r)}\left( j \right)} 
	\label{eqRpr}
	\end{equation}
with
	\begin{equation}
	A_{k}^{(p,r)}\left( t \right) = \sum_{j'=1}^{M}{X^{(p,r)}\left( t+j'-1 \right)E_{k}^{(p,r)}\left( j' \right)}, 
	\end{equation}
and $\left(M_t, L_t, U_t \right)$ are defined after Equation~(12) of \citet{ghil_etal:02a} as follows\footnote{Please notice that in Equation~(\ref{eqRpr}) we use the normalization factor $M_t$ instead of $1/M_t$ as it was in Equation~(11) of \citet{ghil_etal:02a}, because we note a misprint in Equation~(12) of \citet{ghil_etal:02a}, where it should be $\left(1/M_t, L_t, U_t \right)$ to correctly normalize their Equation~(11).}:
	\begin{equation}
	\left( M_t,L_t,U_t \right) = \left\{ \begin{array}{ll}
	\left( \frac{1}{t},1,t \right) & 1 \leq t \leq M-1, \\
	\left( \frac{1}{M}, 1, M \right) &  M \leq t \leq \mathcal{N}', \\
	\left( \frac{1}{\mathcal{N}-t+1}, t-\mathcal{N}+M, M \right) &  \mathcal{N}'+1 \leq t \leq \mathcal{N}, \\
	\end{array}\right.
	\end{equation}
with $\mathcal{N}'=\mathcal{N}-M+1$.

Having computed $R^{(p,r)}\left( t \right)$, we calculate the mean square difference $Q^{(p,r)}$,
	\begin{equation}
	Q^{(p,r)}=\sum_{t \in \mathcal{A}}^{}\left( R^{(p,r)} \left( t \right) - X \left( t \right) / S\left(p,r\right) \right)^2,
	\label{eqAQpr}
	\end{equation}
and assign the new values to the elements of set $\mathcal{B}$ using $\forall_{t \in \mathcal{B}}X^{(p)}\left( t \right) = R^{(p,r)}\left( t \right)$.

We normalize all elements from sets $\mathcal{A}$ and $\mathcal{B}$ in the following way:
	\begin{equation}
	X^{(p,r+1)}\left( t \right) = \frac{X^{(p,r)}\left( t \right)}{\mathrm{stdDev}\left( X^{(p,r)}\left( t \right) \right)}.
	\end{equation}
With the $\mathrm{stdDev}$ known for each set, we assign a new value to the normalization parameter $S \left(p,r+1\right) = S\left(p,r\right) \cdot \mathrm{stdDev}\left( X^{(p,r)}\left(t\right) \right)$.

All steps described above are repeated iteratively in two loops. The internal one goes over iterator $r$ with iterator $p$ kept constant, and the external one over iterator $p$. The internal loop over $r$ checks the value of $r$, and if $r=1$, then increases $r$ by one to $r=2$ and starts the calculations from Equations~(\ref{eqACijp}) to (\ref{eqAQpr}). If $r>1$ then the condition $Q^{(p,r)}/Q^{(p,r-1)} < 1 - \epsilon$ is checked, and if it is true, then $r$ is increased by one $(r\rightarrow r+1)$, but if not, then $p$ is increased by one $(p\rightarrow p+1)$, $r$ is again set to be $r=1$, and $\forall_{t}X^{(p+1,r=1)}(t)=X^{(p,r)}(t)$. The external loop over $p$ continues as long as $p \leq J$.

The reconstructed (filled) time series of the coefficients of the spherical harmonics are built with
	\begin{equation}
	\tilde{X}\left( t \right)= \left\{ \begin{array}{ll}
	X \left( t \right) & t \in \mathcal{A}, \\
	S^{(p,r)} \cdot X^{(p,r)}\left( t \right) & t \in \mathcal{B}. \\\end{array}\right.
	\end{equation}
At the end we have to reverse the normalization and go back to the real magnitudes of $W_{\ell m}$. To do this we use Equations~(\ref{eqAKappalm}) and (\ref{eqAXlm}):
	\begin{equation}
	W_{\ell m} = \tilde{X} \left( t \right) \mathrm{stdDev} \left( \kappa_{\ell m} \right)+\alpha_{\ell m} f\left( t \right)+\beta_{\ell m},
	\end{equation}
for all elements from $\tilde{X} \left( t \right)$.

%%% %%%%%%%%%%%%%%%%%%%%%%%%%%%%%%%%%%%%%%%%%%%%%%%%%%%%%%%%%%%
%% Bibliography
%
% Using BibTeX
%
 \bibliographystyle{spr-mp-sola}
 \bibliography{iplbib}  

\begin{thebibliography}{75}
% BibTex style file: spr-mp-sola.bst (nameyear), 2015-03-09
\ifx\bisbn     \undefined \def\bisbn  #1{ISBN #1}\fi
\ifx\binits    \undefined \def\binits#1{#1}\fi
\ifx\bauthor   \undefined \def\bauthor#1{#1}\fi
\ifx\batitle   \undefined \def\batitle#1{#1}\fi
\ifx\bjtitle   \undefined \def\bjtitle#1{\textit{#1}}\fi
\ifx\bvolume   \undefined \def\bvolume#1{\textbf{#1}}\fi
\ifx\byear     \undefined \def\byear#1{#1}\fi
\ifx\bissue    \undefined \def\bissue#1{#1}\fi
\ifx\bfpage    \undefined \def\bfpage#1{#1}\fi
\ifx\blpage    \undefined \def\blpage #1{#1}\fi
\ifx\burl      \undefined \def\burl#1{\textsf{#1}}\fi
\ifx\href      \undefined \def\href#1#2{\textsf{#2}}\fi
\ifx\betal     \undefined \def\betal{\textit{et al.}}\fi
\ifx\bctitle   \undefined \def\bctitle#1{#1}\fi
\ifx\beditor   \undefined \def\beditor#1{#1}\fi
\ifx\bbtitle   \undefined \def\bbtitle#1{\textit{#1}}\fi
\ifx\bedition  \undefined \def\bedition#1{#1}\fi
\ifx\bseriesno \undefined \def\bseriesno#1{\textbf{#1}}\fi
\ifx\blocation \undefined \def\blocation#1{#1}\fi
\ifx\bsertitle \undefined \def\bsertitle#1{\textit{#1}}\fi
\ifx\bsnm      \undefined \def\bsnm#1{#1}\fi
\ifx\bsuffix   \undefined \def\bsuffix#1{#1}\fi
\ifx\bparticle \undefined \def\bparticle#1{#1}\fi
\ifx\barticle  \undefined \def\barticle#1{}\fi
\ifx\binstitute  \undefined \def\binstitute#1{#1}\fi
\ifx\bpublisher  \undefined \def\bpublisher#1{#1}\fi
\ifx\doiurl    \undefined
  \def\doiurl#1{\href{http://dx.doi.org/#1}{\textsf{DOI}}}\fi
\ifx\arxivurl  \undefined
  \def\arxivurl#1{\href{http://arxiv.org/abs/#1}{\textsf{arXiv}}}\fi
\ifx\adsurl    \undefined
  \def\adsurl#1{\href{http://adsabs.harvard.edu/abs/#1}{\textsf{ADS}}}\fi
\ifx\botherref \undefined \def\botherref#1{}\fi
\ifx\url       \undefined \def\url#1{\textsf{#1}}\fi
\ifx\bchapter  \undefined \def\bchapter#1{}\fi
\ifx\bbook     \undefined \def\bbook#1{}\fi
\ifx\bcomment  \undefined \def\bcomment#1{#1}\fi
\ifx\oauthor   \undefined \def\oauthor#1{#1}\fi
\ifx\citeauthoryear \undefined\def \citeauthoryear#1{#1}\fi
\ifx\endbibitem\undefined \def\endbibitem{}\fi
\ifx\bconflocation  \undefined \def\bconflocation#1{#1} \fi

\bibitem[\protect\citeauthoryear{{Asai} \textit{et~al.}}{1998}]{asai_etal:98a}
\begin{barticle}
\bauthor{\bsnm{{Asai}}, \binits{K.}},
\bauthor{\bsnm{{Kojima}}, \binits{M.}},
\bauthor{\bsnm{{Tokumaru}}, \binits{M.}},
\bauthor{\bsnm{{Yokobe}}, \binits{A.}},
\bauthor{\bsnm{{Jackson}}, \binits{B.V.}},
\bauthor{\bsnm{{Hick}}, \binits{P.L.}},
\bauthor{\bsnm{{Manoharan}}, \binits{P.K.}}:
\byear{1998},
\batitle{{Heliospheric tomography using interplanetary scintillation
  observations 3. Correlation between speed and electron density fluctuations
  in the solar wind}}.
\bjtitle{\jgr}
\bvolume{103},
\bfpage{1991}.
\doiurl{10.1029/97JA02750}.
\end{barticle}
\endbibitem

\bibitem[\protect\citeauthoryear{{Bame} \textit{et~al.}}{1992}]{bame_etal:92a}
\begin{barticle}
\bauthor{\bsnm{{Bame}}, \binits{S.J.}},
\bauthor{\bsnm{{McComas}}, \binits{D.J.}},
\bauthor{\bsnm{{Barraclough}}, \binits{B.L.}},
\bauthor{\bsnm{{Phillips}}, \binits{J.L.}},
\bauthor{\bsnm{{Sofaly}}, \binits{K.J.}},
\bauthor{\bsnm{{Chavez}}, \binits{J.C.}},
\bauthor{\bsnm{{Goldstein}}, \binits{B.E.}},
\bauthor{\bsnm{{Sakurai}}, \binits{R.K.}}:
\byear{1992},
\batitle{{The ULYSSES solar wind plasma experiment}}.
\bjtitle{\aaps}
\bvolume{92},
\bfpage{237}.
\end{barticle}
\endbibitem

\bibitem[\protect\citeauthoryear{{Bisi} \textit{et~al.}}{2009}]{bisi_etal:09a}
\begin{barticle}
\bauthor{\bsnm{{Bisi}}, \binits{M.M.}},
\bauthor{\bsnm{{Jackson}}, \binits{B.V.}},
\bauthor{\bsnm{{Buffington}}, \binits{A.}},
\bauthor{\bsnm{{Clover}}, \binits{J.M.}},
\bauthor{\bsnm{{Hick}}, \binits{P.P.}},
\bauthor{\bsnm{{Tokumaru}}, \binits{M.}}:
\byear{2009},
\batitle{{Low-resolution STELab IPS 3D reconstructions of the Whole Heliosphere
  Interval and comparison with in-ecliptic solar wind measurements from STEREO
  and Wind instrumentation}}.
\bjtitle{\solphys}
\bvolume{256},
\bfpage{201}.
\doiurl{10.1007/s11207-009-9350-9}.
\end{barticle}
\endbibitem

\bibitem[\protect\citeauthoryear{{Bisi} \textit{et~al.}}{2010}]{bisi_etal:10d}
\begin{barticle}
\bauthor{\bsnm{{Bisi}}, \binits{M.M.}},
\bauthor{\bsnm{{Jackson}}, \binits{B.V.}},
\bauthor{\bsnm{{Fallows}}, \binits{R.A.}},
\bauthor{\bsnm{{Dorrian}}, \binits{G.D.}},
\bauthor{\bsnm{{Manoharan}}, \binits{P.K.}},
\bauthor{\bsnm{{Clover}}, \binits{J.M.}},
\bauthor{\bsnm{{Hick}}, \binits{P.P.}},
\bauthor{\bsnm{{Buffington}}, \binits{A.}},
\bauthor{\bsnm{{Breen}}, \binits{A.R.}},
\bauthor{\bsnm{{Tokumaru}}, \binits{M.}}:
\byear{2010},
\batitle{{Solar wind and CME studies of the inner heliosphere using IPS data
  from STELab, ORT, and EISCAT}}.
\bjtitle{\advGeosci}
\bvolume{21},
\bfpage{33}.
\end{barticle}
\endbibitem

\bibitem[\protect\citeauthoryear{Bzowski
  \textit{et~al.}}{2003}]{bzowski_etal:03a}
\begin{barticle}
\bauthor{\bsnm{Bzowski}, \binits{M.}},
\bauthor{\bsnm{M{\"a}kinen}, \binits{T.}},
\bauthor{\bsnm{Kyr{\"o}l{\"a}}, \binits{E.}},
\bauthor{\bsnm{Summanen}, \binits{T.}},
\bauthor{\bsnm{Qu{\`e}merais}, \binits{E.}}:
\byear{2003},
\batitle{Latitudinal structure and north-south asymmetry of the solar wind from
  {L}yman-$\alpha$ remote sensing by {SWAN}}.
\bjtitle{\aap}
\bvolume{408},
\bfpage{1165}.
\end{barticle}
\endbibitem

\bibitem[\protect\citeauthoryear{Bzowski
  \textit{et~al.}}{2013}]{bzowski_etal:13a}
\begin{bchapter}
\bauthor{\bsnm{Bzowski}, \binits{M.}},
\bauthor{\bsnm{Sok{\'{o}}{\l}}, \binits{J.M.}},
\bauthor{\bsnm{Tokumaru}, \binits{M.}},
\bauthor{\bsnm{Fujiki}, \binits{K.}},
\bauthor{\bsnm{Quemerais}, \binits{E.}},
\bauthor{\bsnm{Lallement}, \binits{R.}},
\bauthor{\bsnm{Ferron}, \binits{S.}},
\bauthor{\bsnm{Bochsler}, \binits{P.}},
\bauthor{\bsnm{McComas}, \binits{D.J.}}:
\byear{2013},
\bctitle{Solar parameters for modeling interplanetary background}.
In: \beditor{\bsnm{Bonnet}, \binits{R.M.}},
\beditor{\bsnm{Qu{\'e}merais}, \binits{E.}},
\beditor{\bsnm{Snow}, \binits{M.}} (eds.)
\bbtitle{{Cross-Calibration of Past and Present Far {UV} Spectra of Solar
  Objects and the Heliosphere}},
\bpublisher{{Springer Science+Business Media}},
\blocation{{New York}},
\bfpage{67}.
\end{bchapter}
\endbibitem

\bibitem[\protect\citeauthoryear{{Coles} and
  {Kaufman}}{1978}]{coles_kaufman:78a}
\begin{barticle}
\bauthor{\bsnm{{Coles}}, \binits{W.A.}},
\bauthor{\bsnm{{Kaufman}}, \binits{J.J.}}:
\byear{1978},
\batitle{{Solar wind velocity estimation from multi-station IPS}}.
\bjtitle{Radio Sci.}
\bvolume{13},
\bfpage{591}.
\doiurl{10.1029/RS013i003p00591}.
\end{barticle}
\endbibitem

\bibitem[\protect\citeauthoryear{{Coles} and {Maagoe}}{1972}]{coles_maagoe:72a}
\begin{barticle}
\bauthor{\bsnm{{Coles}}, \binits{W.A.}},
\bauthor{\bsnm{{Maagoe}}, \binits{S.}}:
\byear{1972},
\batitle{{Solar-wind velocity from IPS observations}}.
\bjtitle{\jgr}
\bvolume{77},
\bfpage{5622}.
\doiurl{10.1029/JA077i028p05622}.
\end{barticle}
\endbibitem

\bibitem[\protect\citeauthoryear{{Coles}
  \textit{et~al.}}{1980}]{coles_etal:80a}
\begin{barticle}
\bauthor{\bsnm{{Coles}}, \binits{W.A.}},
\bauthor{\bsnm{{Rickett}}, \binits{B.J.}},
\bauthor{\bsnm{{Rumsey}}, \binits{V.H.}},
\bauthor{\bsnm{{Kaufman}}, \binits{J.J.}},
\bauthor{\bsnm{{Turley}}, \binits{D.G.}},
\bauthor{\bsnm{{Ananthakrishnan}}, \binits{S.}},
\bauthor{\bsnm{{Armstrong}}, \binits{J.W.}},
\bauthor{\bsnm{{Harmons}}, \binits{J.K.}},
\bauthor{\bsnm{{Scott}}, \binits{S.L.}},
\bauthor{\bsnm{{Sime}}, \binits{D.G.}}:
\byear{1980},
\batitle{{Solar cycle changes in the polar solar wind}}.
\bjtitle{\nat}
\bvolume{286},
\bfpage{239}.
\doiurl{10.1038/286239a0}.
\end{barticle}
\endbibitem

\bibitem[\protect\citeauthoryear{{Coles}
  \textit{et~al.}}{1995}]{coles_etal:95a}
\begin{barticle}
\bauthor{\bsnm{{Coles}}, \binits{W.A.}},
\bauthor{\bsnm{{Grall}}, \binits{R.R.}},
\bauthor{\bsnm{{Klinglesmith}}, \binits{M.T.}},
\bauthor{\bsnm{{Bourgois}}, \binits{G.}}:
\byear{1995},
\batitle{{Solar cycle changes in the level of compressive microturbulence near
  the Sun}}.
\bjtitle{\jgr}
\bvolume{100},
\bfpage{17069}.
\doiurl{10.1029/95JA00896}.
\end{barticle}
\endbibitem

\bibitem[\protect\citeauthoryear{{Czechowski}, {Grygorczuk}, and
  {McComas}}{2015}]{czechowski_etal:15a}
\begin{barticle}
\bauthor{\bsnm{{Czechowski}}, \binits{A.}},
\bauthor{\bsnm{{Grygorczuk}}, \binits{J.}},
\bauthor{\bsnm{{McComas}}, \binits{D.J.}}:
\byear{2015},
\batitle{{Heliosphere in strong magnetic field as a source of energetic neutral
  atoms}}.
\bjtitle{\aap}
\bvolume{000},
\bfpage{000}.
\doiurl{000}.
\end{barticle}
\endbibitem

\bibitem[\protect\citeauthoryear{{Dudok de Wit}}{2011}]{dudokdewit:11b}
\begin{barticle}
\bauthor{\bsnm{{Dudok de Wit}}, \binits{T.}}:
\byear{2011},
\batitle{{A method for filling gaps in solar irradiance and solar proxy data}}.
\bjtitle{\aap}
\bvolume{533},
\bfpage{A29}.
\doiurl{10.1051/0004-6361/201117024}.
\end{barticle}
\endbibitem

\bibitem[\protect\citeauthoryear{{Fujiki}
  \textit{et~al.}}{2014}]{fujiki_etal:14a}
\begin{barticle}
\bauthor{\bsnm{{Fujiki}}, \binits{K.}},
\bauthor{\bsnm{{Washimi}}, \binits{H.}},
\bauthor{\bsnm{{Hayashi}}, \binits{K.}},
\bauthor{\bsnm{{Zank}}, \binits{G.P.}},
\bauthor{\bsnm{{Tokumaru}}, \binits{M.}},
\bauthor{\bsnm{{Tanaka}}, \binits{T.}},
\bauthor{\bsnm{{Florinski}}, \binits{V.}},
\bauthor{\bsnm{{Kubo}}, \binits{Y.}}:
\byear{2014},
\batitle{{MHD analysis of the velocity oscillations in the outer heliosphere}}.
\bjtitle{\grl}
\bvolume{41},
\bfpage{1420}.
\doiurl{10.1002/2014GL059391}.
\end{barticle}
\endbibitem

\bibitem[\protect\citeauthoryear{{Funsten}
  \textit{et~al.}}{2015}]{funsten_etal:15a}
\begin{barticle}
\bauthor{\bsnm{{Funsten}}, \binits{H.O.}},
\bauthor{\bsnm{{Bzowski}}, \binits{M.}},
\bauthor{\bsnm{{Cai}}, \binits{D.M.}},
\bauthor{\bsnm{{Dayeh}}, \binits{M.}},
\bauthor{\bsnm{{DeMajistre}}, \binits{R.}},
\bauthor{\bsnm{{Frisch}}, \binits{P.C.}},
\bauthor{\bsnm{{Herrikhuisen}}, \binits{J.}},
\bauthor{\bsnm{{Hidgon}}, \binits{D.M.}},
\bauthor{\bsnm{{Janzen}}, \binits{P.}},
\bauthor{\bsnm{{Larsen}}, \binits{B.A.}},
\bauthor{\bsnm{{Livadiotis}}, \binits{G.}},
\bauthor{\bsnm{{McComas}}, \binits{D.J.}},
\bauthor{\bsnm{{M{\"o}bius}}, \binits{E.}},
\bauthor{\bsnm{{Reese}}, \binits{C.S.}},
\bauthor{\bsnm{{Roelof}}, \binits{E.C.}},
\bauthor{\bsnm{{Reisenfeld}}, \binits{D.B.}},
\bauthor{\bsnm{{Schwadron}}, \binits{N.A.}},
\bauthor{\bsnm{{Zirnstein}}, \binits{E.J.}}:
\byear{2015},
\batitle{{Symmetry of the IBEX Ribbon of enhanced energetic neutral atom (ENA)
  flux}}.
\bjtitle{\apj}
\bvolume{799},
\bfpage{68}.
\doiurl{10.1088/0004-637X/799/1/68}.
\end{barticle}
\endbibitem

\bibitem[\protect\citeauthoryear{{Gapper}
  \textit{et~al.}}{1982}]{gapper_etal:82a}
\begin{barticle}
\bauthor{\bsnm{{Gapper}}, \binits{G.R.}},
\bauthor{\bsnm{{Hewish}}, \binits{A.}},
\bauthor{\bsnm{{Purvis}}, \binits{A.}},
\bauthor{\bsnm{{Duffett-Smith}}, \binits{P.J.}}:
\byear{1982},
\batitle{{Observing interplanetary disturbances from the ground}}.
\bjtitle{\nat}
\bvolume{296},
\bfpage{633}.
\doiurl{10.1038/296633a0}.
\end{barticle}
\endbibitem

\bibitem[\protect\citeauthoryear{{Ghil} \textit{et~al.}}{2002}]{ghil_etal:02a}
\begin{barticle}
\bauthor{\bsnm{{Ghil}}, \binits{M.}},
\bauthor{\bsnm{{Allen}}, \binits{M.R.}},
\bauthor{\bsnm{{Dettinger}}, \binits{M.D.}},
\bauthor{\bsnm{{Ide}}, \binits{K.}},
\bauthor{\bsnm{{Kondrashov}}, \binits{D.}},
\bauthor{\bsnm{{Mann}}, \binits{M.E.}},
\bauthor{\bsnm{{Robertson}}, \binits{A.W.}},
\bauthor{\bsnm{{Saunders}}, \binits{A.}},
\bauthor{\bsnm{{Tian}}, \binits{Y.}},
\bauthor{\bsnm{{Varadi}}, \binits{F.}},
\bauthor{\bsnm{{Yiou}}, \binits{P.}}:
\byear{2002},
\batitle{{Advanced spectral methods for climatic time series}}.
\bjtitle{\rgp}
\bvolume{40},
\bfpage{1003}.
\doiurl{10.1029/2000RG000092}.
\end{barticle}
\endbibitem

\bibitem[\protect\citeauthoryear{{Grygorczuk}, {Czechowski}, and
  {Grzedzielski}}{2015}]{grygorczuk_etal:15a}
\begin{botherref}
\oauthor{\bsnm{{Grygorczuk}}, \binits{J.}},
\oauthor{\bsnm{{Czechowski}}, \binits{A.}},
\oauthor{\bsnm{{Grzedzielski}}, \binits{S.}}:
2015,
{Approximate mirror symmetry in heliospheric plasma flow explains VOYAGER 2
  observations}.
\textit{ArXiv e-prints}.
\end{botherref}
\endbibitem

\bibitem[\protect\citeauthoryear{{Grzedzielski}, {Swaczyna}, and
  {Bzowski}}{2013}]{grzedzielski_etal:13a}
\begin{barticle}
\bauthor{\bsnm{{Grzedzielski}}, \binits{S.}},
\bauthor{\bsnm{{Swaczyna}}, \binits{P.}},
\bauthor{\bsnm{{Bzowski}}, \binits{M.}}:
\byear{2013},
\batitle{{Heavy coronal ions in the heliosphere. II. Expected fluxes of
  energetic neutral He atoms from the heliosheath}}.
\bjtitle{\aap}
\bvolume{549},
\bfpage{A76+}.
\doiurl{10.1051/004-6361/201220104}.
\end{barticle}
\endbibitem

\bibitem[\protect\citeauthoryear{{Grzedzielski}
  \textit{et~al.}}{2014}]{grzedzielski_etal:14a}
\begin{barticle}
\bauthor{\bsnm{{Grzedzielski}}, \binits{S.}},
\bauthor{\bsnm{{Swaczyna}}, \binits{P.}},
\bauthor{\bsnm{{Czechowski}}, \binits{A.}},
\bauthor{\bsnm{{Hilchenbach}}, \binits{M.}}:
\byear{2014},
\batitle{{Solar wind He pickup ions as source of tens-of-keV/n neutral He atoms
  observed by the HSTOF/SOHO detector}}.
\bjtitle{\aap}
\bvolume{563},
\bfpage{A134}.
\doiurl{10.1051/0004-6361/201322927}.
\end{barticle}
\endbibitem

\bibitem[\protect\citeauthoryear{{Heerikhuisen}
  \textit{et~al.}}{2014}]{heerikhuisen_etal:14a}
\begin{barticle}
\bauthor{\bsnm{{Heerikhuisen}}, \binits{J.}},
\bauthor{\bsnm{{Zirnstein}}, \binits{E.J.}},
\bauthor{\bsnm{{Funsten}}, \binits{H.O.}},
\bauthor{\bsnm{{Pogorelov}}, \binits{N.V.}},
\bauthor{\bsnm{{Zank}}, \binits{G.P.}}:
\byear{2014},
\batitle{{The effect of new interstellar medium parameters on the heliosphere
  and energetic neutral atoms from the interstellar boundary}}.
\bjtitle{\apj}
\bvolume{784},
\bfpage{73}.
\doiurl{10.1088/0004-637X/784/1/73}.
\end{barticle}
\endbibitem

\bibitem[\protect\citeauthoryear{{Hess Webber}
  \textit{et~al.}}{2014}]{hess-webber_etal:14a}
\begin{botherref}
\oauthor{\bsnm{{Hess Webber}}, \binits{S.A.}},
\oauthor{\bsnm{{Karna}}, \binits{N.}},
\oauthor{\bsnm{{Pesnell}}, \binits{W.D.}},
\oauthor{\bsnm{{Kirk}}, \binits{M.S.}}:
2014,
{Areas of polar coronal holes from 1996 through 2010}.
\textit{\solphys}.
\doiurl{10.1007/s11207-014-0564-0}.
\end{botherref}
\endbibitem

\bibitem[\protect\citeauthoryear{{Hewish}, {Scott}, and
  {Wills}}{1964}]{hewish_etal:64a}
\begin{barticle}
\bauthor{\bsnm{{Hewish}}, \binits{A.}},
\bauthor{\bsnm{{Scott}}, \binits{P.F.}},
\bauthor{\bsnm{{Wills}}, \binits{D.}}:
\byear{1964},
\batitle{Interplanetary scintillation of small diameter radio sources}.
\bjtitle{\nat}
\bvolume{203},
\bfpage{1214}.
\doiurl{10.1038/2031214a0}.
\end{barticle}
\endbibitem

\bibitem[\protect\citeauthoryear{{Hick} and {Jackson}}{2004}]{hick_jackson:04a}
\begin{bchapter}
\bauthor{\bsnm{{Hick}}, \binits{P.P.}},
\bauthor{\bsnm{{Jackson}}, \binits{B.V.}}:
\byear{2004},
\bctitle{{Heliospheric tomography: an algorithm for the reconstruction of the
  3D solar wind from remote sensing observations}}.
In: \beditor{\bsnm{{Fineschi}}, \binits{S.}},
\beditor{\bsnm{{Gummin}}, \binits{M.A.}} (eds.)
\bbtitle{Telescopes and Instrumentation for Solar Astrophysics},
\bsertitle{\spie}
\bseriesno{5171},
\bfpage{287}.
\doiurl{10.1117/12.513122}.
\end{bchapter}
\endbibitem

\bibitem[\protect\citeauthoryear{{Houminer}}{1971}]{houminer:71a}
\begin{barticle}
\bauthor{\bsnm{{Houminer}}, \binits{Z.}}:
\byear{1971},
\batitle{{Corotating plasma streams revealed by interplanetary scintillation}}.
\bjtitle{Nature \psc}
\bvolume{231},
\bfpage{165}.
\doiurl{10.1038/physci231165a0}.
\end{barticle}
\endbibitem

\bibitem[\protect\citeauthoryear{{Houminer} and
  {Hewish}}{1972}]{houminer_hewish:72a}
\begin{barticle}
\bauthor{\bsnm{{Houminer}}, \binits{Z.}},
\bauthor{\bsnm{{Hewish}}, \binits{A.}}:
\byear{1972},
\batitle{{Long-lived sectors of enhanced density irregularities in the solar
  wind}}.
\bjtitle{\pss}
\bvolume{20},
\bfpage{1703}.
\doiurl{10.1016/0032-0633(72)90192-4}.
\end{barticle}
\endbibitem

\bibitem[\protect\citeauthoryear{{Houminer} and
  {Hewish}}{1974}]{houminer_hewish:74a}
\begin{barticle}
\bauthor{\bsnm{{Houminer}}, \binits{Z.}},
\bauthor{\bsnm{{Hewish}}, \binits{A.}}:
\byear{1974},
\batitle{{Correlation of interplanetary scintillation and spacecraft plasma
  density measurements}}.
\bjtitle{\pss}
\bvolume{22},
\bfpage{1041}.
\doiurl{10.1016/0032-0633(74)90173-1}.
\end{barticle}
\endbibitem

\bibitem[\protect\citeauthoryear{{Jackson} and {Hick}}{2004}]{jackson_hick:04a}
\begin{bchapter}
\bauthor{\bsnm{{Jackson}}, \binits{B.V.}},
\bauthor{\bsnm{{Hick}}, \binits{P.P.}}:
\byear{2004},
\bctitle{{Three-dimensional tomography of interplanetary disturbances}}.
In: \beditor{\bsnm{{Gary}}, \binits{D.E.}},
\beditor{\bsnm{{Keller}}, \binits{C.U.}} (eds.)
\bbtitle{Solar and Space Weather Radiophysics}
\bseriesno{314},
\bpublisher{Kluwer Academic Publishers},
\blocation{Dordecht},
\bfpage{355}.
\doiurl{10.1007/1-4020-2814-8_17}.
\end{bchapter}
\endbibitem

\bibitem[\protect\citeauthoryear{{Jackson}
  \textit{et~al.}}{1997}]{jackson_etal:97a}
\begin{barticle}
\bauthor{\bsnm{{Jackson}}, \binits{B.V.}},
\bauthor{\bsnm{{Hick}}, \binits{P.L.}},
\bauthor{\bsnm{{Kojima}}, \binits{M.}},
\bauthor{\bsnm{{Yokobe}}, \binits{A.}}:
\byear{1997},
\batitle{{Heliospheric tomography using interplanetary scintillation
  observations}}.
\bjtitle{\asr}
\bvolume{20},
\bfpage{23}.
\doiurl{10.1016/S0273-1177(97)00474-2}.
\end{barticle}
\endbibitem

\bibitem[\protect\citeauthoryear{Jackson
  \textit{et~al.}}{1998}]{jackson_etal:98a}
\begin{barticle}
\bauthor{\bsnm{Jackson}, \binits{B.V.}},
\bauthor{\bsnm{Hick}, \binits{P.L.}},
\bauthor{\bsnm{Kojima}, \binits{M.}},
\bauthor{\bsnm{Yokobe}, \binits{A.}}:
\byear{1998},
\batitle{{Heliospheric tomography using interplanetary scintillation
  observations 1. Combined Nagoya and Cambridge data}}.
\bjtitle{\jgr}
\bvolume{103}(\bissue{A6}),
\bfpage{12049}.
\end{barticle}
\endbibitem

\bibitem[\protect\citeauthoryear{{Jackson}
  \textit{et~al.}}{2003}]{jackson_etal:03a}
\begin{bchapter}
\bauthor{\bsnm{{Jackson}}, \binits{B.V.}},
\bauthor{\bsnm{{Hick}}, \binits{P.P.}},
\bauthor{\bsnm{{Buffington}}, \binits{A.}},
\bauthor{\bsnm{{Kojima}}, \binits{M.}},
\bauthor{\bsnm{{Tokumaru}}, \binits{M.}},
\bauthor{\bsnm{{Fujiki}}, \binits{K.}},
\bauthor{\bsnm{{Ohmi}}, \binits{T.}},
\bauthor{\bsnm{{Yamashita}}, \binits{M.}}:
\byear{2003},
\bctitle{{Time-dependent tomography of hemispheric features using
  interplanetary scintillation (IPS) remote-sensing observations}}.
In: \beditor{\bsnm{Velli}, \binits{M.}},
\beditor{\bsnm{Bruno}, \binits{R.}},
\beditor{\bsnm{Malara}, \binits{F.}},
\beditor{\bsnm{Bucci}, \binits{B.}} (eds.)
\bbtitle{Solar Wind Ten},
\bsertitle{\aip}
\bseriesno{679},
\bfpage{75}.
\doiurl{10.1063/1.1618545}.
\end{bchapter}
\endbibitem

\bibitem[\protect\citeauthoryear{{Jackson}
  \textit{et~al.}}{2010}]{jackson_etal:10a}
\begin{barticle}
\bauthor{\bsnm{{Jackson}}, \binits{B.V.}},
\bauthor{\bsnm{{Hick}}, \binits{P.P.}},
\bauthor{\bsnm{{Buffington}}, \binits{A.}},
\bauthor{\bsnm{{Bisi}}, \binits{M.M.}},
\bauthor{\bsnm{{Clover}}, \binits{J.M.}},
\bauthor{\bsnm{{Tokumaru}}, \binits{M.}}:
\byear{2010},
\batitle{{Solar Mass Ejection Imager (SMEI) and interplanetary scintillation
  (IPS) 3D-reconstructions of the inner heliosphere}}.
\bjtitle{\advGeosci}
\bvolume{21},
\bfpage{339}.
\end{barticle}
\endbibitem

\bibitem[\protect\citeauthoryear{{Jackson}
  \textit{et~al.}}{2011}]{jackson_etal:11b}
\begin{barticle}
\bauthor{\bsnm{{Jackson}}, \binits{B.V.}},
\bauthor{\bsnm{{Hick}}, \binits{P.P.}},
\bauthor{\bsnm{{Buffington}}, \binits{A.}},
\bauthor{\bsnm{{Bisi}}, \binits{M.M.}},
\bauthor{\bsnm{{Clover}}, \binits{J.M.}},
\bauthor{\bsnm{{Tokumaru}}, \binits{M.}},
\bauthor{\bsnm{{Kojima}}, \binits{M.}},
\bauthor{\bsnm{{Fujiki}}, \binits{K.}}:
\byear{2011},
\batitle{{Three-dimensional reconstruction of heliospheric structure using
  iterative tomography: A review}}.
\bjtitle{\jastp}
\bvolume{73},
\bfpage{1214}.
\doiurl{10.1016/j.jastp.2010.10.007}.
\end{barticle}
\endbibitem

\bibitem[\protect\citeauthoryear{Jackson
  \textit{et~al.}}{2015}]{jackson_etal:15a}
\begin{barticle}
\bauthor{\bsnm{Jackson}, \binits{B.V.}},
\bauthor{\bsnm{Odstricil}, \binits{D.}},
\bauthor{\bsnm{Yu}, \binits{H.S.}},
\bauthor{\bsnm{Hick}, \binits{P.P.}},
\bauthor{\bparticle{an} \bsnm{J.~C.~{Nejia-Ambris}}, \binits{A.B.}},
\bauthor{\bsnm{Kim}, \binits{J.C.}},
\bauthor{\bsnm{Hong}, \binits{S.}},
\bauthor{\bsnm{Kim}, \binits{Y.}},
\bauthor{\bsnm{Han}, \binits{J.}},
\bauthor{\bsnm{Tokumaru}, \binits{M.}}:
\byear{2015},
\batitle{{The UCSD kinematic IPS solar wind boundary and its use in the ENLIL
  3-D MHD prediction model}}.
\bjtitle{Space Weather}
\bvolume{13}.
\doiurl{10.1002/2014SW00130}.
\end{barticle}
\endbibitem

\bibitem[\protect\citeauthoryear{{Kakinuma}}{1977}]{kakinuma:77a}
\begin{bchapter}
\bauthor{\bsnm{{Kakinuma}}, \binits{T.}}:
\byear{1977},
\bctitle{{Observations of interplanetary scintillation - Solar wind velocity
  measurements}}.
In: \beditor{\bsnm{{Shea}}, \binits{M.A.}},
\beditor{\bsnm{{Smart}}, \binits{D.F.}},
\beditor{\bsnm{{Wu}}, \binits{S.T.}} (eds.)
\bbtitle{Study of Travelling Interplanetary Phenomena 1997},
\bsertitle{Astrophysics and Space Science Library}
\bseriesno{71},
\bpublisher{D.Reidel},
\blocation{Dordrecht},
\bfpage{101}.
\end{bchapter}
\endbibitem

\bibitem[\protect\citeauthoryear{{Kasper}
  \textit{et~al.}}{2012}]{kasper_etal:12a}
\begin{barticle}
\bauthor{\bsnm{{Kasper}}, \binits{J.C.}},
\bauthor{\bsnm{{Stevens}}, \binits{M.L.}},
\bauthor{\bsnm{{Korreck}}, \binits{K.E.}},
\bauthor{\bsnm{{Maruca}}, \binits{B.A.}},
\bauthor{\bsnm{{Kiefer}}, \binits{K.K.}},
\bauthor{\bsnm{{Schwadron}}, \binits{N.A.}},
\bauthor{\bsnm{{Lepri}}, \binits{S.T.}}:
\byear{2012},
\batitle{{Evolution of the relationships between helium abundance, minor ion
  charge state, and solar wind speed over the solar cycle}}.
\bjtitle{\apj}
\bvolume{745},
\bfpage{162}.
\doiurl{10.1088/0004-637X/745/2/162}.
\end{barticle}
\endbibitem

\bibitem[\protect\citeauthoryear{{Katushkina}, {Izmodenov}, and
  {Alexashov}}{2015}]{katushkina_etal:15a}
\begin{barticle}
\bauthor{\bsnm{{Katushkina}}, \binits{O.A.}},
\bauthor{\bsnm{{Izmodenov}}, \binits{V.V.}},
\bauthor{\bsnm{{Alexashov}}, \binits{D.B.}}:
\byear{2015},
\batitle{{Direction of interstellar hydrogen flow in the heliosphere:
  theoretical modelling and comparison with SOHO/SWAN data}}.
\bjtitle{\mnras}
\bvolume{446},
\bfpage{2929}.
\doiurl{10.1093/mnras/stu2218}.
\end{barticle}
\endbibitem

\bibitem[\protect\citeauthoryear{Katushkina
  \textit{et~al.}}{2013}]{katushkina_etal:13a}
\begin{botherref}
\oauthor{\bsnm{Katushkina}, \binits{O.A.}},
\oauthor{\bsnm{Izmodenov}, \binits{V.V.}},
\oauthor{\bsnm{Qu{\'e}merais}, \binits{E.}},
\oauthor{\bsnm{Sok{\'o}{\l}}, \binits{J.M.}}:
2013,
Heliolatitudinal and time variations of the solar wind mass flux: Inferences
  from the backscattered solar lyman-alpha intensity maps.
\textit{\jgr},
1.
\doiurl{10.1002/jgra.50303}.
\end{botherref}
\endbibitem

\bibitem[\protect\citeauthoryear{{Katushkina}
  \textit{et~al.}}{2014}]{katushkina_etal:14c}
\begin{barticle}
\bauthor{\bsnm{{Katushkina}}, \binits{O.A.}},
\bauthor{\bsnm{{Izmodenov}}, \binits{V.V.}},
\bauthor{\bsnm{{Wood}}, \binits{B.E.}},
\bauthor{\bsnm{{McMullin}}, \binits{D.R.}}:
\byear{2014},
\batitle{{Neutral interstellar helium parameters based on Ulysses/GAS and
  IBEX-Lo observations: What are the reasons for the differences?}}
\bjtitle{\apj}
\bvolume{789},
\bfpage{80}.
\doiurl{10.1088/0004-637X/789/1/80}.
\end{barticle}
\endbibitem

\bibitem[\protect\citeauthoryear{{Kim} \textit{et~al.}}{2014}]{kim_etal:14a}
\begin{barticle}
\bauthor{\bsnm{{Kim}}, \binits{T.K.}},
\bauthor{\bsnm{{Pogorelov}}, \binits{N.V.}},
\bauthor{\bsnm{{Borovikov}}, \binits{S.N.}},
\bauthor{\bsnm{{Jackson}}, \binits{B.V.}},
\bauthor{\bsnm{{Yu}}, \binits{H.-S.}},
\bauthor{\bsnm{{Tokumaru}}, \binits{M.}}:
\byear{2014},
\batitle{{MHD heliosphere with boundary conditions from a tomographic
  reconstruction using interplanetary scintillation data}}.
\bjtitle{J. Geophys. Res. (Space Phys.)}
\bvolume{119},
\bfpage{7981}.
\doiurl{10.1002/2013JA019755}.
\end{barticle}
\endbibitem

\bibitem[\protect\citeauthoryear{{King} and
  {Papitashvili}}{2005}]{king_papitashvili:05}
\begin{barticle}
\bauthor{\bsnm{{King}}, \binits{J.H.}},
\bauthor{\bsnm{{Papitashvili}}, \binits{N.E.}}:
\byear{2005},
\batitle{{Solar wind spatial scales in and comparisons of hourly Wind and ACE
  plasma and magnetic field data}}.
\bjtitle{\jgr}
\bvolume{110},
\bfpage{2104}.
\doiurl{10.1029/2004JA010649}.
\end{barticle}
\endbibitem

\bibitem[\protect\citeauthoryear{{Kojima} and
  {Kakinuma}}{1990}]{kojima_kakinuma:90a}
\begin{barticle}
\bauthor{\bsnm{{Kojima}}, \binits{M.}},
\bauthor{\bsnm{{Kakinuma}}, \binits{T.}}:
\byear{1990},
\batitle{{Solar cycle dependence of global distribution of solar wind speed}}.
\bjtitle{\ssr}
\bvolume{53},
\bfpage{173}.
\doiurl{10.1007/BF00212754}.
\end{barticle}
\endbibitem

\bibitem[\protect\citeauthoryear{Kojima
  \textit{et~al.}}{1998}]{kojima_etal:98a}
\begin{barticle}
\bauthor{\bsnm{Kojima}, \binits{M.}},
\bauthor{\bsnm{Tokumaru}, \binits{M.}},
\bauthor{\bsnm{Watanabe}, \binits{H.}},
\bauthor{\bsnm{Yokobe}, \binits{A.}},
\bauthor{\bsnm{Asai}, \binits{K.}},
\bauthor{\bsnm{Jackson}, \binits{B.V.}},
\bauthor{\bsnm{Hick}, \binits{P.L.}}:
\byear{1998},
\batitle{Heliospheric tomography using interplanetary scintillation
  observations 2. {L}atitude and heliocentric distance dependence of solar wind
  structure at 0.1--1~{AU}}.
\bjtitle{\jgr}
\bvolume{103},
\bfpage{1981}.
\end{barticle}
\endbibitem

\bibitem[\protect\citeauthoryear{{Kojima}
  \textit{et~al.}}{2004}]{kojima_etal:04b}
\begin{bchapter}
\bauthor{\bsnm{{Kojima}}, \binits{M.}},
\bauthor{\bsnm{{Fujiki}}, \binits{K.-I.}},
\bauthor{\bsnm{{Hirano}}, \binits{M.}},
\bauthor{\bsnm{{Tokumaru}}, \binits{M.}},
\bauthor{\bsnm{{Ohmi}}, \binits{T.}},
\bauthor{\bsnm{{Hakamada}}, \binits{K.}}:
\byear{2004},
\bctitle{{Solar wind properties from IPS observations}}.
In: \beditor{\bsnm{{Poletto}}, \binits{G.}},
\beditor{\bsnm{{Suess}}, \binits{S.T.}} (eds.)
\bbtitle{The Sun and the Heliosphere as an Integrated System}
\bseriesno{317},
\bpublisher{Kluwer Academic Publishers},
\blocation{Dordrecht},
\bfpage{147}.
\end{bchapter}
\endbibitem

\bibitem[\protect\citeauthoryear{{Kondrashov} and
  {Ghil}}{2006}]{kondrashov_ghil:06a}
\begin{barticle}
\bauthor{\bsnm{{Kondrashov}}, \binits{D.}},
\bauthor{\bsnm{{Ghil}}, \binits{M.}}:
\byear{2006},
\batitle{{Spatio-temporal filling of missing points in geophysical data sets}}.
\bjtitle{Nonlinear Proc. Geophys.}
\bvolume{13},
\bfpage{151}.
\end{barticle}
\endbibitem

\bibitem[\protect\citeauthoryear{{Kondrashov}, {Shprits}, and
  {Ghil}}{2010}]{kondrashov_etal:10a}
\begin{barticle}
\bauthor{\bsnm{{Kondrashov}}, \binits{D.}},
\bauthor{\bsnm{{Shprits}}, \binits{Y.}},
\bauthor{\bsnm{{Ghil}}, \binits{M.}}:
\byear{2010},
\batitle{{Gap filling of solar wind data by singular spectrum analysis}}.
\bjtitle{\grl}
\bvolume{37},
\bfpage{15101}.
\doiurl{10.1029/2010GL044138}.
\end{barticle}
\endbibitem

\bibitem[\protect\citeauthoryear{{Kondrashov}
  \textit{et~al.}}{2014}]{kondrashov_etal:14a}
\begin{barticle}
\bauthor{\bsnm{{Kondrashov}}, \binits{D.}},
\bauthor{\bsnm{{Denton}}, \binits{R.}},
\bauthor{\bsnm{{Shprits}}, \binits{Y.Y.}},
\bauthor{\bsnm{{Singer}}, \binits{H.J.}}:
\byear{2014},
\batitle{{Reconstruction of gaps in the past history of solar wind
  parameters}}.
\bjtitle{\grl}
\bvolume{41},
\bfpage{2702}.
\doiurl{10.1002/2014GL059741}.
\end{barticle}
\endbibitem

\bibitem[\protect\citeauthoryear{Lallement, Bertaux, and
  Kurt}{1985}]{lallement_etal:85a}
\begin{barticle}
\bauthor{\bsnm{Lallement}, \binits{R.}},
\bauthor{\bsnm{Bertaux}, \binits{J.L.}},
\bauthor{\bsnm{Kurt}, \binits{V.G.}}:
\byear{1985},
\batitle{Solar wind decrease at high heliographic latitudes detected from
  {P}rognoz interplanetary {L}yman {a}lpha mapping}.
\bjtitle{\jgr}
\bvolume{90},
\bfpage{1413}.
\end{barticle}
\endbibitem

\bibitem[\protect\citeauthoryear{{Le Chat}, {Issautier}, and
  {Meyer-Vernet}}{2012}]{leChat_etal:12a}
\begin{barticle}
\bauthor{\bsnm{{Le Chat}}, \binits{G.}},
\bauthor{\bsnm{{Issautier}}, \binits{K.}},
\bauthor{\bsnm{{Meyer-Vernet}}, \binits{N.}}:
\byear{2012},
\batitle{{The solar wind energy flux}}.
\bjtitle{\solphys}
\bvolume{279},
\bfpage{197}.
\doiurl{10.1007/s11207-012-9967-y}.
\end{barticle}
\endbibitem

\bibitem[\protect\citeauthoryear{{Manoharan}}{1993}]{manoharan:93b}
\begin{barticle}
\bauthor{\bsnm{{Manoharan}}, \binits{P.K.}}:
\byear{1993},
\batitle{{Three-dimensional structure of the solar wind: Variation of density
  with the solar cycle}}.
\bjtitle{\solphys}
\bvolume{148},
\bfpage{153}.
\doiurl{10.1007/BF00675541}.
\end{barticle}
\endbibitem

\bibitem[\protect\citeauthoryear{{McComas}, Gosling, and
  Skoug}{2000}]{mccomas_etal:00a}
\begin{barticle}
\bauthor{\bsnm{{McComas}}, \binits{D.J.}},
\bauthor{\bsnm{Gosling}, \binits{J.T.}},
\bauthor{\bsnm{Skoug}, \binits{R.M.}}:
\byear{2000},
\batitle{Ulysses observations of the irregularly structured mid-latitude solar
  wind during the approach to solar maximum}.
\bjtitle{\grl}
\bvolume{27},
\bfpage{2437}.
\end{barticle}
\endbibitem

\bibitem[\protect\citeauthoryear{{McComas}
  \textit{et~al.}}{2000}]{mccomas_etal:00b}
\begin{barticle}
\bauthor{\bsnm{{McComas}}, \binits{D.J.}},
\bauthor{\bsnm{Barraclough}, \binits{B.L.}},
\bauthor{\bsnm{Funsten}, \binits{H.O.}},
\bauthor{\bsnm{Gosling}, \binits{J.T.}},
\bauthor{\bsnm{Santiago-Munoz}},
\bauthor{\bsnm{Goldstein}, \binits{B.E.}},
\bauthor{\bsnm{Neugebauer}, \binits{M.}},
\bauthor{\bsnm{Riley}, \binits{P.}},
\bauthor{\bsnm{Balogh}, \binits{A.}}:
\byear{2000},
\batitle{Solar wind observations over {U}lysses first full polar orbit}.
\bjtitle{\jgr}
\bvolume{105},
\bfpage{10419}.
\end{barticle}
\endbibitem

\bibitem[\protect\citeauthoryear{{McComas}
  \textit{et~al.}}{2006}]{mccomas_etal:06a}
\begin{barticle}
\bauthor{\bsnm{{McComas}}, \binits{D.J.}},
\bauthor{\bsnm{{Elliott}}, \binits{H.A.}},
\bauthor{\bsnm{{Gosling}}, \binits{J.T.}},
\bauthor{\bsnm{{Skoug}}, \binits{R.M.}}:
\byear{2006},
\batitle{{Ulysses observations of very different heliospheric structure during
  the declining phase of solar activity cycle 23}}.
\bjtitle{\grl}
\bvolume{330},
\bfpage{L09102}.
\doiurl{10.1029/2006GL025915}.
\end{barticle}
\endbibitem

\bibitem[\protect\citeauthoryear{{McComas}
  \textit{et~al.}}{2008}]{mccomas_etal:08a}
\begin{barticle}
\bauthor{\bsnm{{McComas}}, \binits{D.J.}},
\bauthor{\bsnm{Ebert}, \binits{R.W.}},
\bauthor{\bsnm{Elliot}, \binits{H.A.}},
\bauthor{\bsnm{Goldstein}, \binits{B.E.}},
\bauthor{\bsnm{Gosling}, \binits{J.T.}},
\bauthor{\bsnm{Schwadron}, \binits{N.A.}},
\bauthor{\bsnm{Skoug}, \binits{R.M.}}:
\byear{2008},
\batitle{Weaker solar wind from the polar coronal holes and the whole {S}un}.
\bjtitle{\grl}
\bvolume{35},
\bfpage{L18103}.
\doiurl{10.1029/2008GL034896}.
\end{barticle}
\endbibitem

\bibitem[\protect\citeauthoryear{{McComas}
  \textit{et~al.}}{2009a}]{mccomas_etal:09c}
\begin{barticle}
\bauthor{\bsnm{{McComas}}, \binits{D.J.}},
\bauthor{\bsnm{{Allegrini}}, \binits{F.}},
\bauthor{\bsnm{{Bochsler}}, \binits{P.}},
\bauthor{\bsnm{{Bzowski}}, \binits{M.}},
\bauthor{\bsnm{{Christian}}, \binits{E.R.}},
\bauthor{\bsnm{{Crew}}, \binits{G.B.}},
\bauthor{\bsnm{{DeMajistre}}, \binits{R.}},
\bauthor{\bsnm{{Fahr}}, \binits{H.}},
\bauthor{\bsnm{{Fichtner}}, \binits{H.}},
\bauthor{\bsnm{{Frisch}}, \binits{P.C.}},
\bauthor{\bsnm{{Funsten}}, \binits{H.O.}},
\bauthor{\bsnm{{Fuselier}}, \binits{S.A.}},
\bauthor{\bsnm{{Gloeckler}}, \binits{G.}},
\bauthor{\bsnm{{Gruntman}}, \binits{M.}},
\bauthor{\bsnm{{Heerikhuisen}}, \binits{J.}},
\bauthor{\bsnm{{Izmodenov}}, \binits{V.}},
\bauthor{\bsnm{{Janzen}}, \binits{P.}},
\bauthor{\bsnm{{Knappenberger}}, \binits{P.}},
\bauthor{\bsnm{{Krimigis}}, \binits{S.}},
\bauthor{\bsnm{{Kucharek}}, \binits{H.}},
\bauthor{\bsnm{{Lee}}, \binits{M.}},
\bauthor{\bsnm{{Livadiotis}}, \binits{G.}},
\bauthor{\bsnm{{Livi}}, \binits{S.}},
\bauthor{\bsnm{{MacDowall}}, \binits{R.J.}},
\bauthor{\bsnm{{Mitchell}}, \binits{D.}},
\bauthor{\bsnm{{M{\"o}bius}}, \binits{E.}},
\bauthor{\bsnm{{Moore}}, \binits{T.}},
\bauthor{\bsnm{{Pogorelov}}, \binits{N.V.}},
\bauthor{\bsnm{{Reisenfeld}}, \binits{D.}},
\bauthor{\bsnm{{Roelof}}, \binits{E.}},
\bauthor{\bsnm{{Saul}}, \binits{L.}},
\bauthor{\bsnm{{Schwadron}}, \binits{N.A.}},
\bauthor{\bsnm{{Valek}}, \binits{P.W.}},
\bauthor{\bsnm{{Vanderspek}}, \binits{R.}},
\bauthor{\bsnm{{Wurz}}, \binits{P.}},
\bauthor{\bsnm{{Zank}}, \binits{G.P.}}:
\byear{2009}a,
\batitle{{Global observations of the interstellar interaction from the
  Interstellar Boundary Explorer (IBEX)}}.
\bjtitle{Science}
\bvolume{326},
\bfpage{959}.
\doiurl{10.1126/science.1180906}.
\end{barticle}
\endbibitem

\bibitem[\protect\citeauthoryear{{McComas}
  \textit{et~al.}}{2009b}]{mccomas_etal:09a}
\begin{barticle}
\bauthor{\bsnm{{McComas}}, \binits{D.J.}},
\bauthor{\bsnm{{Allegrini}}, \binits{F.}},
\bauthor{\bsnm{{Bochsler}}, \binits{P.}},
\bauthor{\bsnm{{Bzowski}}, \binits{M.}},
\bauthor{\bsnm{{Collier}}, \binits{M.}},
\bauthor{\bsnm{{Fahr}}, \binits{H.}},
\bauthor{\bsnm{{Fichtner}}, \binits{H.}},
\bauthor{\bsnm{{Frisch}}, \binits{P.}},
\bauthor{\bsnm{{Funsten}}, \binits{H.O.}},
\bauthor{\bsnm{{Fuselier}}, \binits{S.A.}},
\bauthor{\bsnm{{Gloeckler}}, \binits{G.}},
\bauthor{\bsnm{{Gruntman}}, \binits{M.}},
\bauthor{\bsnm{{Izmodenov}}, \binits{V.}},
\bauthor{\bsnm{{Knappenberger}}, \binits{P.}},
\bauthor{\bsnm{{Lee}}, \binits{M.}},
\bauthor{\bsnm{{Livi}}, \binits{S.}},
\bauthor{\bsnm{{Mitchell}}, \binits{D.}},
\bauthor{\bsnm{{M{\"o}bius}}, \binits{E.}},
\bauthor{\bsnm{{Moore}}, \binits{T.}},
\bauthor{\bsnm{{Pope}}, \binits{S.}},
\bauthor{\bsnm{{Reisenfeld}}, \binits{D.}},
\bauthor{\bsnm{{Roelof}}, \binits{E.}},
\bauthor{\bsnm{{Scherrer}}, \binits{J.}},
\bauthor{\bsnm{{Schwadron}}, \binits{N.}},
\bauthor{\bsnm{{Tyler}}, \binits{R.}},
\bauthor{\bsnm{{Wieser}}, \binits{M.}},
\bauthor{\bsnm{{Witte}}, \binits{M.}},
\bauthor{\bsnm{{Wurz}}, \binits{P.}},
\bauthor{\bsnm{{Zank}}, \binits{G.}}:
\byear{2009}b,
\batitle{{IBEX -- Interstellar Boundary Explorer}}.
\bjtitle{\ssr}
\bvolume{146},
\bfpage{11}.
\doiurl{10.1007/s11214-009-9499-4}.
\end{barticle}
\endbibitem

\bibitem[\protect\citeauthoryear{{McComas}
  \textit{et~al.}}{2012}]{mccomas_etal:12c}
\begin{barticle}
\bauthor{\bsnm{{McComas}}, \binits{D.J.}},
\bauthor{\bsnm{{Dayeh}}, \binits{M.A.}},
\bauthor{\bsnm{{Allegrini}}, \binits{F.}},
\bauthor{\bsnm{{Bzowski}}, \binits{M.}},
\bauthor{\bsnm{{DeMajistre}}, \binits{R.}},
\bauthor{\bsnm{{Fujiki}}, \binits{K.}},
\bauthor{\bsnm{{Funsten}}, \binits{H.O.}},
\bauthor{\bsnm{{Fuselier}}, \binits{S.A.}},
\bauthor{\bsnm{{Gruntman}}, \binits{M.}},
\bauthor{\bsnm{{Janzen}}, \binits{P.H.}},
\bauthor{\bsnm{{Kubiak}}, \binits{M.A.}},
\bauthor{\bsnm{{Kucharek}}, \binits{H.}},
\bauthor{\bsnm{{Livadiotis}}, \binits{G.}},
\bauthor{\bsnm{{M{\"o}bius}}, \binits{E.}},
\bauthor{\bsnm{{Reisenfeld}}, \binits{D.B.}},
\bauthor{\bsnm{{Reno}}, \binits{M.}},
\bauthor{\bsnm{{Schwadron}}, \binits{N.A.}},
\bauthor{\bsnm{{Sok{\'o}{\l}}}, \binits{J.M.}},
\bauthor{\bsnm{{Tokumaru}}, \binits{M.}}:
\byear{2012},
\batitle{{The first three years of IBEX observations and our evolving
  heliosphere}}.
\bjtitle{\apjs}
\bvolume{203},
\bfpage{1}.
\doiurl{10.1088/0067-0049/203/1/1}.
\end{barticle}
\endbibitem

\bibitem[\protect\citeauthoryear{{McComas}
  \textit{et~al.}}{2013}]{mccomas_etal:13b}
\begin{barticle}
\bauthor{\bsnm{{McComas}}, \binits{D.J.}},
\bauthor{\bsnm{Angold}, \binits{N.}},
\bauthor{\bsnm{Elliott}, \binits{H.A.}},
\bauthor{\bsnm{Livadiotis}, \binits{G.}},
\bauthor{\bsnm{Schwadron}, \binits{N.A.}},
\bauthor{\bsnm{Skoug}, \binits{R.M.}},
\bauthor{\bsnm{Smith}, \binits{C.W.}}:
\byear{2013},
\batitle{Weakest solar wind of the space age and the current ``mini'' solar
  maximum}.
\bjtitle{\apj}
\bvolume{779},
\bfpage{2}.
\end{barticle}
\endbibitem

\bibitem[\protect\citeauthoryear{{McComas}
  \textit{et~al.}}{2014}]{mccomas_etal:14b}
\begin{barticle}
\bauthor{\bsnm{{McComas}}, \binits{D.J.}},
\bauthor{\bsnm{Allegrini}, \binits{F.}},
\bauthor{\bsnm{Bzowski}, \binits{M.}},
\bauthor{\bsnm{Dayeh}, \binits{M.A.}},
\bauthor{\bsnm{{Demajistre}}, \binits{R.}},
\bauthor{\bsnm{Funsten}, \binits{H.O.}},
\bauthor{\bsnm{Fuselier}, \binits{S.A.}},
\bauthor{\bsnm{Gruntman}, \binits{M.}},
\bauthor{\bsnm{Janzen}, \binits{P.H.}},
\bauthor{\bsnm{Kubiak}, \binits{M.A.}},
\bauthor{\bsnm{Kucharek}, \binits{H.}},
\bauthor{\bsnm{{M{\"o}bius}}, \binits{E.}},
\bauthor{\bsnm{Reisenfeld}, \binits{D.B.}},
\bauthor{\bsnm{Schwadron}, \binits{N.A.}},
\bauthor{\bsnm{{Sok{\'o}{\l}}}, \binits{J.M.}},
\bauthor{\bsnm{Tokumaru}, \binits{M.}}:
\byear{2014},
\batitle{{IBEX: The first five years (2009-2013)}}.
\bjtitle{\apjs}
\bvolume{231},
\bfpage{28}.
\doiurl{10.1088/0067-0049/213/2/20}.
\end{barticle}
\endbibitem

\bibitem[\protect\citeauthoryear{{Provornikova}
  \textit{et~al.}}{2014}]{provornikova_etal:14a}
\begin{barticle}
\bauthor{\bsnm{{Provornikova}}, \binits{E.}},
\bauthor{\bsnm{{Opher}}, \binits{M.}},
\bauthor{\bsnm{{Izmodenov}}, \binits{V.V.}},
\bauthor{\bsnm{{Richardson}}, \binits{J.D.}},
\bauthor{\bsnm{{Toth}}, \binits{G.}}:
\byear{2014},
\batitle{{Plasma flows in the heliosheath along the Voyager 1 and 2
  trajectories due to effects of the 11 yr solar cycle}}.
\bjtitle{\apj}
\bvolume{794},
\bfpage{29}.
\doiurl{10.1088/0004-637X/794/1/29}.
\end{barticle}
\endbibitem

\bibitem[\protect\citeauthoryear{Richardson
  \textit{et~al.}}{2008}]{richardson_etal:08b}
\begin{barticle}
\bauthor{\bsnm{Richardson}, \binits{J.D.}},
\bauthor{\bsnm{Kasper}, \binits{J.C.}},
\bauthor{\bsnm{Wang}, \binits{C.}},
\bauthor{\bsnm{Belcher}, \binits{J.W.}},
\bauthor{\bsnm{Lazarus}, \binits{A.J.}}:
\byear{2008},
\batitle{{Cool heliosheath plasma and deceleration of the upstream solar wind
  at the termination shock}}.
\bjtitle{\nat}
\bvolume{454},
\bfpage{63}.
\doiurl{10.1038/nature07024}.
\end{barticle}
\endbibitem

\bibitem[\protect\citeauthoryear{{Schwadron} and
  {McComas}}{2013}]{schwadron_mccomas:13a}
\begin{barticle}
\bauthor{\bsnm{{Schwadron}}, \binits{N.A.}},
\bauthor{\bsnm{{McComas}}, \binits{D.J.}}:
\byear{2013},
\batitle{{Spatial retention of ions producing the IBEX ribbon}}.
\bjtitle{\apj}
\bvolume{764},
\bfpage{92}.
\doiurl{10.1088/0004-637X/764/1/92}.
\end{barticle}
\endbibitem

\bibitem[\protect\citeauthoryear{{Sok\'{o}{\l}}
  \textit{et~al.}}{2013}]{sokol_etal:13a}
\begin{barticle}
\bauthor{\bsnm{{Sok\'{o}{\l}}}, \binits{J.M.}},
\bauthor{\bsnm{{Bzowski}}, \binits{M.}},
\bauthor{\bsnm{{Tokumaru}}, \binits{M.}},
\bauthor{\bsnm{{Fujiki}}, \binits{K.}},
\bauthor{\bsnm{{McComas}}, \binits{D.J.}}:
\byear{2013},
\batitle{Heliolatitude and time variations of solar wind structure from in-situ
  measurements and interplanetary scintillation observations}.
\bjtitle{\solphys}
\bvolume{285},
\bfpage{167}.
\doiurl{10.1007/s11207-012-9993-9}.
\end{barticle}
\endbibitem

\bibitem[\protect\citeauthoryear{{Stone}
  \textit{et~al.}}{2005}]{stone_etal:05a}
\begin{barticle}
\bauthor{\bsnm{{Stone}}, \binits{E.C.}},
\bauthor{\bsnm{{Cummings}}, \binits{A.C.}},
\bauthor{\bsnm{{McDonald}}, \binits{F.B.}},
\bauthor{\bsnm{{Heikkila}}, \binits{B.C.}},
\bauthor{\bsnm{{Lal}}, \binits{N.}},
\bauthor{\bsnm{{Webber}}, \binits{W.R.}}:
\byear{2005},
\batitle{{Voyager 1 explores the termination shock region and the heliosheath
  beyond}}.
\bjtitle{Science}
\bvolume{309},
\bfpage{2017}.
\doiurl{10.1126/science.1117684}.
\end{barticle}
\endbibitem

\bibitem[\protect\citeauthoryear{{Stone}
  \textit{et~al.}}{2008}]{stone_etal:08a}
\begin{barticle}
\bauthor{\bsnm{{Stone}}, \binits{E.C.}},
\bauthor{\bsnm{{Cummings}}, \binits{A.C.}},
\bauthor{\bsnm{{McDonald}}, \binits{F.B.}},
\bauthor{\bsnm{{Heikkila}}, \binits{B.C.}},
\bauthor{\bsnm{{Lal}}, \binits{N.}},
\bauthor{\bsnm{{Webber}}, \binits{W.R.}}:
\byear{2008},
\batitle{An asymmetric solar wind termination shock}.
\bjtitle{\nat}
\bvolume{454},
\bfpage{71}.
\doiurl{10.1038/nature07022}.
\end{barticle}
\endbibitem

\bibitem[\protect\citeauthoryear{{Swaczyna}, {Grzedzielski}, and
  {Bzowski}}{2014}]{swaczyna_etal:14a}
\begin{barticle}
\bauthor{\bsnm{{Swaczyna}}, \binits{P.}},
\bauthor{\bsnm{{Grzedzielski}}, \binits{S.}},
\bauthor{\bsnm{{Bzowski}}, \binits{M.}}:
\byear{2014},
\batitle{{Assessment of energetic neutral He atom intensities expected from the
  IBEX ribbon}}.
\bjtitle{\apj}
\bvolume{782},
\bfpage{106}.
\doiurl{10.1088/0004-637X/782/2/106}.
\end{barticle}
\endbibitem

\bibitem[\protect\citeauthoryear{{Tappin}}{1986}]{tappin:86a}
\begin{barticle}
\bauthor{\bsnm{{Tappin}}, \binits{S.J.}}:
\byear{1986},
\batitle{{Interplanetary scintillation and plasma density}}.
\bjtitle{\pss}
\bvolume{34},
\bfpage{93}.
\doiurl{10.1016/0032-0633(86)90106-6}.
\end{barticle}
\endbibitem

\bibitem[\protect\citeauthoryear{{Tapping}}{2013}]{tapping:13a}
\begin{barticle}
\bauthor{\bsnm{{Tapping}}, \binits{K.F.}}:
\byear{2013},
\batitle{The 10.7 cm solar radio flux ($f_{10.7}$)}.
\bjtitle{Space Weather}
\bvolume{11},
\bfpage{1}.
\doiurl{10.1002/swe.20064}.
\end{barticle}
\endbibitem

\bibitem[\protect\citeauthoryear{{Tokumaru}, {Fujiki}, and
  {Iju}}{2015}]{tokumaru_etal:15a}
\begin{barticle}
\bauthor{\bsnm{{Tokumaru}}, \binits{M.}},
\bauthor{\bsnm{{Fujiki}}, \binits{K.}},
\bauthor{\bsnm{{Iju}}, \binits{T.}}:
\byear{2015},
\batitle{{North-south asymmetry in global distribution of the solar wind speed
  during {1985--2013}}}.
\bjtitle{J. Geophys. Res. (Space Phys.)}
\bvolume{120}.
\doiurl{10.1002/2014JA020765}.
\end{barticle}
\endbibitem

\bibitem[\protect\citeauthoryear{{Tokumaru}, {Kojima}, and
  {Fujiki}}{2010}]{tokumaru_etal:10a}
\begin{barticle}
\bauthor{\bsnm{{Tokumaru}}, \binits{M.}},
\bauthor{\bsnm{{Kojima}}, \binits{M.}},
\bauthor{\bsnm{{Fujiki}}, \binits{K.}}:
\byear{2010},
\batitle{{Solar cycle evolution of the solar wind speed distribution from 1985
  to 2008}}.
\bjtitle{\jgr}
\bvolume{115},
\bfpage{A04102}.
\doiurl{10.1029/2009JA014628}.
\end{barticle}
\endbibitem

\bibitem[\protect\citeauthoryear{{Tokumaru}, {Kojima}, and
  {Fujiki}}{2012}]{tokumaru_etal:12b}
\begin{barticle}
\bauthor{\bsnm{{Tokumaru}}, \binits{M.}},
\bauthor{\bsnm{{Kojima}}, \binits{M.}},
\bauthor{\bsnm{{Fujiki}}, \binits{K.}}:
\byear{2012},
\batitle{{Long-term evolution in the global distribution of solar wind speed
  and density fluctuations during {1997--2009}}}.
\bjtitle{J. Geophys. Res. (Space Phys.)}
\bvolume{117},
\bfpage{6108}.
\doiurl{10.1029/2011JA017379}.
\end{barticle}
\endbibitem

\bibitem[\protect\citeauthoryear{{Tokumaru}
  \textit{et~al.}}{2000}]{tokumaru_etal:00a}
\begin{barticle}
\bauthor{\bsnm{{Tokumaru}}, \binits{M.}},
\bauthor{\bsnm{{Kojima}}, \binits{M.}},
\bauthor{\bsnm{{Ishida}}, \binits{Y.}},
\bauthor{\bsnm{{Yokobe}}, \binits{A.}},
\bauthor{\bsnm{{Ohmi}}, \binits{T.}}:
\byear{2000},
\batitle{{Large-scale structure of solar wind turbulence near solar activity
  minimum}}.
\bjtitle{Adv. Space Res.}
\bvolume{25},
\bfpage{1943}.
\doiurl{10.1016/S0273-1177(99)00630-4}.
\end{barticle}
\endbibitem

\bibitem[\protect\citeauthoryear{{Tokumaru}
  \textit{et~al.}}{2007}]{tokumaru_etal:07a}
\begin{barticle}
\bauthor{\bsnm{{Tokumaru}}, \binits{M.}},
\bauthor{\bsnm{{Kojima}}, \binits{M.}},
\bauthor{\bsnm{{Fujiki}}, \binits{K.}},
\bauthor{\bsnm{{Yamashita}}, \binits{M.}},
\bauthor{\bsnm{{Jackson}}, \binits{B.V.}}:
\byear{2007},
\batitle{{The source and propagation of the interplanetary disturbance
  associated with the full-halo coronal mass ejection on 28 October 2003}}.
\bjtitle{J. Geophys. Res. (Space Phys.)}
\bvolume{112},
\bfpage{5106}.
\doiurl{10.1029/2006JA012043}.
\end{barticle}
\endbibitem

\bibitem[\protect\citeauthoryear{{van der Holst}
  \textit{et~al.}}{2010}]{vanderHolst_etal:10a}
\begin{barticle}
\bauthor{\bsnm{{van der Holst}}, \binits{B.}},
\bauthor{\bsnm{{Manchester}}, \binits{W.B.} \bsuffix{IV}},
\bauthor{\bsnm{{Frazin}}, \binits{R.A.}},
\bauthor{\bsnm{{V{\'a}squez}}, \binits{A.M.}},
\bauthor{\bsnm{{T{\'o}th}}, \binits{G.}},
\bauthor{\bsnm{{Gombosi}}, \binits{T.I.}}:
\byear{2010},
\batitle{{A data-driven, two-temperature solar wind model with Alfv{\'e}n
  waves}}.
\bjtitle{\apj}
\bvolume{725},
\bfpage{1373}.
\doiurl{10.1088/0004-637X/725/1/1373}.
\end{barticle}
\endbibitem

\bibitem[\protect\citeauthoryear{{Washimi}
  \textit{et~al.}}{2011}]{washimi_etal:11a}
\begin{barticle}
\bauthor{\bsnm{{Washimi}}, \binits{H.}},
\bauthor{\bsnm{{Zank}}, \binits{G.P.}},
\bauthor{\bsnm{{Hu}}, \binits{Q.}},
\bauthor{\bsnm{{Tanaka}}, \binits{T.}},
\bauthor{\bsnm{{Munakata}}, \binits{K.}},
\bauthor{\bsnm{{Shinagawa}}, \binits{H.}}:
\byear{2011},
\batitle{{Realistic and time-varying outer heliospheric modelling}}.
\bjtitle{\mnras}
\bvolume{416},
\bfpage{1475}.
\doiurl{10.1111/j.1365-2966.2011.19144.x}.
\end{barticle}
\endbibitem

\bibitem[\protect\citeauthoryear{{Zirnstein}
  \textit{et~al.}}{2013}]{zirnstein_etal:13a}
\begin{barticle}
\bauthor{\bsnm{{Zirnstein}}, \binits{E.J.}},
\bauthor{\bsnm{{Heerikhuisen}}, \binits{J.}},
\bauthor{\bsnm{{McComas}}, \binits{D.J.}},
\bauthor{\bsnm{{Schwadron}}, \binits{N.A.}}:
\byear{2013},
\batitle{{Simulating the Compton-Getting effect for hydrogen flux measurements:
  Implications for IBEX-Hi and -Lo observations}}.
\bjtitle{\apj}
\bvolume{778},
\bfpage{112}.
\doiurl{10.1088/0004-637X/778/2/112}.
\end{barticle}
\endbibitem

\end{thebibliography}
%
% Without BibTeX 
% \begin{thebibliography}{}
% \bibitem[\protect\citeauthoryear{Author}{Year}]{key}
%   <bibliographical entry>
%
% \bibitem[\protect\citeauthoryear{}{}]{}
%   
%  
% \end{thebibliography}

\end{article} 
\end{document}